\documentclass[numberedappendix,onecolumn]{emulateapj}

\usepackage{natbib}
\usepackage{rotating, bm}

\renewcommand{\vec}[1]{\bm{#1}}
\newcommand{\La}{Ly\,$\alpha$}
\newcommand{\Lb}{Ly\,$\beta$}

\newcommand{\Ha}{H$\alpha$}
\newcommand{\pruh}[1]{\overline{#1}}
\newcommand{\threej}[6]{\left(\begin{array}{ccc}#1&#2&#3\\#4&#5&#6\\\end{array}\right)}

\shorttitle{Scattering Polarization of Hydrogen Lines}
\shortauthors{Ji\v{r}\'{\i} \v{S}t\v{e}p\'an and Javier Trujillo Bueno}

\begin{document}

\title{Scattering Polarization of Hydrogen Lines in Weakly Magnetized Stellar Atmospheres
I. Formulation and Application to Isothermal Models.}

\author{Ji\v{r}\'{\i} \v{S}t\v{e}p\'an\altaffilmark{\dag} and Javier Trujillo Bueno\altaffilmark{*}}

\affil{Instituto de Astrof\'{\i}sica de Canarias, V\'\i a L\'actea s/n, E-38205 La Laguna, Tenerife, Spain}
\affil{Departamento de Astrof\'{\i}sica, Universidad de La Laguna (ULL), E-38206 La Laguna, Tenerife, Spain}

\altaffiltext{\dag}{Associate Scientist at Astronomical Institute ASCR, v.v.i., Ond\v{r}ejov, Czech Republic}
\altaffiltext{*}{Consejo Superior de Investigaciones Cient\'{\i}ficas (Spain)}

\email{stepan@iac.es, jtb@iac.es}

\begin{abstract}
Although the spectral lines of hydrogen contain valuable information on the physical properties of a variety of astrophysical plasmas, including the upper solar chromosphere, relatively little is known about their scattering polarization signals whose modification via the Hanle effect may be exploited for magnetic field diagnostics. Here we report on a basic theoretical investigation of the linear polarization produced by scattering processes and the Hanle effect in Ly$\alpha$, Ly$\beta$ and H$\alpha$ taking into account multilevel radiative transfer effects in an isothermal stellar atmosphere model, the fine-structure of the hydrogen levels, as well as the impact of collisions with electrons and protons. The main aim of this first paper is to elucidate the key physical mechanisms that control the emergent fractional linear polarization in the three lines, as well as its sensitivity to the perturbers density and to the strength and structure of micro-structured and deterministic magnetic fields. To this end, we apply an efficient radiative transfer code we have developed for performing numerical simulations of the Hanle effect in multilevel systems with overlapping line transitions. For low density plasmas such as that of the upper solar chromosphere collisional depolarization is caused mainly by collisional transitions between the fine-structure levels of $n=3$, so that it is virtually insignificant for Ly$\alpha$ but important for Ly$\beta$ and H$\alpha$. We show the impact of the Hanle effect on the three lines taking into account the radiative transfer coupling between the different hydrogen line transitions. For example, we demonstrate that the linear polarization profile of the H$\alpha$ line is sensitive to the presence of magnetic field gradients in the line core formation region and that in solar-like chromospheres selective absorption of polarization components does not play any significant role on the emergent scattering polarization.
\end{abstract}



\section{Introduction}
\label{sec:intro}

This paper is the first of a series aimed at elucidating the physics and the diagnostic potential of the linear polarization produced by scattering processes in some hydrogen lines formed in stellar atmospheres (i.e., \Ha, \La\/ and \Lb), taking into account their sensitivity to magnetic fields via the Hanle effect. This is of particular interest for the exploration of magnetic fields in the outer solar atmosphere \citep[e.g.,][]{jtb05ESASP}, but also for facilitating the understanding of the spectral line polarization observed in other astrophysical systems \citep[e.g.,][]{harrington07} and the development of suitable polarized radiation diagnostics.

The true physical origin of the scattering line polarization is the presence of atomic level polarization (population imbalances and quantum coherences between the sublevels of degenerate atomic levels), caused by the absorption of anisotropic radiation \citep[e.g.,][]{ll04}. Typically, in weakly magnetized atmospheres anisotropic radiation pumping produces atomic level alignment (i.e., the populations of sublevels with different values of $|M|$ are unequal, with $M$ the magnetic quantum number) and the ensuing selective emission and/or selective absorption of polarization components give rise to linear polarization in the emergent spectral line radiation. In principle, atomic level orientation may also be induced (i.e., an atomic excitation such that the sublevels with $M>0$ are differently populated from those with $M<0$), but since its production requires very especial circumstances (e.g., a significant net circular polarization in the incident radiation) here we assume that the atomic levels are aligned, but not oriented.

In the atmospheres of the Sun and of other stars the hydrogen lines considered here are optically thick. Therefore, in this investigation we take into account radiative transfer effects in given atmospheric models. The numerical solution is obtained by solving jointly the statistical equilibrium equations for the multipolar components of the atomic density matrix corresponding to each level of total angular momentum $j$ and the Stokes vector transfer equation for each of the allowed transitions in the multilevel atomic model under consideration. This so-called non-LTE problem of the 2nd kind \citep{ll04} can be solved through the application of very efficient iterative methods and accurate formal solvers of the Stokes-vector transfer equation \citep{jtb99spw,jtb03}, as shown by \citet{mansosainz03,mansosainz10} for the Ca\,{\sc ii} IR triplet and by \citet{stepan07protons} and \citet{stepanjtb10asym} for the \Ha\/ line.

\begin{figure}
\epsscale{0.5}
\plotone{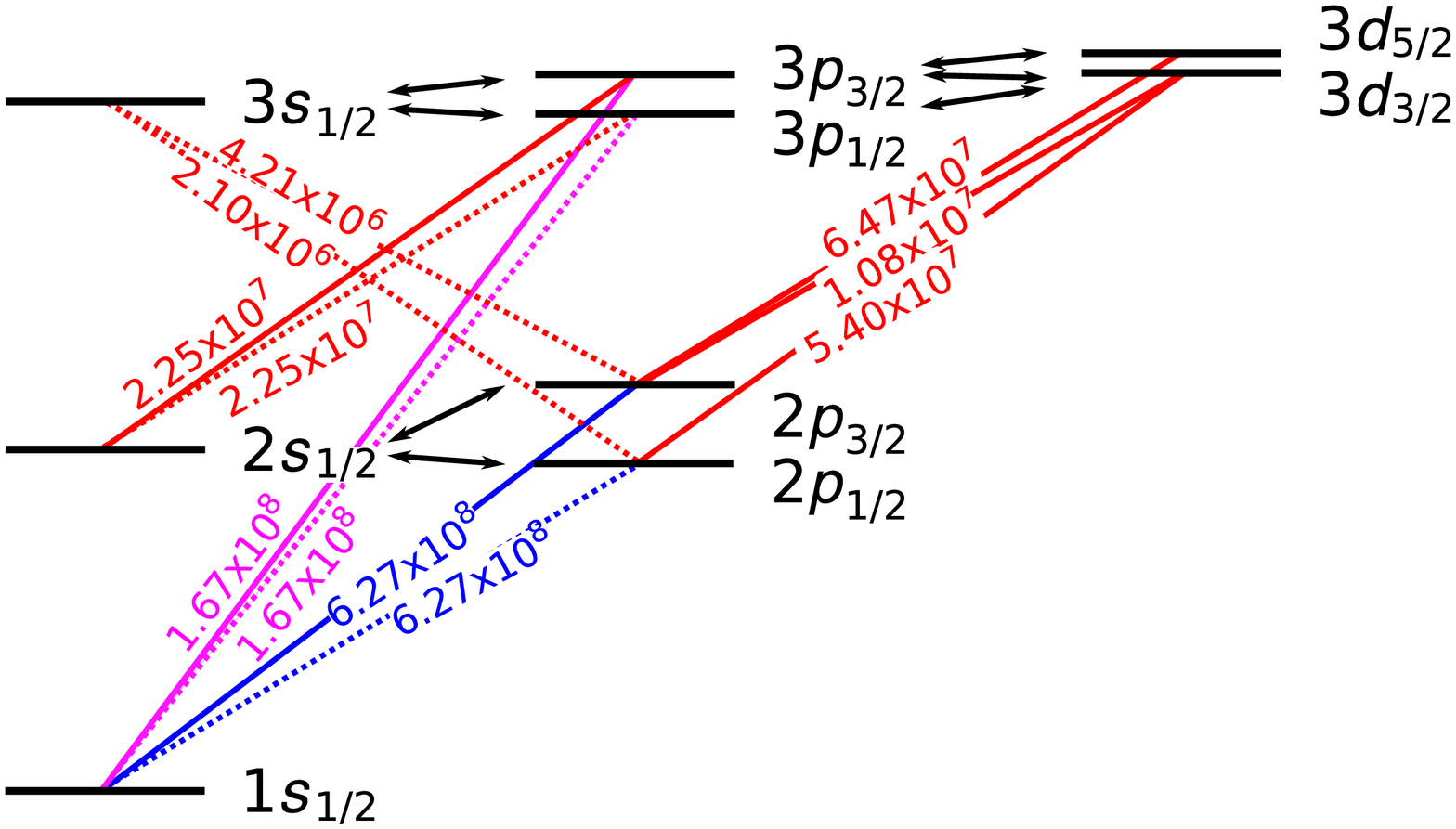}
\epsscale{1.0}
\caption{
Grotrian diagram of a three $n$-level hydrogen model atom (not in scale). The components of \La, \Lb, and \Ha\/ are indicated along with their experimental Einstein $A_{ul}$ coefficients (in s$^{-1}$). The solid lines correspond to the transitions involving a polarizable upper level. Such transitions play a dominant role in the scattering polarization problem of the above mentioned lines. The arrows indicate the collisional dipolar transitions, which play a depolarizing role.
}
\label{fig:grd}
\end{figure}

The problem of scattering polarization in hydrogen lines has attracted the attention of various researchers before, but concerning mainly the diagnostic problem of solar prominences. For example, \citet{landi87a} used a simplified radiative transfer model to estimate the expected linear polarization in \Ha\/ and H$\beta$ in quiescent prominences having a non-negligible optical thickness in \Ha. The authors assumed that the $I$-component of the source function corresponding to each hydrogen line is constant within the prominence, which was schematized as an ``infinitely sharp slab'' standing vertically over the solar surface. The solution of the radiative transfer equation for each hydrogen line transition, assuming that the ensuing source function is constant, allowed them to compute the spectral line intensity at each height within the slab and the radiative rates that enter the statistical equilibrium equations for the multipolar components of the atomic density matrix. The numerical solution of the resulting linear set of equations allowed them to obtain a zero-order estimate of the $I$, $Q$ and $U$ emissivities and to compute the ensuing emergent Stokes profiles.

Concerning the Lyman lines, earlier investigations considered always {\em the optically thin} case of the solar corona observed off-the-limb at large distances above it, either neglecting the effects of integration along the line of sight \citep{bommier78} or taking them into account assuming given coronal magnetic field models \citep{fineschi93,raouafi09,derouich10,khan11}. Another interesting optically-thin investigation is that of \citet{casini06} who expanded the work of \citet{casini05} and of \citet{favati87} by considering the effects of turbulent microscopic electric fields on the scattering polarization of hydrogen lines. Their work was based on the quasi-static approximation for describing the interaction of charged particles with the radiating hydrogen atoms, which might not be suitable for quantifying the significance of possible polarization signatures caused by random electric fields in the low density plasma of solar chromospheric structures \citep[e.g.,][]{stehle90}. 

When the line of sight (LOS) points to the solar disk the observed spectral line radiation in the Lyman lines does not originate in the corona, but in the optically-thick plasma of the solar transition region (or in that of a filament, if such a plasma structure happens to be located along the LOS). Therefore, the possibility of having measurable scattering polarization in the Lyman lines when doing on-disk observations is of great diagnostic interest, because of its sensitivity to the Hanle effect in the solar transition plasma. In their paper on the scientific case for spectropolarimetry from space \cite{jtb05ESASP} argued that {\em on-disk} observations in the strongest lines of the Lyman series should show scattering polarization signals, even in forward scattering geometry at the solar disk center when in the presence of an inclined magnetic field. That tentative conclusion was based on a simplified model, but in a forthcoming letter and in paper-2 of this series we will show through radiative transfer calculations in semi-empirical and hydrodynamical models of the solar atmosphere that measurable scattering polarization in the Lyman lines is indeed expected \citep[see][for a brief overview]{jtb11spw6}.

The radiative transfer problem of scattering polarization and the Hanle effect in hydrogen lines is, however, rather complicated and it has not been systematically studied before. Solving this type of radiative transfer problems requires finding the self-consistent solution of the atomic level polarization produced by radiatively induced population imbalances and quantum coherences, taking into account the Hanle effect in multilevel atomic systems with overlapping line transitions. To this end, we have developed and applied the very efficient numerical methods outlined in Appendix~\ref{app:methods}, with which we can investigate a variety of interesting radiative transfer problems in solar and stellar physics. In this first paper we focus on the case of an exponentially stratified isothermal model atmosphere, an apparently simple but very suitable model to elucidate the physics of formation of hydrogen lines in an optically thick medium as the extended solar atmosphere. It is interesting to note that the bulk of the solar chromosphere is essentially isothermal because the dissipation of mechanical energy there is taken up by latent heat of ionization, and is rapidly lost by radiation \citep{judge06}. Therefore, for some spectral lines (e.g., the solar \Ha\/ line) the isothermal atmosphere model is not as unreasonable as one might think at first sight. Moreover, it might also be a reasonable choice for a first estimation of the scattering polarization amplitudes of the Lyman lines in filaments embedded in the solar corona. However, the isothermal atmosphere model is certainly unsuitable for predicting the linear polarization signals of \La\/ and \Lb\/ produced by scattering processes in the solar transition region plasma itself. As mentioned above, the aim of this first paper is not to make predictions on the scattering polarization amplitudes of the hydrogen lines, but to gain physical insight to facilitate the understanding of a complicated radiative transfer problem of great diagnostic potential.

At present, all the above-mentioned physical ingredients can only be taken into account within the framework of the quantum theory of spectral line polarization \citep[e.g.,][]{ll04}, which treats the scattering line polarization phenomenon as the temporal succession of 1st-order absorption and re-emission processes, interpreted as statistically independent events. This complete redistribution (CRD) approximation is suitable for subordinate lines like \Ha\/ \citep[see][]{stepanjtb10asym} and for estimating the line-center scattering polarization amplitudes of \La\/ and \Lb\/, but it cannot be used for modeling the wings of the scattering polarization profiles of the same lines \citep[e.g.,][and more references therein]{sampoorna10b}.

This paper is organized as follows. After describing in \S\ref{sec:models} the atmospheric and atomic model, we discuss in \S\ref{sec:equation} the transfer equation for the Stokes parameters and the statistical equilibrium equations for the elements of the atomic density matrix. Section \ref{sec:non-mag} discusses in great detail our results for the unmagnetized reference case, with emphasis on the observational signatures caused by the radiatively induced polarization of the hydrogen levels. Similar discussions can be found in \S\ref{sec:iso-uni-mag} for the magnetized case, distinguishing between micro-structured and deterministic magnetic fields. Our conclusions are summarized in \S\ref{sec:concl}, while three appendices provide information on the numerical method of solution, on the role of collisions and on the equations that correspond to the micro-structured magnetic field case.


\section{The atmospheric and the atomic model}
\label{sec:models}

\begin{figure}
\begin{center}$
\begin{array}{ccc}
\includegraphics[width=2.3in]{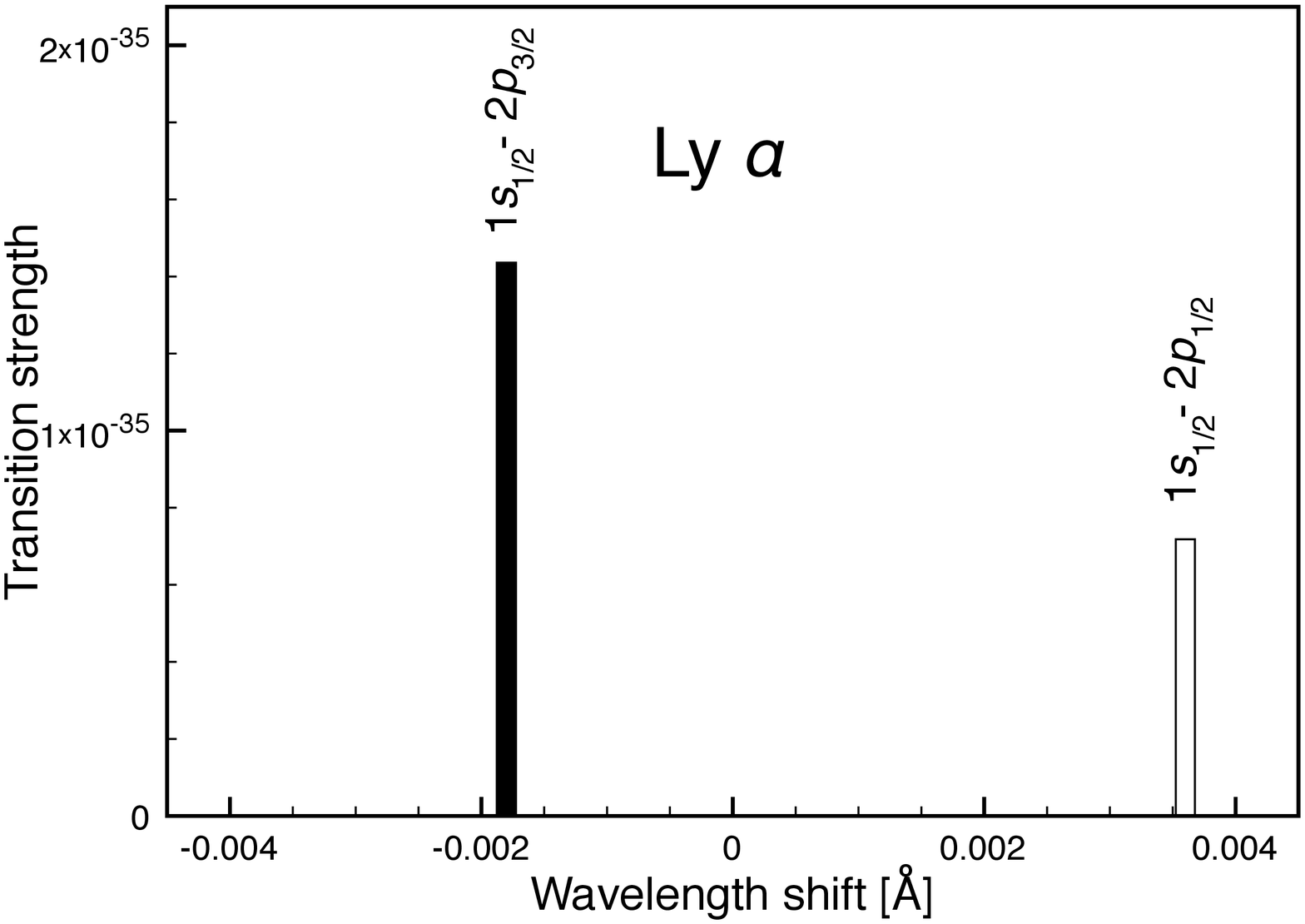} &
\includegraphics[width=2.3in]{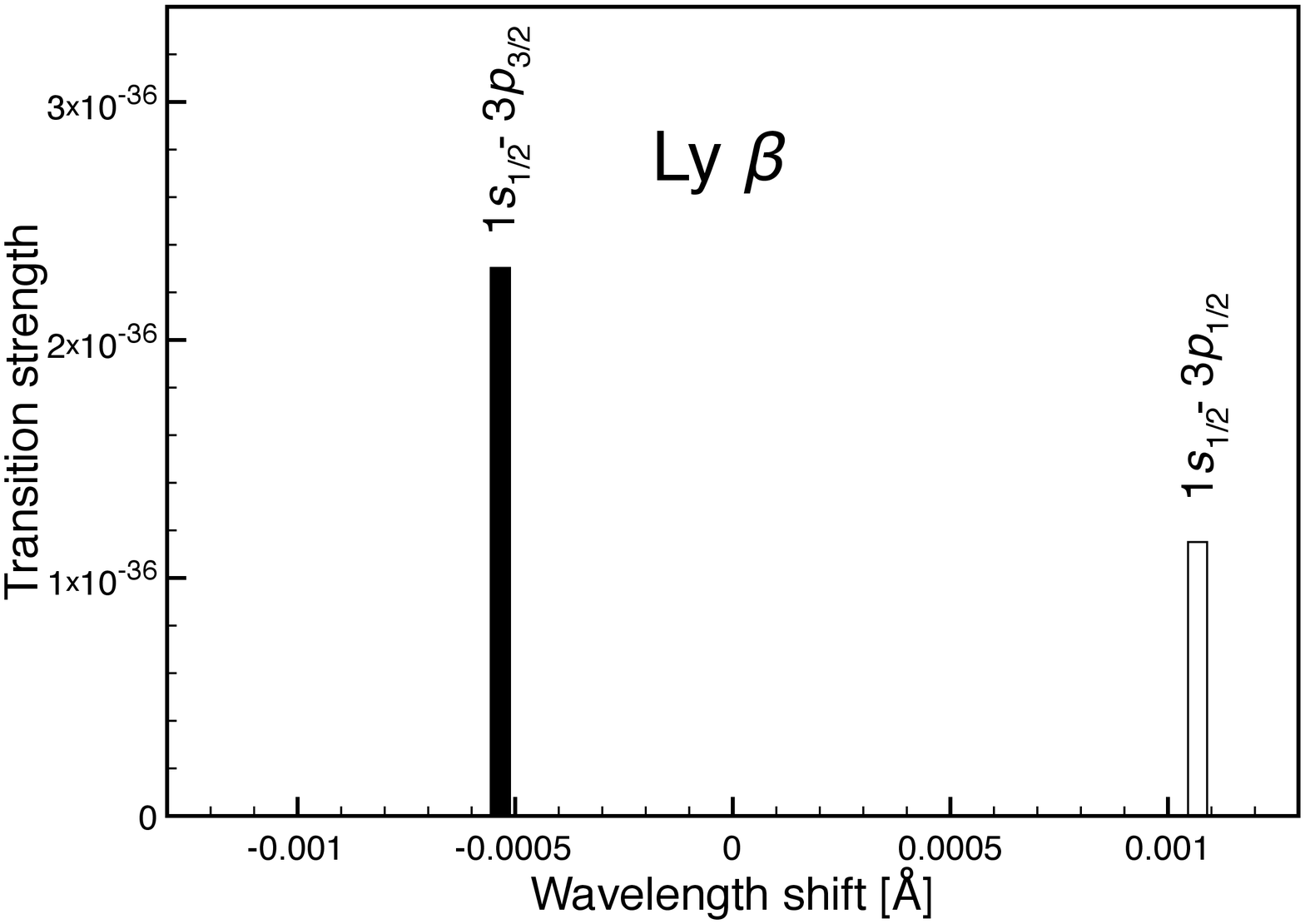} &
\includegraphics[width=2.3in]{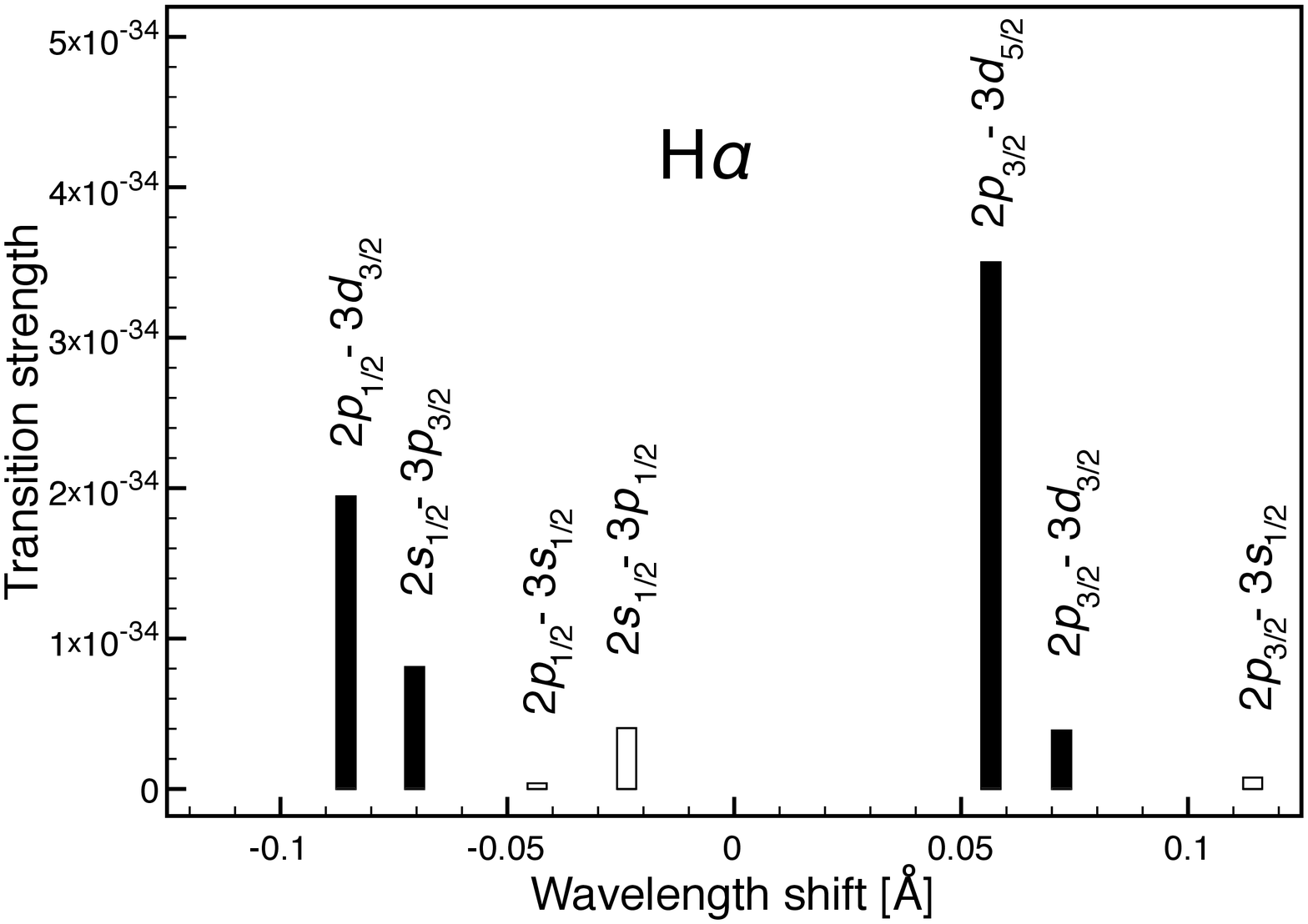}
\end{array}$
\end{center}
\caption{
Displacements of the line components from the line central wavelength for \La\/ ({\it left}), \Lb\/ ({\it middle}), and \Ha\/ ({\it right}). The full bars indicate the transitions with a polarizable upper level (playing a major active role in the line polarization) while the empty bars correspond to  transitions with an unpolarizable upper level.
}
\label{fig:lst}
\end{figure}

\begin{figure*}
\plottwo{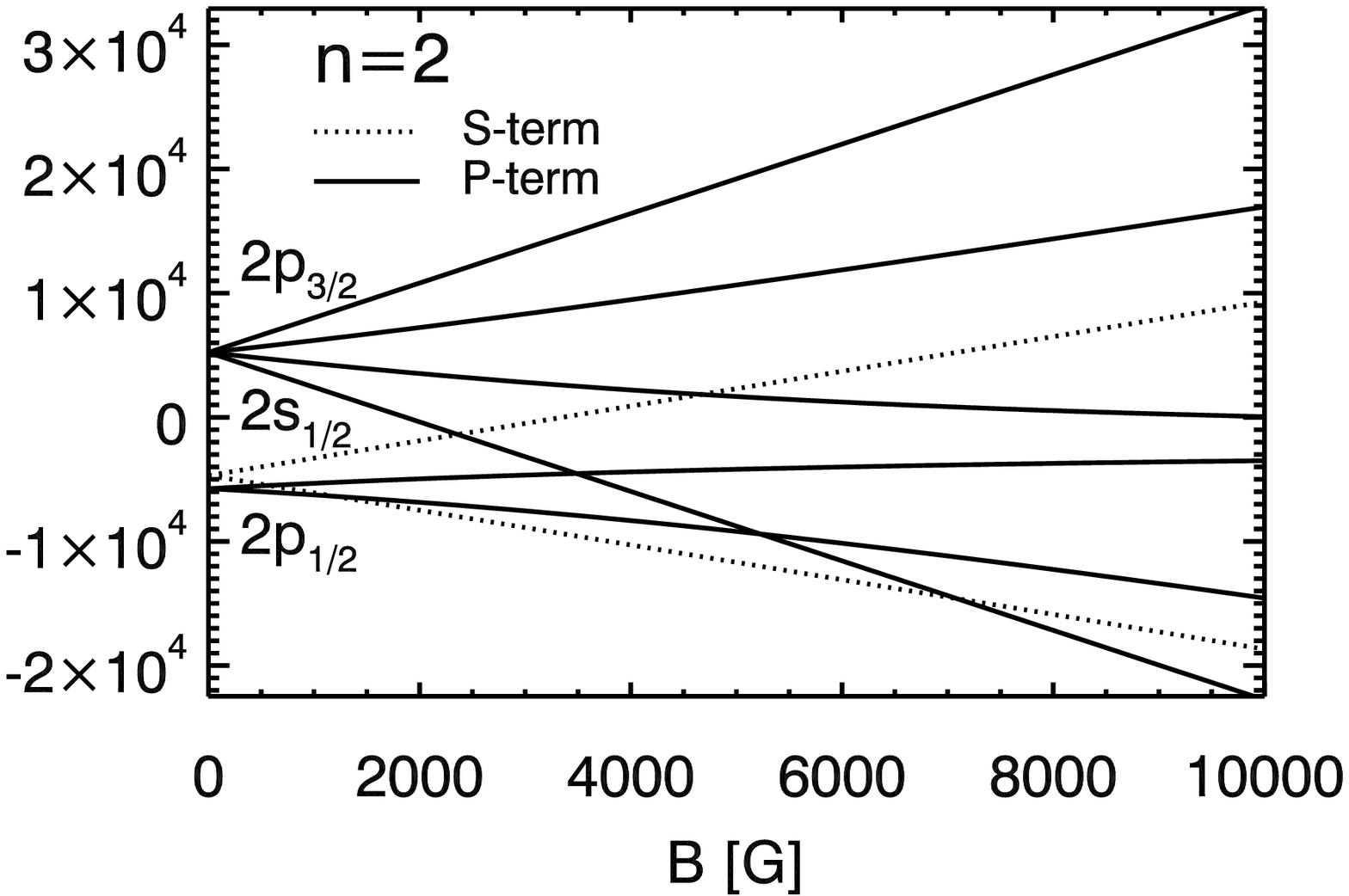}{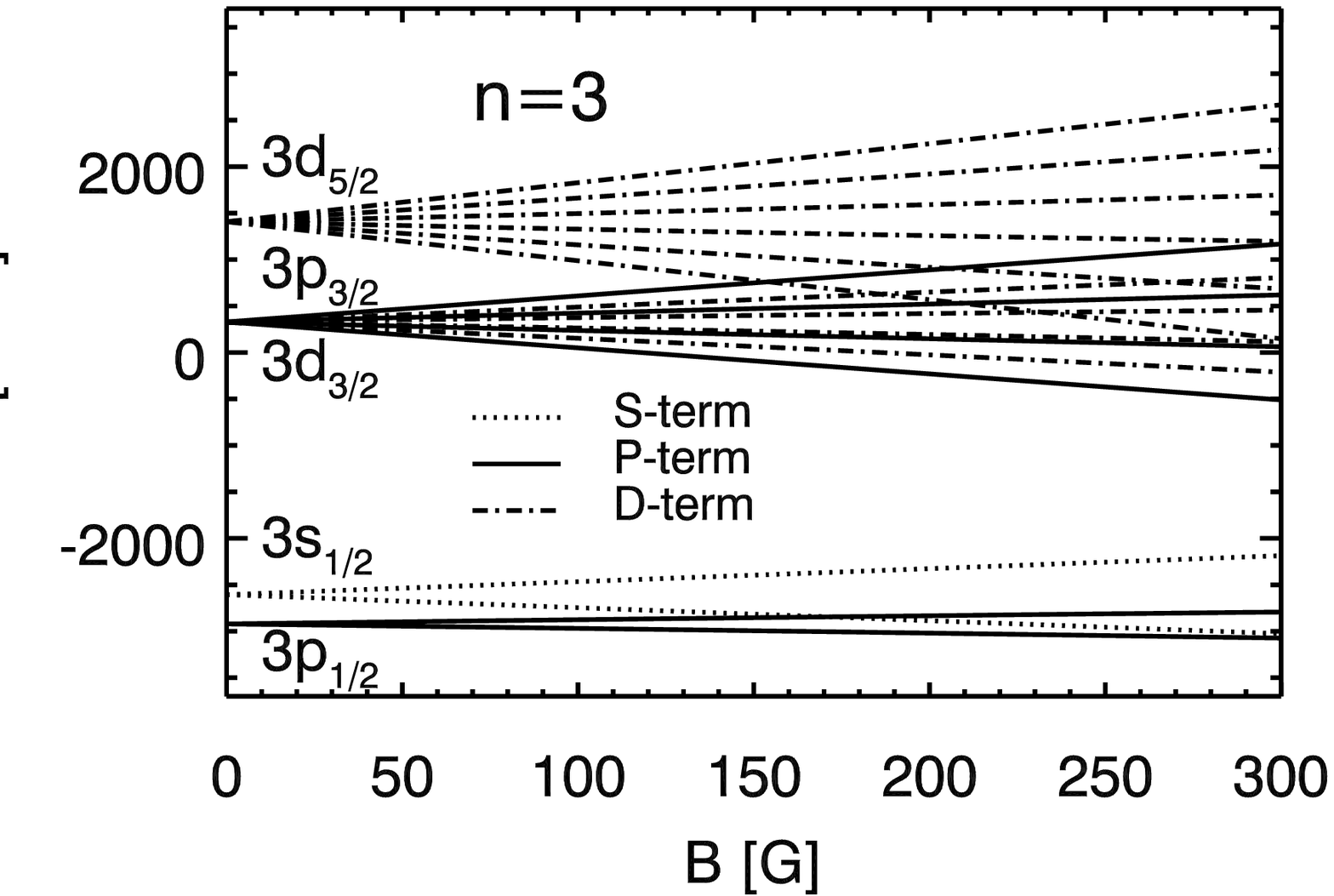}
\caption{
The energy levels of $n=2$ ({\it left}) and $n=3$ ({\it right}) as a function of the magnetic field strength. Note that the natural width of the $2p_{1/2}$ and $2p_{3/2}$ levels, $\Gamma=A_{ul}/2\pi$, is $100\,$MHz and the largest natural widths among the $n=3$ sublevels are $\Gamma=27$\,MHz (those of $3p_{1/3}$ and $3p_{3/2}$). In both cases, the natural width is much smaller than separation of the relevant levels. This remains valid unless the perturber's density is very large ($N_{\rm pert}\gtrsim 10^{13}\,{\rm cm^{-3}}$).
}
\label{fig:split}
\end{figure*}

We consider an isothermal, one-dimensional, plane-parallel, semi-infinite, exponentially stratified solar model atmosphere. The geometrical extension of the model atmosphere is 2500\,km and its total optical thickness is $4.5\times 10^{10}$ at the center of the \La\/ line, $7.1\times 10^{9}$ at the center of \Lb\/ and $1.2\times 10^{7}$ at the \Ha\/ line center. The kinetic temperature is $T=10^4$\,K. For simplicity, the calculations are carried out using constant values of the electron and proton densities (denoted respectively by $n_{\rm e}$ and $n_{\rm p}$). Our main aim is to study in depth the non-LTE problem of the formation, in this apparently simple stellar atmospheric model, of the linear polarization profiles of the \La, \Lb, and \Ha\/ lines of hydrogen in the absence and in the presence of  ``weak" magnetic fields (e.g., with magnetic strengths $B{\lesssim}100$ G). 

Under such circumstances the linear polarization of \La, \Lb\/, and \Ha\/ is fully dominated by atomic level polarization and the Hanle effect, with no significant impact of the transverse Zeeman effect, because the wavelength shifts (produced by the Zeeman effect) between the $\pi$ ($\Delta{M}=M_u-M_l=0$) and ${\sigma}_{b,r}$ ($\Delta{M}={\pm}1$) components ($M$ being the magnetic quantum number) are only a very small fraction of the width of the profiles. In this first paper we include only the thermal broadening of the spectral lines and we do not consider any background continuum opacity. All the ingredients needed for radiative transfer modeling of spectro-polarimetric observations (e.g., the Stark broadening effect) will be included in our forthcoming papers.

Figure~\ref{fig:grd} shows the hydrogen atomic model we have used in this study, indicating with the usual notation the $n\ell j$ values of each level (with $n$ being the principal quantum number, $\ell$ the orbital quantum number and $j$ the total angular momentum). The model atom includes the 9 fine-structure levels of the first three $n$-levels of neutral hydrogen, between which 11 allowed radiative transitions take place: 2 for \La, 2 for \Lb\/ and 7 for \Ha. The energies of these levels and the Einstein $A_{ul}$ coefficients of the allowed transitions have been taken from the NIST online database\footnote{\url{http://www.nist.gov/physlab/data/asd.cfm}}. We have checked that similar results for the fractional linear polarization of these lines are obtained when adding the fine-structure levels of the $n=4$ level. Fig.~\ref{fig:lst} shows the strengths and wavelength positions of the \La\/ (left panel), \Lb\/ (middle panel) and \Ha\/ (right panel) components. Although in laboratory experiments such individual fine-structure components can be resolved using Doppler-free saturation spectroscopy \citep{schawlow82}, they are however blended in stellar atmospheres. The hyperfine structure of the hydrogen levels is expected to play a minor role on the scattering polarization (e.g., see \citet{bommier82} for the case of \La\/ and \citet{bommier86a} for the case of \Ha).

Table~\ref{tab:critical} gives the Land\'e factors of the levels of Fig.~\ref{fig:grd} that can carry atomic alignment (i.e., those with $j>1/2$), calculated using the LS coupling approximation. The same table shows also the Hanle critical field ($B_{H}$, in gauss) of each of such $j$-levels (i.e., the magnetic strength for which the level's Zeeman splitting equals its natural width) without accounting for the quenching effect of collisions. Note that
\begin{equation}
B_{H}\,{=}\,{{1.137{\times}10^{-7}}\over{t_{\rm life}g_j}},
\label{eq:bhcritical}
\end{equation}
where $t_{\rm life}$ and $g_j$ are, respectively, the lifetime (in seconds) and the Land\'e factor of the $j$ level under consideration. In this paper we neglect quantum coherences between the magnetic sublevels pertaining to different $j$-levels. For the zero-field case this should be a reasonable approximation because the energy separation between such levels is larger than their natural width, even when we take into account that the lifetime of the levels is reduced by collisions with protons and electrons (hereafter the perturbers) at plasma densities not very much larger than those of the quiet solar chromosphere. In the presence of magnetic fields the ensuing level crossings and repulsions may have a significant impact on the above-mentioned $j$-$j^{'}$ interferences. Fig.~\ref{fig:split} shows the energy splitting of each $j$-level due to the presence of a magnetic field, both for the $n=2$ levels (left panel) and for the $n=3$ levels (right panel). It is very important to point out that the selection rule $\Delta\ell={\pm}1$ for radiative transitions inhibits couplings between levels $n\ell j$ and $n(\ell\pm 1)j'$. Therefore, we only have to worry about the possibility of level crossings and repulsions between levels $n\ell j$ and $n\ell j'$ and between levels $n\ell j$ and $n(\ell\pm 2)j'$. As seen in the figure, it seems reasonable to neglect them for the magnetic strengths considered in this paper (i.e., $B\lesssim 100$ G).

\begin{deluxetable}{ccc}
\tablewidth{0pt}
\tablecaption{Critical field of the hydrogen levels.\label{tab:critical}}
\tablehead{\colhead{Level} & \colhead{Land\'e factor} & \colhead{$B_H$ [G]}}
\startdata
2$p_{3/2}$ & 1.33 & 53.5 \\
3$p_{3/2}$ & 1.33 & 16.2 \\
3$d_{3/2}$ & 0.80 & 9.2 \\
3$d_{5/2}$ & 1.20 & 6.1 \\
\enddata
\end{deluxetable}


\section{The equations}
\label{sec:equation}

The aim of this section is to provide a brief summary on the equations we have solved for the radiative transfer investigation of this paper. Information on the numerical methods of solution can be found in Appendix~\ref{app:methods}, while Appendix~\ref{app:microturb} describes the efficient strategy we apply for dealing with micro-structured magnetic fields when considering multilevel systems. 


\subsection{The multipolar components of the atomic density matrix}
\label{ssec:densm}

We quantify the excitation state of each $n\ell j$ level by means of the multipolar components of the atomic density matrix \citep{omont77}, denoted here with the symbol $\rho^K_Q(j)$ (with $K$ and $Q$ integer numbers such that $0{\le}K{\le}2j$ and $-K{\le}Q{\le}K$). The $\rho^K_Q$ elements with $Q=0$ are {\it real} numbers given by linear combinations of the populations, $\rho_j(M,M)$, of the various Zeeman sublevels $n\ell jM$ corresponding to the level of total angular momentum $j$. The total population of the atomic level is quantified by ${\sqrt{2j+1}}{\rho^0_0}$, while the population imbalances among the Zeeman sublevels are quantified by $\rho^K_0$ (e.g., $\rho^2_0(j=\frac 32)=\frac 12(\rho_j(\frac 32,\frac 32)+\rho_j(-\frac 32,-\frac 32)-\rho_j(\frac 12,\frac 12)-\rho_j(-\frac 12,-\frac 12))$). However, the $\rho^K_Q$ elements with $Q {\ne} 0$ are {\it complex} numbers given by linear combinations of the coherences, $\rho_j(M,M')$, between Zeeman sublevels whose magnetic quantum numbers differ by $Q$. In fact, since the density operator is Hermitian, we have that for each spherical statistical tensor component $\rho^K_Q$ with $Q>0$, there exists another component with $Q<0$ given by $\rho^K_{-Q}=(-1)^Q[\rho^K_{Q}]^*$, with `$*$' denoting the complex conjugate. In quiet regions of the solar atmosphere it is a good approximation to assume that the net circular polarization of the incident radiation is zero; therefore, the odd-$K$ $\rho^K_Q$ elements (i.e., the orientation components) vanish. Therefore, the 9-level hydrogen model of Fig.~\ref{fig:grd} requires 38 $\rho^K_Q$ elements to fully quantify the atomic excitation state. 

It is important to note that in this paper the $\rho^K_Q(j)$ elements are defined in a reference system whose $Z$ axis (i.e., the quantization axis of total angular momentum) is along the stellar radius vector (i.e., the vertical direction in the case of a plane-parallel atmosphere).


\subsection{The transfer equation for the Stokes parameters}

In general, the transfer of polarized radiation is described by the following vectorial transfer equation \citep[e.g.,][]{ll04}
\begin{equation}
\frac{d}{ds}\vec I=\vec\epsilon-\vec K\vec I\,,
\label{eq:rte}
\end{equation}
where $s$ is the geometrical distance along the ray under consideration, $\vec I=(I,Q,U,V)^{\rm T}$ is the wavelength-dependent Stokes parameters, $\vec\epsilon=(\epsilon_I,\epsilon_Q,\epsilon_U,\epsilon_V)^{\rm T}$ is the emission vector resulting from spontaneous emission events, and $\vec K$ is the $4\times 4$ propagation matrix whose coefficients are ${\eta_I}$ (the diagonal element), ${\eta_Q}, {\eta_U}, {\eta_V}$ (which account for dichroism) and ${\rho_Q},{\rho_U},{\rho_V}$ (which describe the anomalous dispersion).

The general expressions of the components of the emission vector $\vec\epsilon$ and of the propagation matrix ${\vec K}$ are very involved and they will not be written here\footnote{For the case of non-overlapping transitions see \citet{mansosainz10}, who studied the radiative transfer problem of scattering polarization and the Hanle effect in the IR triplet of ionized calcium considering multilevel atomic models.} \citep[see Eqs. 7.15 in][]{ll04}. In the Hanle effect regime considered in this paper they are given in terms of the local values of the $\rho^K_Q(j)$ density-matrix elements corresponding, respectively, to the upper ($j=j_u$) and lower ($j=j_l$) levels of the line transition under consideration. We point out that only the alignment components with $K=2$ contribute to $\epsilon_Q$, $\epsilon_U$, $\eta_Q$, $\eta_U$, $\rho_Q$ and $\rho_U$. The $\epsilon_V$, $\eta_V$ and $\rho_V$ expressions depend on the orientation components with $K=1$, but in our case the odd-$K$ $\rho^K_Q$ elements vanish because we assume that there is no net circular polarization in the incident radiation (see \S\ref{ssec:densm}). In solar-like atmospheres the expressions for $\epsilon_I$ and $\eta_I$ are dominated by the $\rho^0_0$ components (the overall population of the upper and lower levels, respectively), but they also depend on the $K=2$ components. The frequency dependence of all such radiative transfer coefficients is established by the Voigt ($\phi_{\nu}$) and Faraday-Voigt ($\psi_{\nu}$) profiles, whose dependence on the magnetic quantum number $M$ we have neglected here arguing that for $B{\lesssim}100$ G the Zeeman splittings of the hydrogen levels of Fig.~\ref{fig:grd} are much smaller than the thermal widths of the spectral lines under investigation. This is a very good approximation for calculating the linear polarization profiles of \La, \Lb\/ and \Ha, because in the quiet solar atmosphere the contribution of the transverse Zeeman effect is insignificant compared with that caused by atomic level alignment and its modification by the Hanle effect. As mentioned above $\rho^1_Q=0$, so that there is no contribution of atomic level orientation to $\epsilon_V$, $\eta_V$ and $\rho_V$. Therefore, the only contribution of atomic alignment to the emergent circular polarization comes from the (second-order) term $-{\rho_U}Q+{\rho_Q}U$ of the transfer equation for Stokes $V$, but such a contribution is negligible compared with that produced by the longitudinal Zeeman effect which is not considered in this paper. Likewise, the second-order terms $-{\rho_V}U+{\rho_U}V$ and $-{\rho_Q}V+{\rho_V}Q$ of the transfer equations for $Q$ and $U$, respectively, do not play any significant role in the weakly polarized atmosphere.

In our calculations of the linear polarization of hydrogen lines in the model atmosphere we always  solve the full transfer equation~(\ref{eq:rte}) via the DELOPAR method proposed by \cite{jtb03}, but it is useful to note that the following Eddington-Barbier approximate formula can be used to estimate the emergent fractional linear polarization at the center of a strong line transition \citep{jtb03}
\begin{equation}
\frac QI\,\approx\,{3\over{2\sqrt{2}}}(1-\mu^2)
[w_{j_uj_\ell}^{(2)}\,\sigma^2_0({j_u})\,-\,w^{(2)}_{j_lj_u}\,\sigma^2_0({j_l})]\,,
\label{eq:qieddb}
\end{equation}
where $\sigma^2_0=\rho^2_0/\rho^0_0$ must be evaluated at height in the model atmosphere where the line-center optical distance along the line of sight (LOS) is unity and $w^{(2)}_{jj'}$ is a numerical coefficient that depends only on the level's angular momentum values $j$ and $j'$ \citep[see Table~10.1 in][]{ll04}.

The first term in the square brackets of Eq.~(\ref{eq:qieddb}) is due to {\it selective emission} of polarization components, caused by the population imbalances of the upper level, while the second term is due to {\it selective absorption} of polarization components (or ``zero-field dichroism"), caused by the population imbalances of the lower level. It can be shown \citep[e.g.,][]{jtb99} that for resonance line transitions in a weakly anisotropic medium like that of the solar atmosphere the $\sigma^2_0(j_u)$ values of the above expression can be estimated from the following approximate expression
\begin{equation}
\sigma^2_0(j_u)\,{\approx}\,w^{(2)}_{j_uj_l}\,{{J^2_0}\over{J^0_0}}\,,
\label{eq:s20j20}
\end{equation}
where 
\begin{equation}
{{{J}^0_0}}=\int dx 
\oint \frac{d \vec\Omega}{4\pi}\,\phi_x{{I_{x \vec\Omega}}}\,,
\label{eq:j00}
\end{equation}
is the familiar mean radiation field intensity, $\bar J$, of the unpolarized non-LTE problem \citep[e.g.,][]{mihalas78} and 
\begin{equation}
{{{J}^2_0}}=\int dx 
\oint \frac{d \vec{\Omega}}{4\pi}\,
\frac{1}{2\sqrt{2}} \phi_x \left[(3\mu^2-1){{I_{x \vec{\Omega}}}}+
3(\mu^2-1){Q_{x \vec{\Omega}}}\right]\,,
\label{eq:j20}
\end{equation}
with $\phi_x$ being the absorption profile and $x$ the frequency distance from line center measured in units of the Doppler width.


\subsection{The statistical equilibrium equations}

The multipolar components of the atomic density matrix are govened by the following rate equations
\begin{equation}
{{d}\over{dt}}{\rho^K_Q(j)}=
\left[{{d}\over{dt}}{\rho^K_Q(j)}\right]_{\rm Hanle}
+
\left[{{d}\over{dt}}{\rho^K_Q(j)}\right]_{\rm Radiation}+
\left[{{d}\over{dt}}{\rho^K_Q(j)}\right]_{\rm Collisions}\;,
\label{eq:master}
\end{equation}
where the first two terms on the {\it rhs} are the contributions to the rate of change of ${\rho^K_Q(j)}$ due to the Hanle effect and to radiative transitions \citep[see Eq.~7.78 in][]{ll04}, while the last term is that due to collisions \citep[see Eq.~7.101 in][]{ll04}. In order to calculate such $\rho^K_Q$ unknowns at each grid point of the chosen stellar atmosphere model we have assumed statistical equilibrium, i.e.,
\begin{equation}
{{d}\over{dt}}{\rho^K_Q(j)}=0\;.
\end{equation}
We point out that since the resulting system of equations is not linearly independent one of the equations (e.g., the one for the ground-level population) must be substituted by the trace equation of the density matrix, which establishes the conservation of the number of particles
\begin{equation}
{\sum_{j_i}}\sqrt{2j_i+1}{\rho^0_0(j_i)}=1\;.
\end{equation}

Note that the term of Eq.~(\ref{eq:master}) due to radiative transitions includes {\it transfer} rates due to absorption ($T_A$), spontaneous emission ($T_E$) and stimulated emission ($T_S$) {\it from} other levels, and the {\it relaxation} rates due to absorption ($R_A$), spontaneous emission ($R_E$) and stimulated emission ($R_S$) {\it towards} other levels. The explicit expressions for all these transfer and relaxation rates can be found in \S7.2.a of \citet{ll04}. Of particular interest is the transfer rate $T_{\rm A}$ due to {\it absorption} from lower levels. The expression for $T_A$ is
\begin{eqnarray}
T_{\rm A}({\alpha}_lj_l;K_lQ_l \rightarrow {\alpha}j;KQ) =
  (2j_l+1) B({\alpha}_lj_l \rightarrow {\alpha}j)  
   \sum_{K_{\rm r} Q_{\rm r}} (-1)^{K_l+Q_l} \sqrt{3(2K+1)(2K_l+1)(2K_{\rm r}+1)}
   \nonumber \\
   {\times}\,\left\{\begin{array}{ccc}
	j & j_l & 1 \\
	j & j_l & 1 \\
	K & K_l & K_{\rm r} 
	\end{array}\right\} 
	   \threej{K}{K_l}{K_{\rm r}}{-Q}{Q_l}{-Q_{\rm r}}{J}^{K_{\rm r}}_{Q_{\rm r}}
	   (\nu_{\alpha j,\alpha_lj_l}),
\end{eqnarray}
where
\begin{equation}
{J}^{K_{\rm r}}_{Q_{\rm r}}(\nu_{\alpha j,\alpha_lj_l})\,=\,\int d\nu\,\phi_{ij}(\nu-\nu_{\alpha j,\alpha_lj_l})J^{K_{\rm r}}_{Q_{\rm r}}(\nu)\,,
\label{eq:jkqij}
\end{equation}
with $K_{\rm r}=0,1,2$ and $Q_{\rm r}=-K_{\rm r},...,K_{\rm r}$. Note that $J^{K_{\rm r}}_{Q_{\rm r}}(\nu)$ are the monochromatic radiation field tensors, defined as angle averages of the Stokes parameters multiplied by the geometrical irreducible tensors $\mathcal{T}^K_Q$ \citep[see Eq.~5.157 of][]{ll04}, while $J^{K_{\rm r}}_{Q_{\rm r}}(\nu_{\alpha j,\alpha_lj_l})$ are the laboratory frame expressions of the radiation field tensors, with $\phi_{ij}(\nu-\nu_{\alpha j,\alpha_lj_l})$ being the absorption profile of the allowed fine-structure transition under consideration \citep[see their explicit expressions in][]{jtb01}. As mentioned above, $J^0_0$ is the familar $\bar J$-quantity of the standard non-LTE problem. The $J^2_0$ tensor given by Eq.~(\ref{eq:j20}), which in solar-like atmospheres is dominated by the contribution of the Stokes $I$ parameter, quantifies whether the local illumination of the atomic system is preferentially vertical ($J^2_0>0$) or horizontal ($J^2_0<0$). The real and imaginary parts of the $J^2_Q$ tensors (with $Q={\pm 1},{\pm 2}$) quantify the breaking of the axial symmetry of the spectral line radiation through the complex azimuthal exponentials that appear inside the angular integrals. Obviously, a deterministic magnetic field inclined with respect to the symmetry axis of the pumping radiation is needed in order for $J^2_{\pm 1}$ and $J^2_{\pm 2}$ to be non-zero in a plane-parallel or spherically symmetric model atmosphere.


\section{The unmagnetized reference case}
\label{sec:non-mag}

Our three $n$-level hydrogen atomic model has two strong resonant lines (\La\/ and \Lb) and one subordinate line (\Ha). It is well known that coupling of these lines gives rise to effects that cannot be explained using a two-level approximation for the individual lines \citep[e.g.][]{jefferies68,mihalas78}. In particular, coupling of \Ha\/ and \Lb\/ through the common upper level $n=3$ affects the formation of these lines via degradation of the \Lb\/ photons into  \Ha\/ and \La\/ photons. Furthermore, each of the excited $n$-levels of hydrogen consists of several close-lying fine-structure levels. These levels are connected by permitted optical transitions with only few of the fine-structure levels of different $n'$. Since the fine-structure levels of each $n$ level are coupled together by the $\Delta n=0$ collisions, the formation of the lines depends on the rate at which these collisions are able to shuffle the optical electrons among the fine-structure levels of any given $n$ (see Appendix~\ref{sapp:depol}). This process leads to photon conversion among the transitions affecting the line source functions and, in the limiting case of strong collisions, the $n\ell j$ and $n\ell'j'$ levels behave as a single $n$-level. Moreover, these inelastic collisions connecting nearby $j$ levels tend to depolarize the atomic levels and, consequently, modify the scattering polarization of the emergent radiation \citep[cf.,][]{bommier86b,bommier91,sahal96}. As pointed out in Appendix~\ref{app:colls}, elastic collisions between the radiating hydrogen atoms themselves do not play any significant depolarizing role for hydrogen in the low density plasma of the solar chromosphere.

In this section, we discuss the role of several phenomena affecting the formation of the hydrogen lines in our non-magnetized model atmosphere. Special emphasis is given to the role of collisions with the ambient perturbers. Among other effects we discuss, is the fact that the atomic polarization of the $n=3$ levels, due to the anisotropy of the incident \Ha\/ radiation, has a noticeable impact on the \Lb\/ radiation, so that its anisotropy turns out to be significant in the layers where \Lb\/ is optically thick. We point out a possible influence of this effect on the linear polarization of the \Lb\/ wings.


\subsection{Line source functions}
\label{ssec:sourcef}

\begin{figure}
\epsscale{0.7}
\plotone{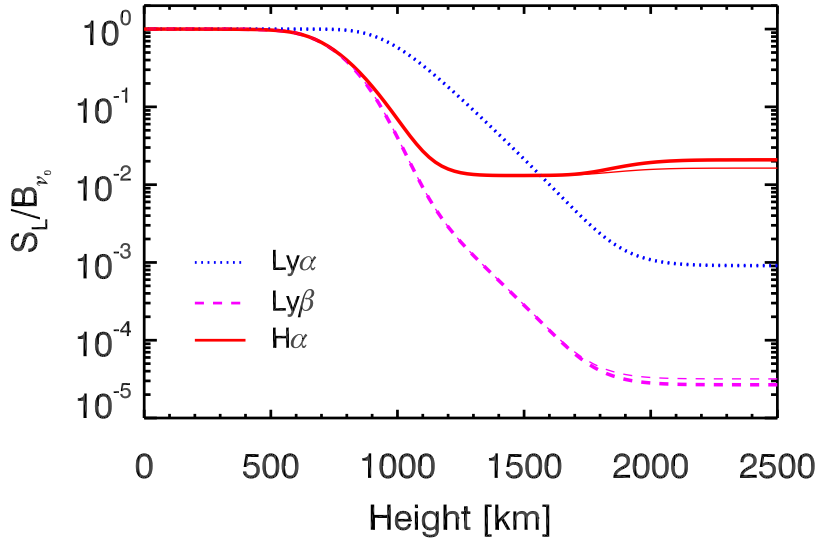}
\epsscale{1.0}
\caption{
The line source functions normalized to the corresponding Planck function. The perturber's density is $N_{\rm pert}=4\times 10^{10}\,{\rm cm}^{-3}$. {\it Thick lines}: all relevant collisional processes are taken into account. {\it Thin lines}: collisions between the fine-structure levels in $n=3$ are ignored.
}
\label{fig:sfpp}
\end{figure}

A detailed discussion of the formation of the unpolarized hydrogen \La---\Lb---\Ha\/ system in an isothermal atmosphere can be found in the literature \citep[e.g.,][]{jefferies68,mihalas78}. Here we briefly review the behavior of the line source functions. In addition, we consider the fine structure of the levels and we discuss the role of the $\Delta n=0$ collisional transitions on the line source functions.

Since the vast majority of the H\,{\sc i} atoms is in the ground state, the optical thickness of the atmosphere is much larger in the resonance lines than in the subordinate line. As a result, \Ha\/ is formed deeper in the model atmosphere and \La\/ and \Lb\/ are optically thick in the layers where \Ha\/ becomes optically thin. Since the levels $3p_{1/2}$ and  $3p_{3/2}$ are common upper levels of \Ha\/ and \Lb, there is an efficient mechanism of conversion of \Lb\/ photons into \Ha\/ photons. This systematic degradation of the \Lb\/ photons is responsible for equality of the \Ha\/ and \Lb\/ source functions (normalized to the Planck function) in the atmospheric layers where the branching ratio $A_{32}/(A_{31}+A_{32})$ is higher than the escape probability of the \Lb\/ photons (where $A_{nn'}$ is the Einstein coefficient of spontaneous emission between levels $n$ and $n'$.) Above such layers, the line source functions become uncoupled \citep{mihalas78}. In conclusion, \Lb\/ is not fully thermalized in relatively deep layers where its optical depth would suggest it. On the other hand, \La\/ is thermalized in such deep layers because the level $n=2$ is not drained by any subordinate line.

If the polarization of the atomic levels is taken into account, then $\epsilon_I$ and $\eta_I$ become angle-dependent \citep[e.g., Eqs.~(7.16) of][]{ll04}. However, if the atomic polarization is small ($|\rho^{K>0}_Q|\ll |\rho^0_0|$) this effect is of relatively small importance. Neglecting stimulated emission and the wavelength separation of the line components, we can obtain an approximate expression for the line intensity source function, $S_I$, in terms of the $K=0$ multipoles of the atomic density matrix,
\begin{eqnarray}
S_I\approx 
\frac{
\sum_{lu} A_{ul}{\sqrt{2j_u+1}}\rho^0_0(u)
}{
\sum_{lu} B_{lu}{\sqrt{2j_l+1}}\rho^0_0(l)
}\,,
\end{eqnarray}
where $u\equiv n_u\ell_u j_u$ and $l\equiv n_l\ell_l j_l$ denote the upper and the lower levels of the line transition,  respectively, while $A_{ul}$ and $B_{lu}$ are the Einstein coefficients for spontaneous emission and for absorption respectively. 

The thick lines in Fig.~\ref{fig:sfpp} show the self-consistently calculated line source functions of the three lines versus height in the model atmosphere. These source functions have been calculated taking into account all the $\Delta n{\ne}0$ and $\Delta n=0$ collisions at a uniform electron and proton density of $4\times 10^{10}\,{\rm cm^{-3}}$. For such a density value, which is typical of the upper chromosphere of the quiet Sun \citep[e.g.][]{fontenla93}, collisional depolarization of the $2p_{3/2}$ level is practically negligible \citep[e.g.,][]{sahal96}. In order to explore possible effects of the $\Delta n=0$ collisions on the line source functions, it is useful to compare the results with a model in which these collisions are neglected. The line source functions of the model with $\Delta n=0$ collisions neglected in the $n=3$ levels are shown by the thin lines of Fig.~\ref{fig:sfpp}. The \Ha\/ source function is slightly smaller in the upper layers compared to the depolarized case, while the source function of \Lb\/ is a bit larger. In the model with $\Delta n=0$ collisions neglected in the $n=3$ levels the populations of the $3p_{1/2}$ and $3p_{3/2}$ levels, which are the upper levels of \Lb, are slightly larger. Likewise, the populations of the $3d_{3/2}$ and $3d_{5/2}$ levels, which play a key role in the \Ha\/ line formation, slightly decrease. The explanation is the following. Collisions between the fine-structure levels tend to balance the populations of the $n$-sublevels. As a result, the $3p$ term populated mainly by the \Lb\/ radiation is collisionally drained in favor of the $3d$ one \citep[this is the well known mechanism of photon conversion discussed by][]{mihalas78}.

To some extent, the quantitative results of this section depend on the collisional rates of the individual inelastic $n\ell j\to n'\ell'j'$ (with $n\neq n'$) transitions. For more details about the collisional rates see \S\ref{ssec:colls} and Appendix~\ref{app:colls}.


\subsection{Anisotropy of the spectral line radiation}
\label{ssec:anis1}

\begin{figure}
\epsscale{0.7}
\plotone{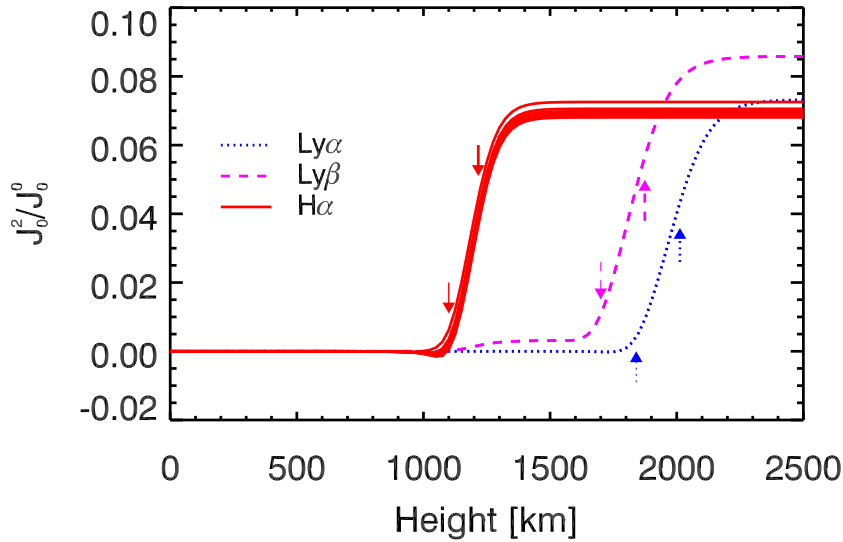}
\epsscale{1.0}
\caption{
Profile-averaged fractional anisotropy of the radiation. The solid lines correspond to the 7 individual components of the \Ha\/ line (cf. Fig.~\ref{fig:lst}). The components of \La\/ and \Lb\/ cannot be distinguished at the resolution of the plot. The vertical arrows indicate the heights where the line-center optical path is unity. The thin arrows correspond to disk-center observations (line of sight inclination $\mu=1$) while the thick arrows correspond to close to the limb observation ($\mu=0.1$). The perturber's density is $N_{\rm pert}=4\times 10^{10}\,{\rm cm}^{-3}$.
}
\label{fig:ja1ra1}
\end{figure}

The atomic level polarization and, consequently, the polarization of the emergent spectral line radiation depends on the symmetry properties of the radiation field in the atmospheric region of line formation \citep[e.g.,][]{jtb01}. In a cylindrically symmetric atmosphere the key quantity is $\mathcal{A}=J^2_0/J^0_0$ (see Eqs.~\ref{eq:j00} and \ref{eq:j20}), which quantifies the degree of anisotropy of the spectral line radiation under consideration. Typically, $\mathcal{A}$ is zero in the deep layers of the atmosphere while it is significant around and above the atmospheric height where the line center optical depth is unity. The sign and size of $\mathcal{A}$ depends on the gradient of $S_I$, with $\mathcal{A}>0$ if the illumination is predominantly vertical and $\mathcal{A}<0$ if it is predominantly horizontal \citep[e.g.,][]{jtb01,ll04}.

Since the hydrogen lines are composed of several overlapping fine-structure transitions, we may expect the  anisotropy of the transitions pertaining to the same line to be similar. Fig.~\ref{fig:ja1ra1} shows the degree of anisotropy of each individual transition as a function of height in the model atmosphere. For each transition the ensuing $\mathcal{A}$ value increases significantly above the height of unity optical depth, due to the lack of incoming photons from the outer layers. The anisotropy of the resonant doublet that produces the \La\/ line remains virtually zero until relatively large heights in the model atmosphere ($\approx 1800$\,km). The anisotropies in both \La\/ components are virtually identical because of their small wavelength separation and because they share the same lower level (see the left panel of Fig.~\ref{fig:lst}). The same applies to the \Lb\/ doublet (middle panel of Fig.~\ref{fig:lst}). The difference with respect to \La\/ is that there is a small but significant anisotropy in \Lb\/ in the intermediate atmospheric layers where the \Lb\/ optical depth is much larger than unity. As seen in Fig.~\ref{fig:ja1ra1}, the \Lb\/ anisotropy in the intermediate layers of the atmosphere increases at the same heights where the anisotropy of \Ha\/ suddenly increases. This behavior will be explained in the following subsection. Fig.~\ref{fig:ja1ra1} shows that the differences among the anisotropies of the different components of the \Ha\/ line are small but noticeable. This is because the separation of the \Ha\/ line components is of the order of 0.1\,\AA\/ (see right panel of Fig.~\ref{fig:lst}) which is comparable to the Doppler width of the line. The individual \Ha\/ components are therefore active in slightly different parts of the line.


\subsection{Atomic level polarization}
\label{ssec:atom1}

\begin{figure*}
\plottwo{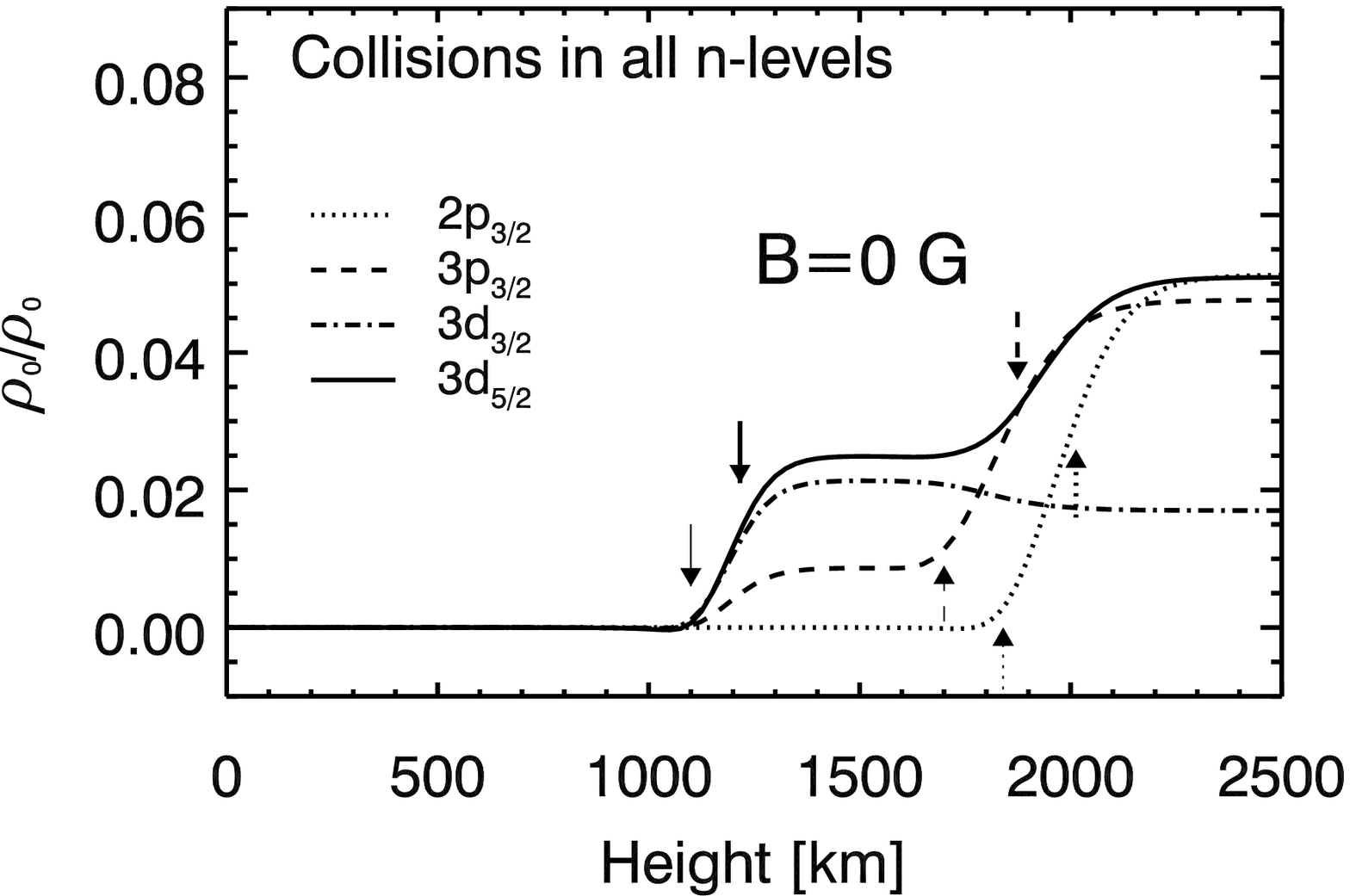}{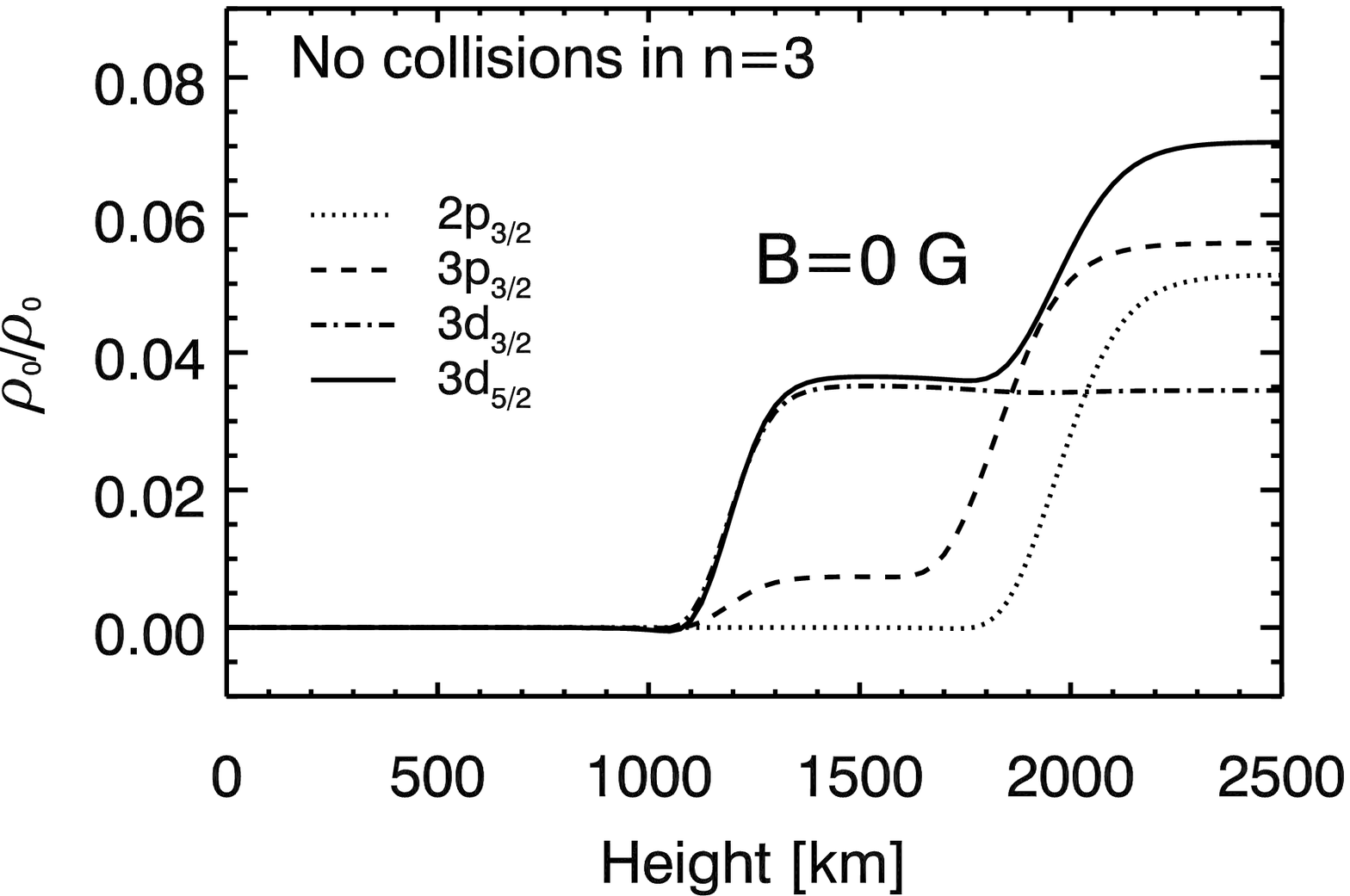}
\plottwo{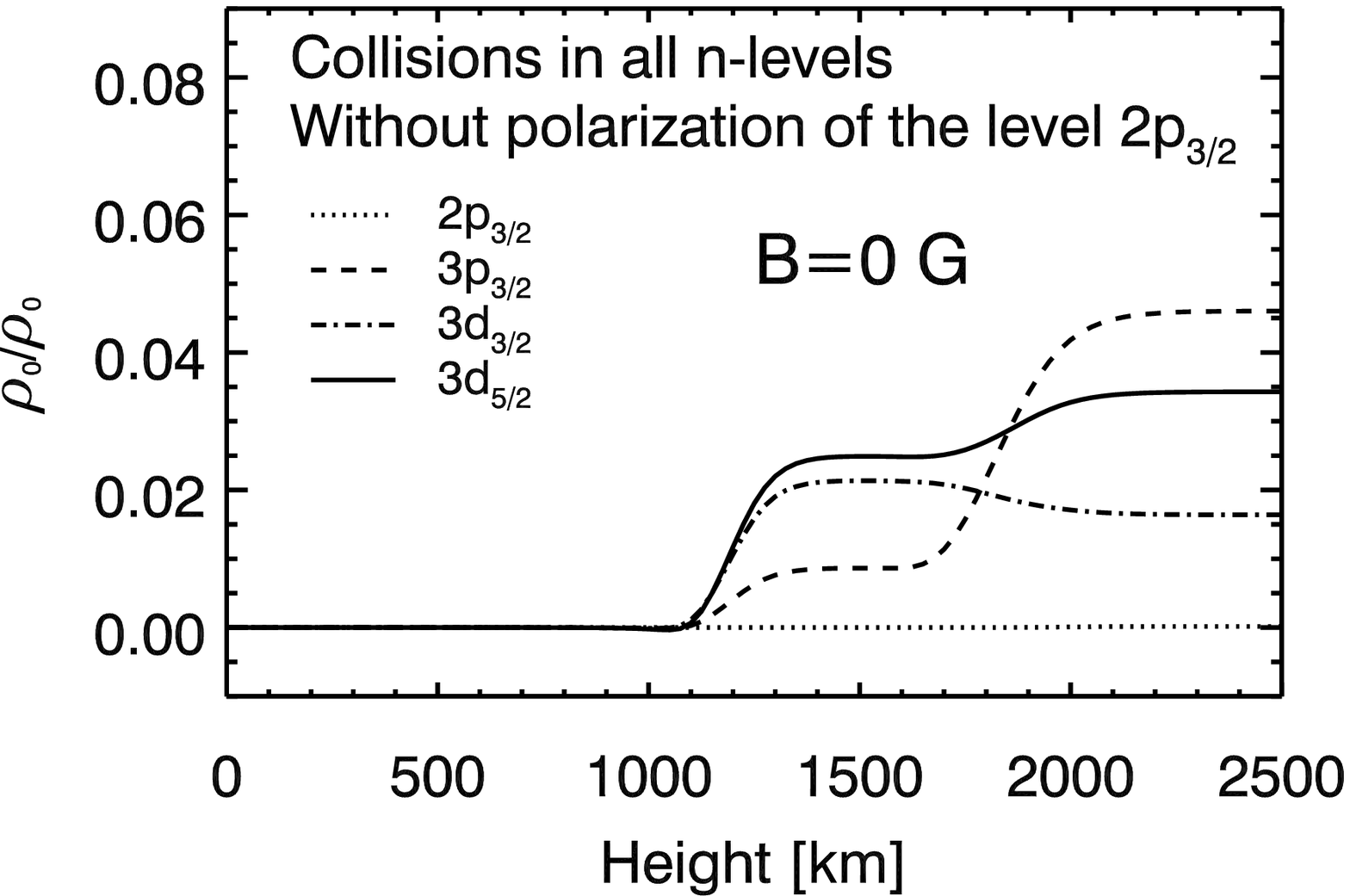}{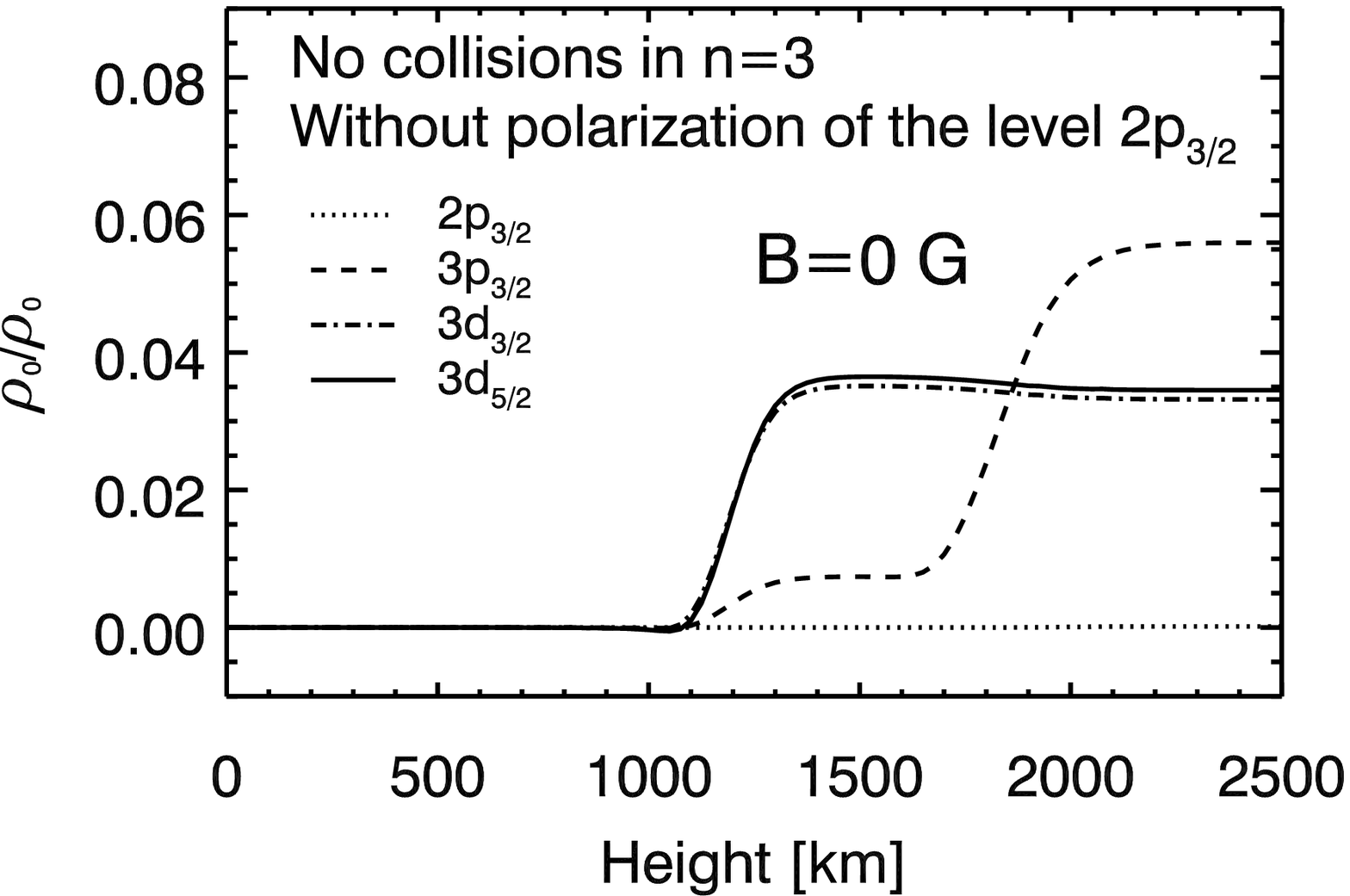}
\caption{
Fractional alignment of the levels and influence of collisions and lower level polarization. {\it Top}: Lower level polarization is taken into account. {\it Bottom}: Lower level polarization is suppressed. {\it Left}: Calculation with a perturbers density $N_{\rm pert}=4\times 10^{10}\,{\rm cm}^{-3}$. {\it Right}: Depolarizing collisions in $n=3$ are neglected.
}
\label{fig:dc1dp1}
\end{figure*}

As mentioned in \S\ref{ssec:densm} our choice for the quantization axis, $Z$, of total angular momentum is along the normal direction to the plane-parallel atmosphere (hereafter, the atmospheric frame). Given that in this section we are considering the unmagnetized case, the ensuing cylindrical symmetry of the atmosphere implies that the multipolar components $\rho^K_{Q\ne{0}}(n\ell j)=0$, so that the state of any given atomic level can be described by its overall  population, $\rho^0_0(n\ell j)$, and the atomic alignment components, $\rho^K_0(n\ell j)$. Moreover, $K>2$ multipoles of the density matrix (i.e., $\rho^K_0(n\ell j)$ with $K=4,6,\dots$) of the levels with $j\ge 5/2$ can be safely neglected in our weakly anisotropic stellar atmosphere model because they are very small compared to the second-rank components. The top left panel of Fig.~\ref{fig:dc1dp1} shows the variation with height of the fractional alignments of the levels $2p_{3/2}$, $3p_{3/2}$, $3d_{3/2}$, and $3d_{5/2}$. In the remaining part of this secti
 on we consider each of such levels separately.

$\mathbf{2p_{3/2}}$. This is the only polarizable level of the \La\/ line. Since the optical thickness in \La\/ is the largest of the hydrogen lines, its anisotropy is negligible up to the uppermost layers of the atmosphere (see Fig.~\ref{fig:ja1ra1}). The fractional alignment of the $2p_{3/2}$ level follows the \La\/ anisotropy as if it were an upper level of a two-level atom \citep[e.g.][]{jtb99}. Because for the chosen density of perturbers ($4\times 10^{10}\,{\rm cm^{-3}}$) the depolarizing collisions for $n=2$ are inefficient, the surface value of the $2p_{3/2}$ fractional alignment approximately corresponds to the theoretical prediction of Eq.~\ref{eq:s20j20}, i.e., $\rho^2_0/\rho^0_0\approx 0.05$ because $w^{(2)}_{3/2,1/2}=0.707$ and $J^2_0/J^0_0\approx 0.07$.

$\mathbf{3p_{3/2}}$. This is a common upper level of \Ha\/ and \Lb. The atomic polarization of this level is affected by optical pumping in both lines. In the layers where the model atmosphere becomes partially transparent to \Ha\/ photons (i.e., 1200 km ${\lesssim}\,h\,{\lesssim}$ 1700 km; see Fig.~\ref{fig:ja1ra1}), the atomic polarization of this level is increased due to the absorption of anisotropic radiation in the \Ha\/ line. The fractional alignment of the $3p_{3/2}$ level is smaller than that of the $3d_{5/2}$ or $3d_{3/2}$ levels because the $3p_{3/2}$ level is strongly populated by the photons of the optically thick \Lb\/ line, which are mostly trapped or degraded into \Ha\/ photons (see the discussion in \S\ref{ssec:sourcef}). At heights $h>1700$ km the anisotropy of the \Lb\/ radiation increases (Fig.~\ref{fig:ja1ra1}) and the $3p_{3/2}$ level is aligned directly by the ensuing anisotropic \Lb\/ pumping. As shown in the top left panel of Fig.~\ref{fig:dc1dp1}, the surface value of $\rho^2_0(3p_{3/2})/\rho^0_0(3p_{3/2})$ lies below the theoretical value ($\approx 0.06$) predicted by the 2-level atom approximation of Eq.~\ref{eq:s20j20}. This is partly because the level is radiatively coupled to $2s_{1/2}$, in addition to $1s_{1/2}$, but mainly because the $\Delta{n}=0$ depolarizing collisions play a more significant role in the $n=3$ than in $n=2$ fine-structure levels.

$\mathbf{3d_{5/2}}$. This is an upper level of \Ha\/ which is not radiatively coupled to the ground level but it is radiatively coupled to the $2p_{3/2}$ level by the strongest transition of the \Ha\/ line (see Fig.~\ref{fig:grd}). The $3d_{5/2}$ level becomes significantly polarized above the atmospheric height where the anisotropy of the \Ha\/ radiation increases (i.e., above $h{\approx}1100$ km). The fractional alignment of the $3d_{5/2}$ level is approximately constant in the intermediate layers of the model atmosphere where \Ha\/ becomes optically thin but the Lyman lines are still opaque (i.e., between 1200 and 1700\,km). The polarization of the $3d_{5/2}$ level is decreased by its strong collisional coupling with the $3p_{3/2}$ level. However, once the \Lb\/ line becomes optically thin and much more anisotropic, its upper level $3p_{3/2}$ is significantly more polarized and the collisional transfer of alignment from $3p_{3/2}$ to $3d_{5/2}$ eventually leads to an increase of the $3d_{5/2}$ polarization (see the lower left panel of Fig.~\ref{fig:dc1dp1}). Since the $3d_{5/2}$ level is strongly radiatively coupled to the $2p_{3/2}$ level, its polarization is, in the uppermost layers of the atmosphere, affected by the atomic alignment of the $2p_{3/2}$ level. Collisional coupling and the role of lower-level polarization in the \Ha\/ line will be discussed in greater detail below.

$\mathbf{3d_{3/2}.}$ This level is radiatively coupled to the $2p_{1/2}$ level and, via a weaker transition, to $2p_{3/2}$. The former transition makes the major contribution to the \Ha\/ line in the blue part of the profile while the later contributes slightly in the red part (see the right panel of Fig.~\ref{fig:lst}). Note that even though the strongest radiative transition involving the $3d_{3/2}$ level is $2p_{1/2}$--$3d_{3/2}$, with $w^{(2)}_{3/2,1/2}=0.707$, the fractional polarization of the $3d_{3/2}$ level throughout the atmosphere is smaller than that of $3d_{5/2}$, whose radiative excitation is solely due to the $2p_{3/2}$--$3d_{5/2}$ transition whose $w^{(2)}_{5/2,3/2}=0.529$. This is because of the $2p_{3/2}$--$3d_{3/2}$ transition with the negative factor $w^{(2)}_{3/2,3/2}=-0.566$ leading to an overall decrease of the upper-level alignment. For a similar reason the alignment of the $3d_{3/2}$ level decreases in the uppermost layers of the atmosphere where $\rho^2_0(3p_{3/2})/\rho^0_0(3p_{3/2})$ rises. The weak collisional transition $3p_{3/2}$--$3d_{3/2}$ has a similar depolarizing effect on $3d_{3/2}$.\footnote{Note, however, that concerning collisional transitions we deal with the transfer of atomic alignment by isotropic collisions while concerning radiative transitions we deal with absorption of anisotropic radiation from levels that do not necessarily have to be polarized \citep[see][for a discussion of the alignment transfer mechanisms in the hydrogen fine-structure levels]{sahal96}.}

\begin{figure*}
\plottwo{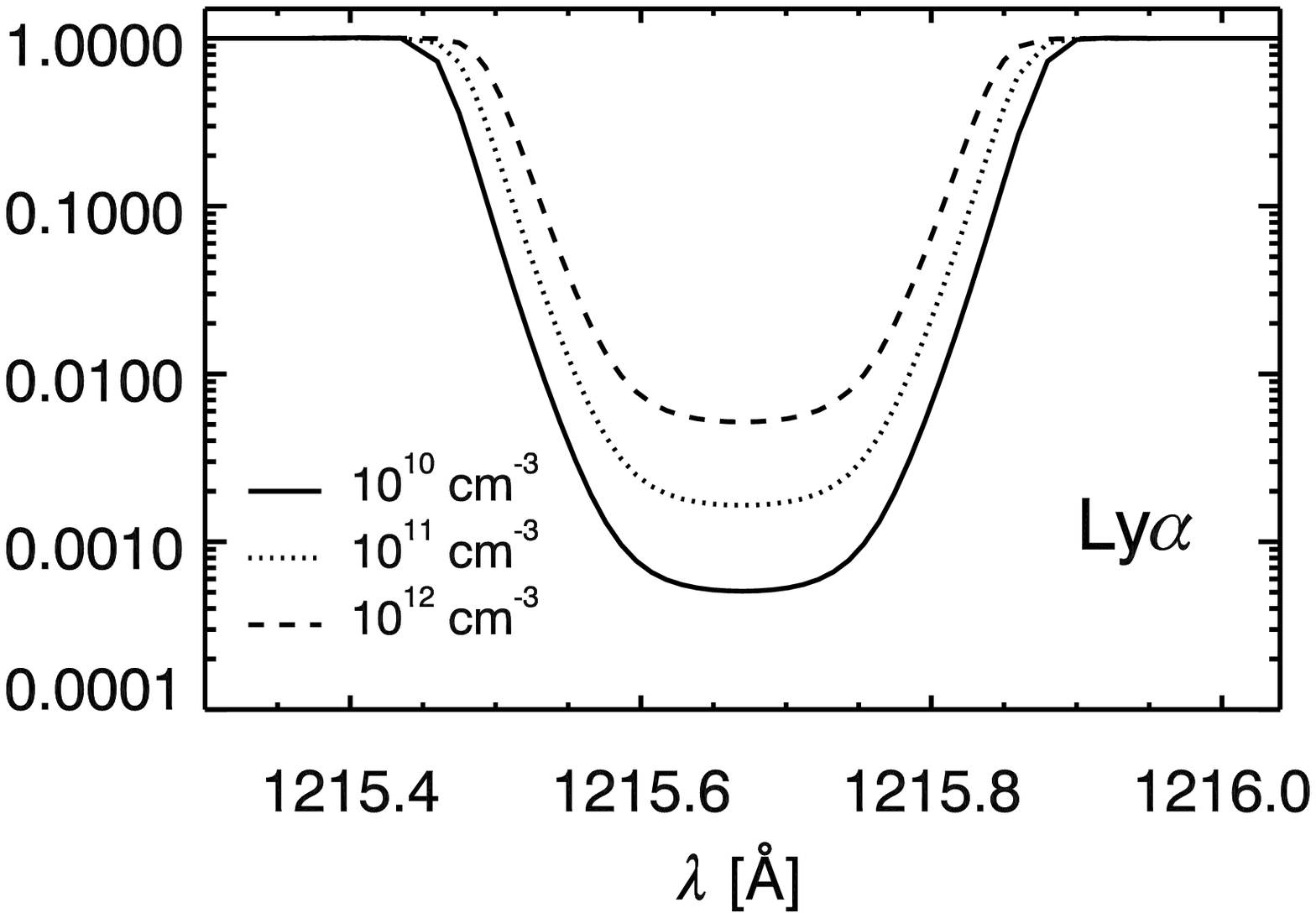}{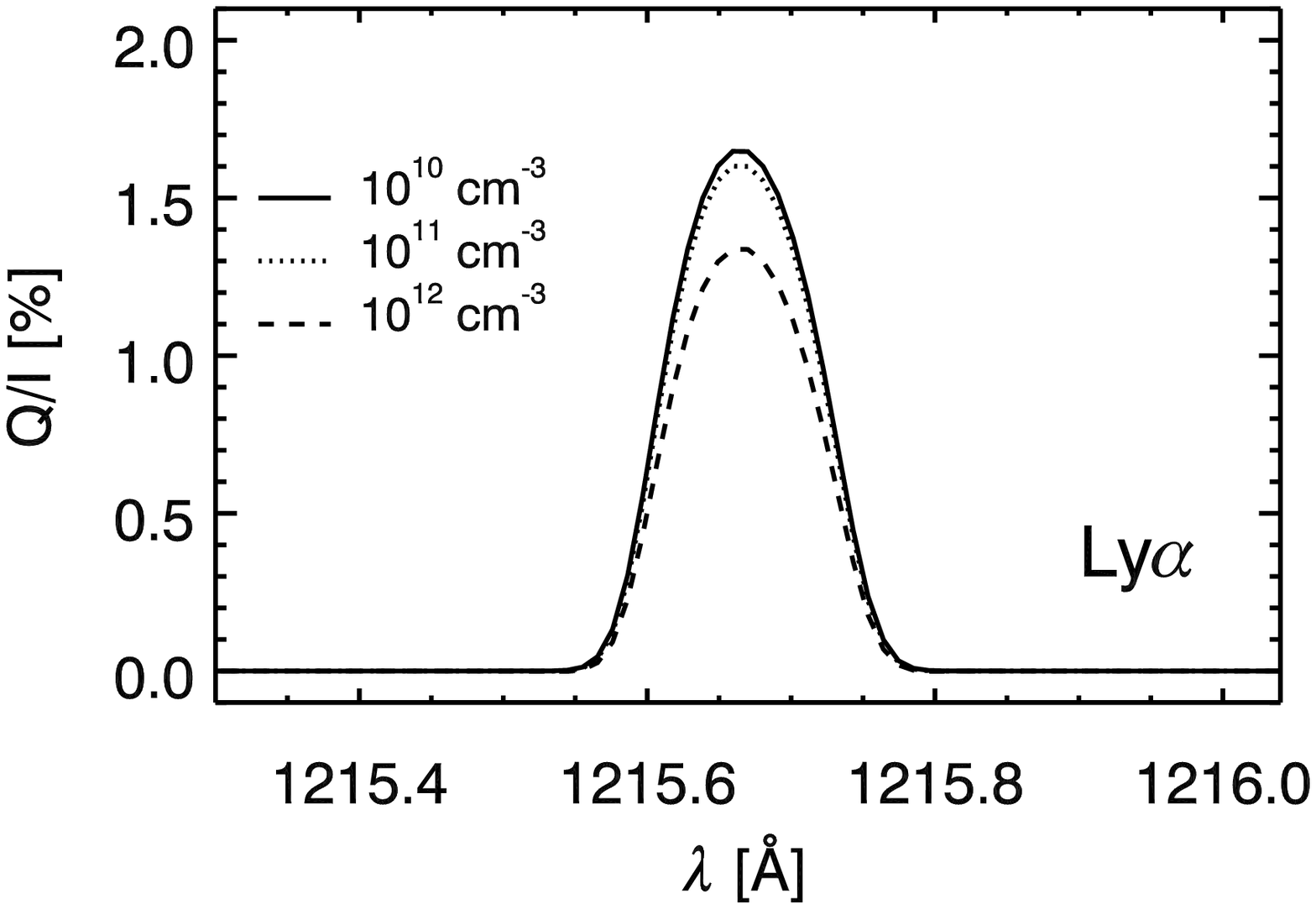}
\plottwo{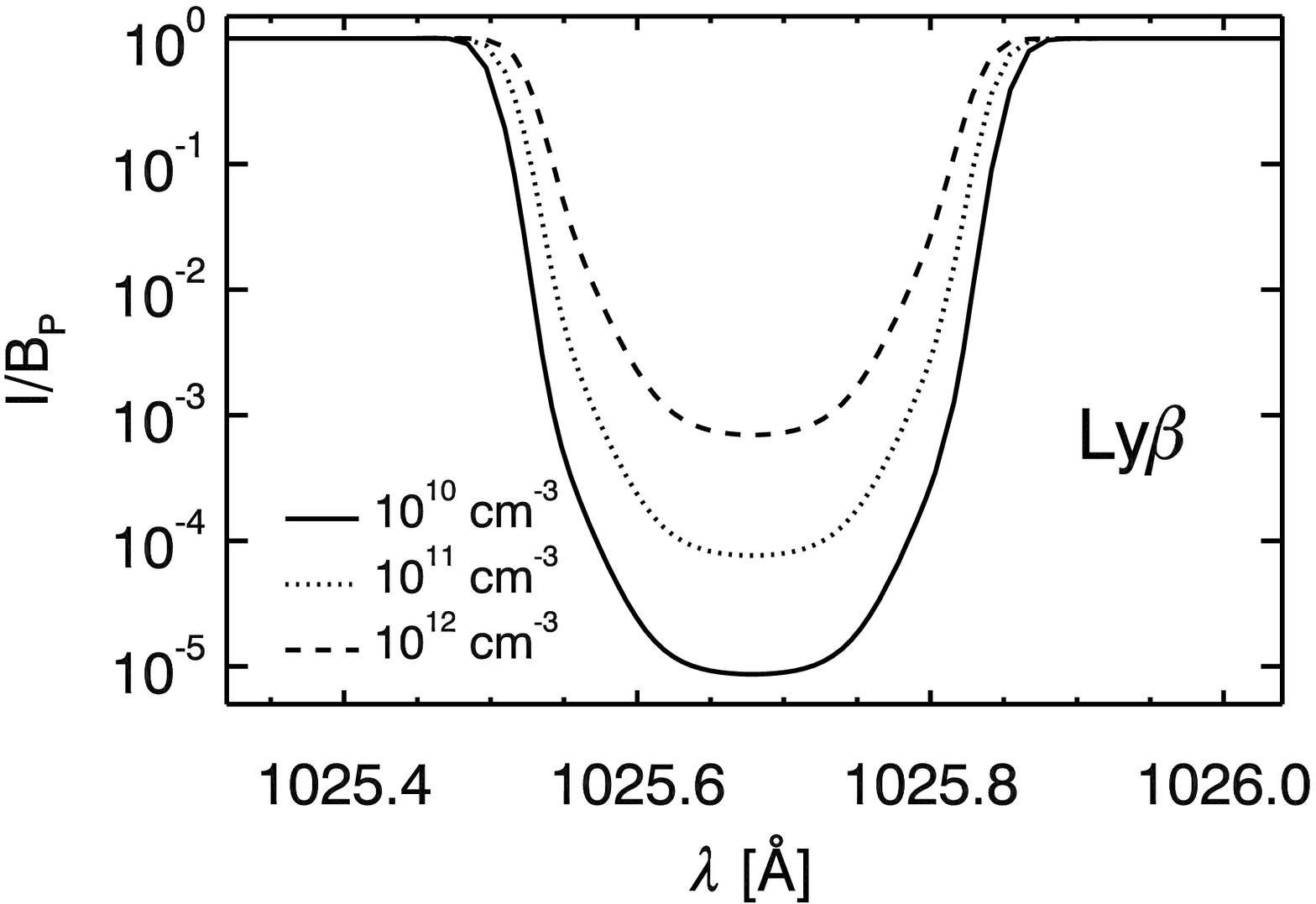}{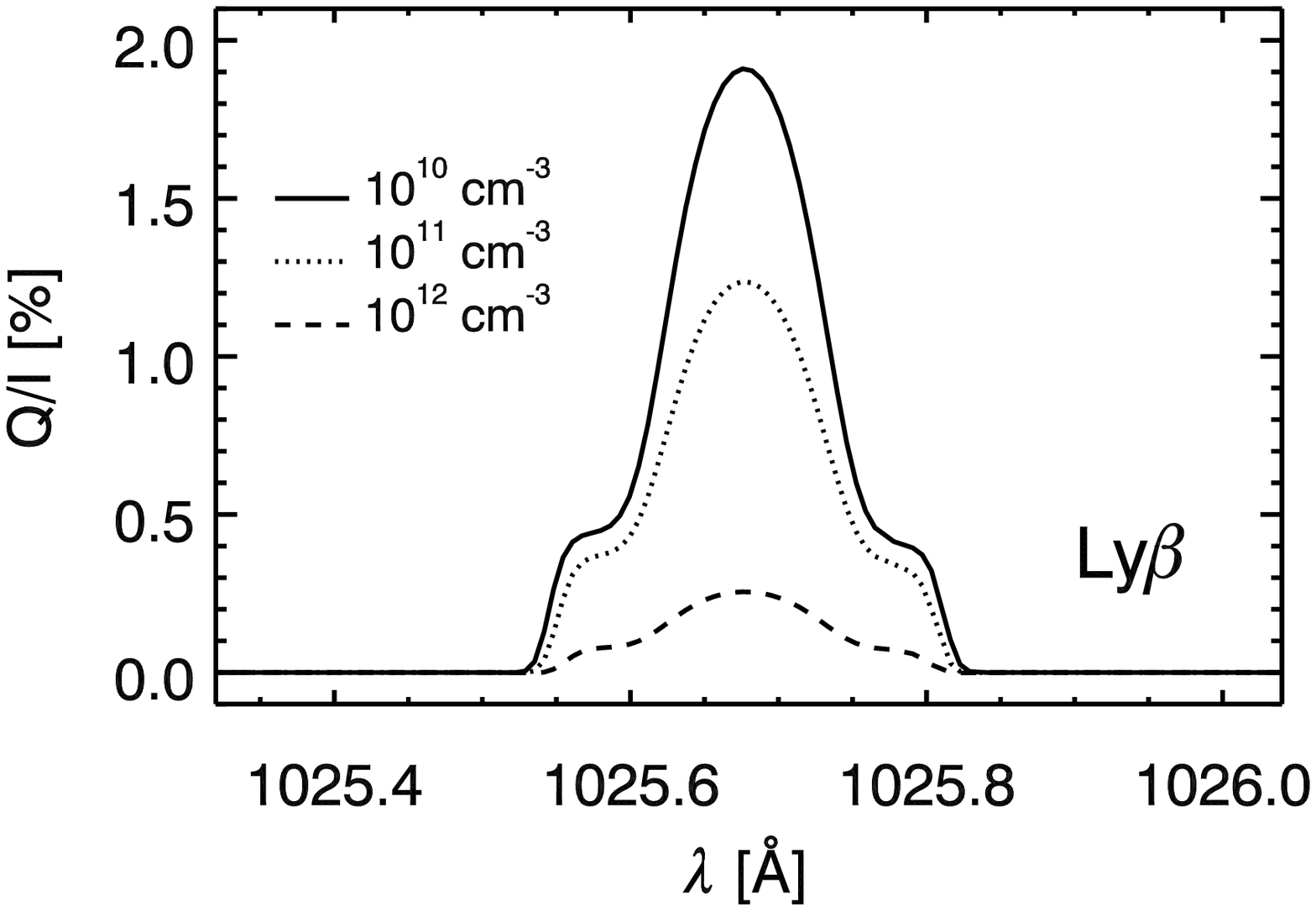}
\plottwo{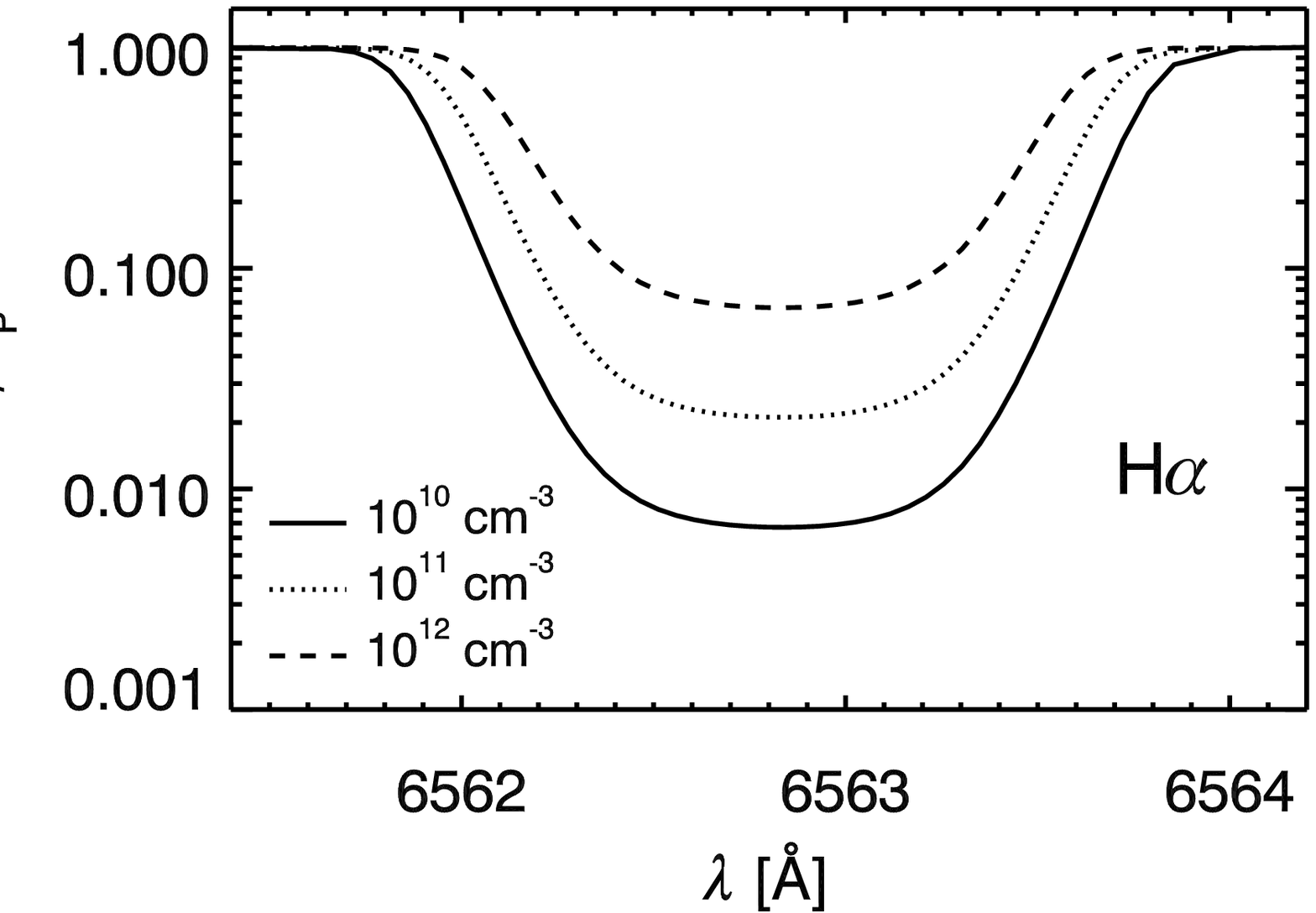}{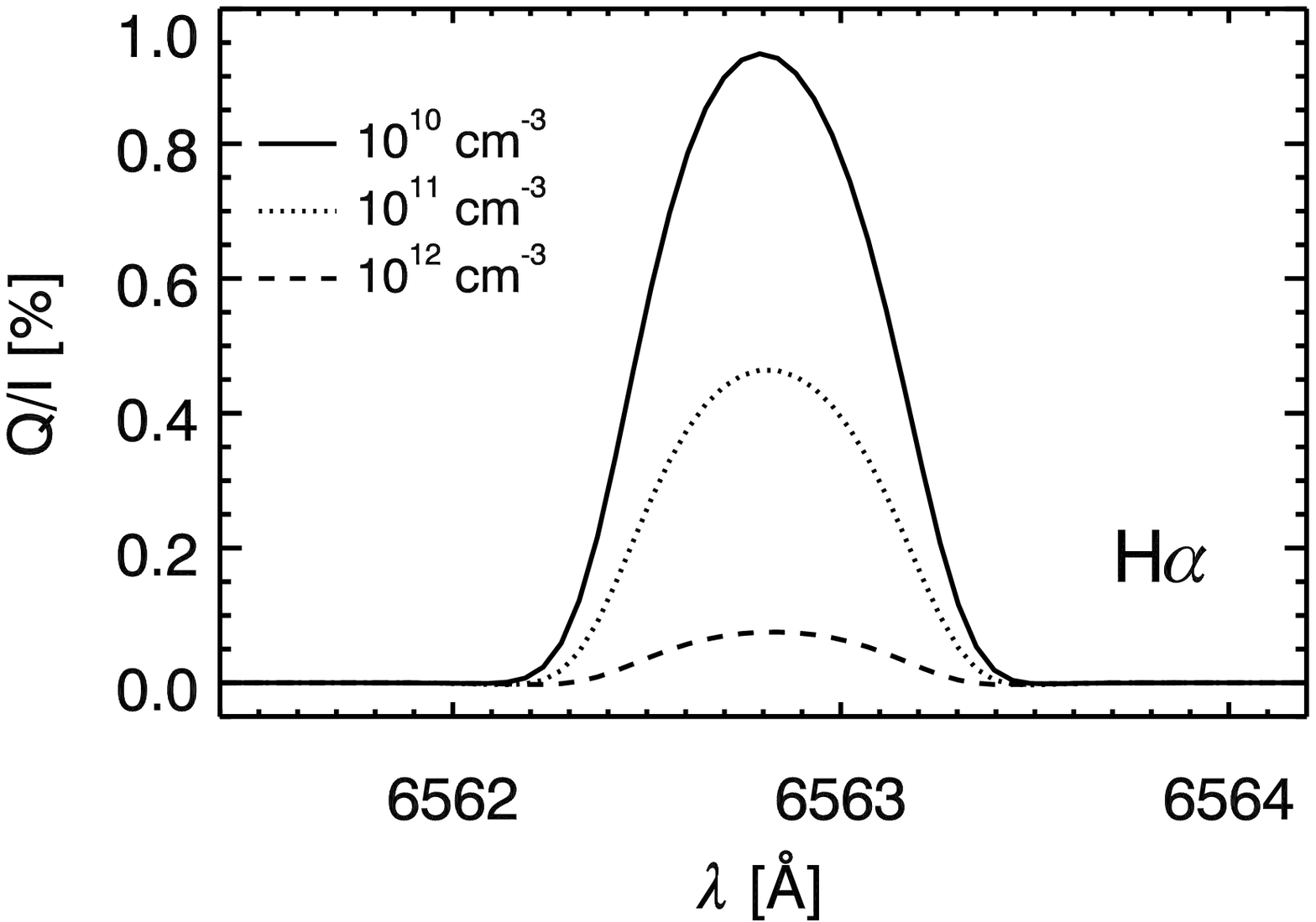}
\caption{
The normalized line intensity ({\it left}) and fractional linear polarization ({\it right}) profiles for a close to the limb LOS ($\mu=0.1$) calculated for several densities of the charged particles. The individual panels correspond to \La\/ ({\it top}), \Lb\/ ({\it middle}), and \Ha\/ ({\it bottom}).
}
\label{fig:clv-nm2}
\end{figure*}

The non-negligible amount of anisotropy in \Lb\/ between 1200 and 1700\,km, where the \Lb\/ optical depth is very large, can be explained by the above-mentioned coupling of the \Lb\/ and \Ha\/ lines. The $3p_{3/2}$ level is pumped by both \Lb\/ and \Ha\/ radiation and it is collisionally coupled to the $3d_{5/2}$ level. The \Ha\/ radiation is no longer isotropic at heights above 1200\,km and the upper levels of \Ha, including $3p_{3/2}$, are aligned. The larger the \Ha\/ anisotropy the greater the 3$p_{3/2}$ fractional alignment. Since the \Lb\/ photons are emitted from the polarized $3p_{3/2}$ level, the emitted \Lb\/ radiation there is sufficiently anisotropic and polarized so as to give rise to a significant contribution to its $J^2_0$ radiation tensor.

The radiation anisotropy in the \Lb\/ line in such relatively deep layers of the atmosphere might have an impact on the polarization of the \Lb\/ line wings. However, in this paper our calculations are restricted to the CRD approximation. Within this approximation, the \Lb\/ anisotropy in such deep layers implies a polarization signal in the line wings (cf. \S\ref{ssec:clv}). Although we cannot make any reliable predictions concerning the scattering polarization in the \Lb\/ wings where the CRD approximation is unsuitable, the mechanism described in the previous paragraph may be important for future PRD theories of multi-level scattering polarization. 

\begin{figure}
\epsscale{0.7}
\plotone{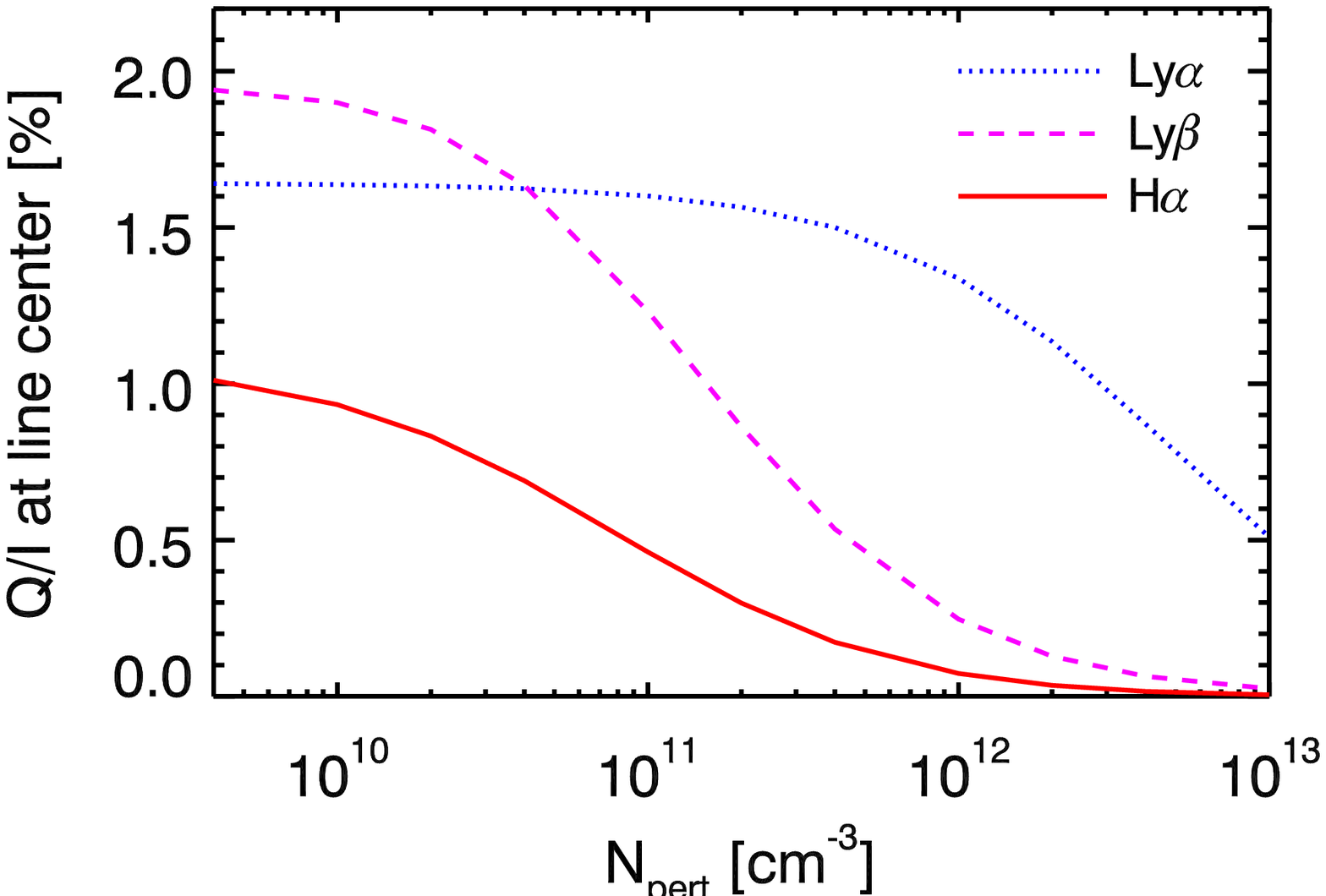}
\epsscale{1.0}
\caption{
$Q/I$ amplitudes for a close to the limb LOS ($\mu=0.1$) in the non-magnetic model atmosphere versus the perturbers' density $N_{\rm pert}$, assuming $N_{\rm e}=N_{\rm p}=N_{\rm pert}$.
}
\label{fig:qinenp}
\end{figure}


\subsection{Influence of lower-level polarization on the H$\alpha$ line}
\label{ssec:dichr}

If the lower level of a line transition is polarized then dichroic effects (i.e., differential absorption of polarization components) may affect the polarization of the emergent radiation, even in the absence of any significant Zeeman splitting \citep{jtb97,jtb99spw,jtb02}. An interesting example of this ``zero-field" dichroism can be found in the scattering polarization of the infrared triplet of Ca\,{\sc ii} \citep{mansosainz03a} and in many  other spectral lines of the second solar spectrum \citep{jtb09rev}. Since the ground level of H\,{\sc i} cannot be aligned when the hyperfine structure of hydrogen is neglected, all the lines of the Lyman series are non-dichroic. On the other hand, the linear polarization of the subordinate lines can, in principle, be modified by selective absorption of polarization components because such lines have always a lower fine-structure level with $j_l\ge 3/2$. In the atomic model of Fig.~\ref{fig:grd} the only lower level that can in principle be aligned is the $2p_{3/2}$ level, which is one of the three lower levels of the \Ha\/ line. Therefore, \Ha\/ is the only spectral line of the atomic model of Fig.~\ref{fig:grd} whose emergent linear polarization could be influenced by selective absorption of polarization components. 

The dichroic transitions of \Ha\/ are $2p_{3/2}$--$3s_{1/2}$, $2p_{3/2}$--$3d_{3/2}$, and $2p_{3/2}$--$3d_{5/2}$. In our isothermal model atmosphere, the polarization of the $2p_{3/2}$ level is only significant in the uppermost layers, where the \La\/ radiation is anisotropic (i.e., above $\approx 1800$\,km, as shown in Fig.~\ref{fig:ja1ra1}). Since \Ha\/ is optically thin in such outer layers of our model atmosphere its emergent $Q/I$ profile is {\it not} influenced by ``zero-field" dichroism. 

It is important to note that, in addition to the possibility of having ``zero-field" dichroism, the presence of lower-level polarization may produce a significant feedback on the atomic polarization of the other levels \citep[e.g., Eqs.~31 and 32 of][]{jtb01}. In order to investigate the impact of the presence of atomic alignment in the $2p_{3/2}$ level on the atomic polarization of the upper levels of the \Ha\/ line, we have solved the very same non-LTE problem of the 2nd kind for the atomic model of Fig. 1 but forcing the lower level $2p_{3/2}$ to be completely unpolarized throughout the model atmosphere. The resulting fractional alignment of the hydrogen levels is shown in the bottom left panel of Fig.~\ref{fig:dc1dp1}. The influence of lower-level polarization can be valorated by comparing the bottom left panel of Fig.~\ref{fig:dc1dp1} with the top left panel of the same figure in which the lower-level polarization is taken into account.

Notice that there is no significant difference below 1800\,km where \La\/ is isotropic and the $2p_{3/2}$ level is unpolarized. Above 1800\,km the only upper level that is seriously affected is $3d_{5/2}$, due to the strong $2p_{3/2}$--$3d_{5/2}$ optical transition. The level 3$p_{3/2}$ cannot be affected directly by the presence of atomic polarization in the $2p_{3/2}$ level because the transition $2p_{3/2}$--$3p_{3/2}$ is forbidden. However, the fractional alignment of this $3p_{3/2}$ level is slightly larger when the atomic polarization of the lower-level is taken into account, because of the strong collisional coupling with the $3d_{5/2}$ and $3d_{3/2}$ levels. The level $3d_{3/2}$ is not  strongly radiatively coupled to $2p_{3/2}$, not even collisionally to the $3p_{3/2}$ level, and it is thus not noticeably affected by the presence of atomic polarization in the $2p_{3/2}$ level.

In conclusion, since the atomic polarization of the $2p_{3/2}$ level only affects the atomic polarization of the upper levels of \Ha\/ in the uppermost layers of the model atmosphere, where \Ha\/ is already optically thin, the emergent $Q/I$ profile of \Ha\/ has nothing to do with lower-level polarization. The same applies to the real solar chromosphere \citep[see][]{stepanjtb10asym}. In different scattering environments it might be possible that the \Ha\/ scattering polarization profile is significantly altered by lower-level polarization. This interesting problem will be carefully addressed in a forthcoming investigation.


\subsection{The role of collisions}
\label{ssec:colls}

In semi-empirical models of the solar chromosphere \citep[e.g.,][]{vernazza81,fontenla93,fontenla07,avrett08} the electron and proton densities typically vary between $10^{10}$ and $10^{11}\,{\rm cm^{-3}}$ in the upper chromosphere, where the line-center features of the hydrogen lines studied here originate. In this section, we discuss the influence of collisions with electrons and protons on the emergent $I$ and $Q/I$ profiles. More information on the inelastic and depolarizing collisional rates of our calculations can be found in Appendix~\ref{app:colls}.

We begin by investigating the impact of collisional transfer between the $n=3$ sublevels on the fractional alignment of the hydrogen atomic levels. Obviously, if collisions among the sublevels of $n=3$ are neglected then the transfer of population and alignment between the ensuing fine structure levels is no longer possible. The top right panel of Fig.~\ref{fig:dc1dp1} shows the results of this numerical experiment. Note that when collisions among the sublevels of $n=3$ are neglected the 3$d_j$ levels are not coupled to the $3p_j$ levels. In contrast to the models in which the $\Delta n=0$ collisions are taken into account, the fractional alignment of the $3d_j$ levels increases in the intermediate layers of the model atmosphere where \Ha\/ starts to become optically thin but the Lyman lines are still opaque.

The bottom right panel of Fig.~\ref{fig:dc1dp1} helps to clarify the effect of the $2p_{3/2}$ polarization on that of the $3d_{5/2}$ level. If the $\Delta n=0$ collisions are neglected in $n=3$ and the polarization of the $2p_{3/2}$ level is artificially suppressed, the polarization of the $3d_{5/2}$ level in the uppermost atmospheric layers is no longer affected by the lower level polarization (compare with the other panels of the same figure). The polarization of the $3d_{5/2}$ and $3d_{3/2}$ levels is due to the anisotropy and polarization of the \Ha\/ radiation and it is practically uncorrelated with the anisotropy and polarization of the \La\/ and \Lb\/ lines.

We now turn to showing in Fig.~\ref{fig:clv-nm2} the emergent line intensity and the fractional linear polarization for increasing values of the perturbers density. Since the model atmosphere is isothermal, the line source functions mostly decrease with height in the atmosphere (see Fig.~\ref{fig:sfpp}) and the emergent intensity profiles appear in absorption. The anisotropy of the spectral line radiation here is dominated by the limb darkening of the outgoing radiation and the emergent $Q/I$ signals are positive (i.e., parallel to the surface of the atmosphere).

Fig.~\ref{fig:clv-nm2} shows that the sensitivity of the $Q/I$ profile of the \La\/ line to collisional depolarization is relatively small (see also Appendix~\ref{sapp:depol}). For densities lower than $10^{11}\,{\rm cm^{-3}}$ there is practically no modification of the emergent $Q/I$ profile. A noticeable decrease in the $Q/I$ line-center amplitude can be seen around $10^{12}\,{\rm cm^{-3}}$, i.e., at a density which is considered to be too high for the uppermost layers of the solar chromosphere.

As shown in Fig.~\ref{fig:ja1ra1} the anisotropy of the spectral line radiation at the atmospheric height where $\tau(\mu=0.1)=1$ is larger for \Lb\/ than for \La. However, whether or not the amplitude of the scattering polarization $Q/I$ profile is larger in \Lb\/ than in \La\/ depends on the perturber's density. The reson is that the \Lb\/ line is very sensitive to collisional depolarization because collisional shuffling of the optical electrons among the fine-structure sublevels is much more efficient in $n=3$ then in $n=2$. Thus, for $N_{\rm pert}=10^{10}\,{\rm cm^{-3}}$ the amplitude of the \Lb\/ polarization is slightly larger than for \La\/ (see Fig.~\ref{fig:clv-nm2}), while for $N_{\rm pert}=10^{12}\,{\rm cm^{-3}}$ is much smaller. 

The most sensitive spectral line to such collisions is \Ha\/ (see the right bottom panel of Fig.~\ref{fig:clv-nm2}). The rate of collisional transitions among the $3\ell j$ levels becomes comparable to the inverse radiative lifetime of the $3d_j$ levels at around a perturbers density of a few times $10^{10}\,{\rm cm^{-3}}$. Note that the source function and anisotropy of the \Ha\/ line are less sensitive to the thermal structure of the chromosphere than \La\/ and \Lb\/ (because \Ha\/ is photo-ionization dominated), and that the \Ha\/ scattering polarization can be significantly reduced by depolarizing collisions with protons if their density exceeds approximately $10^{10}\,{\rm cm^{-3}}$.

The previous results are summarized in Fig.~\ref{fig:qinenp}, which shows the $Q/I$ amplitudes of the three hydrogen lines as a function of electron and proton density.


\subsection{Center-to-limb variation (CLV) of the line profiles}
\label{ssec:clv}

\begin{figure*}
\plottwo{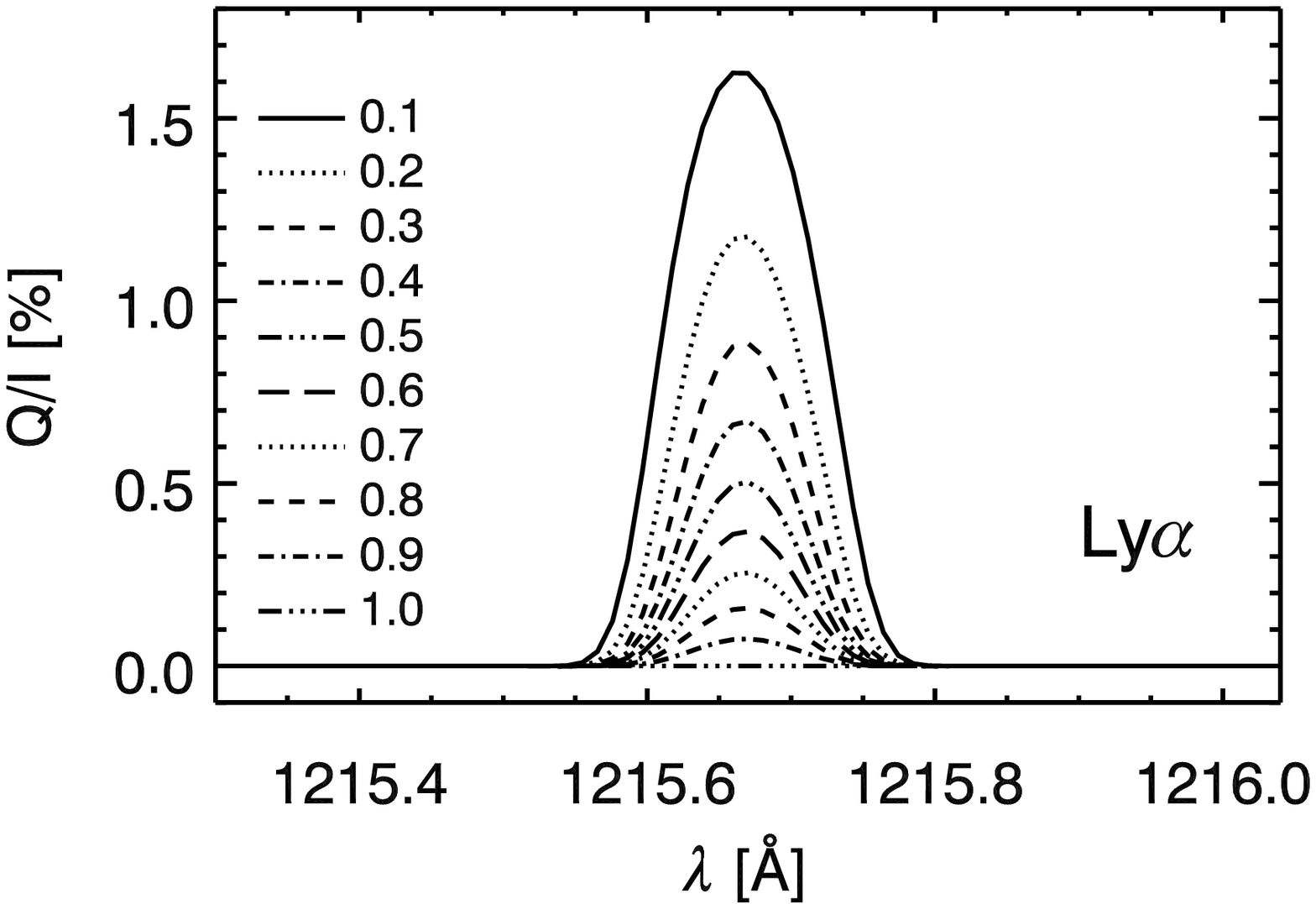}{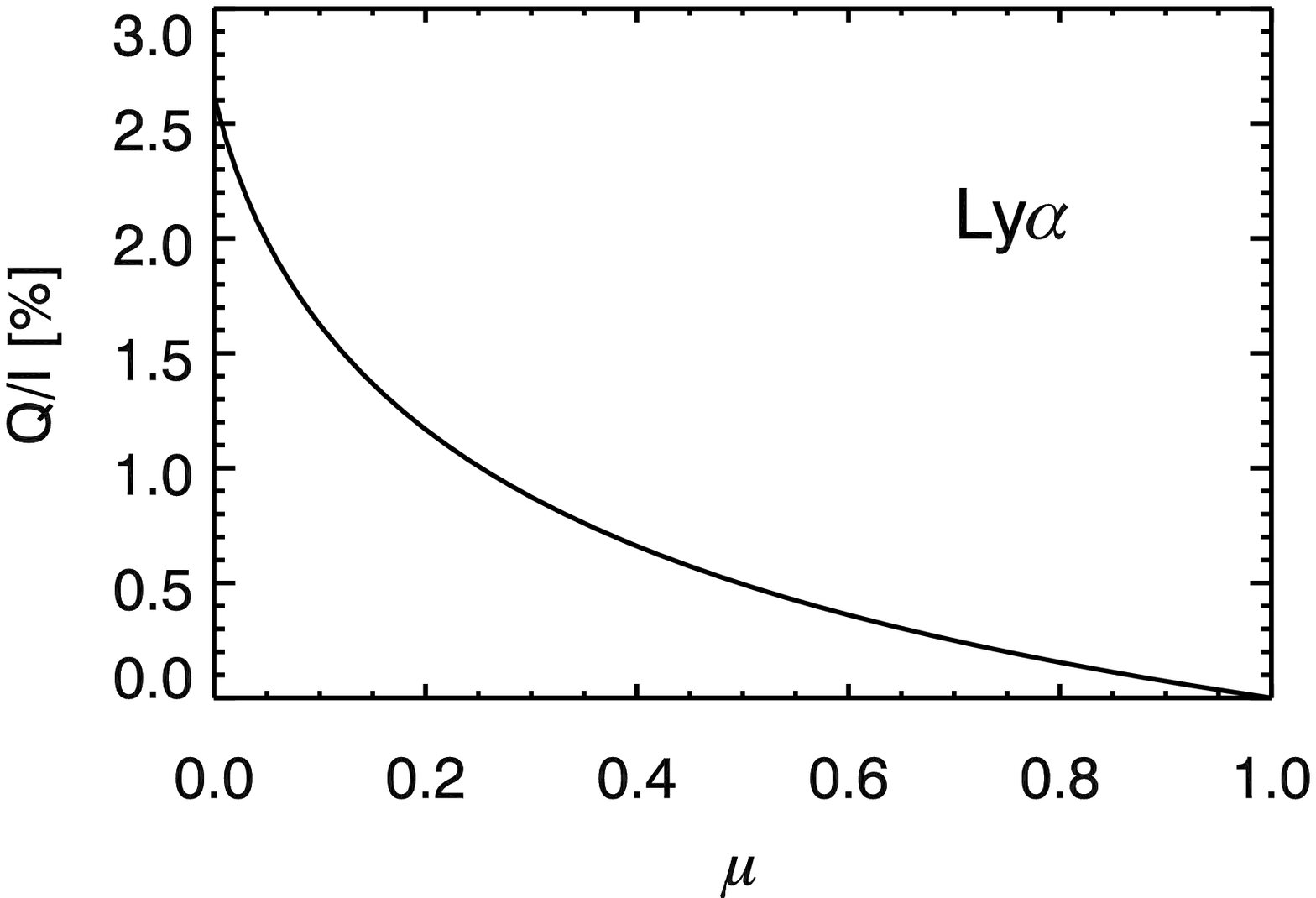}
\plottwo{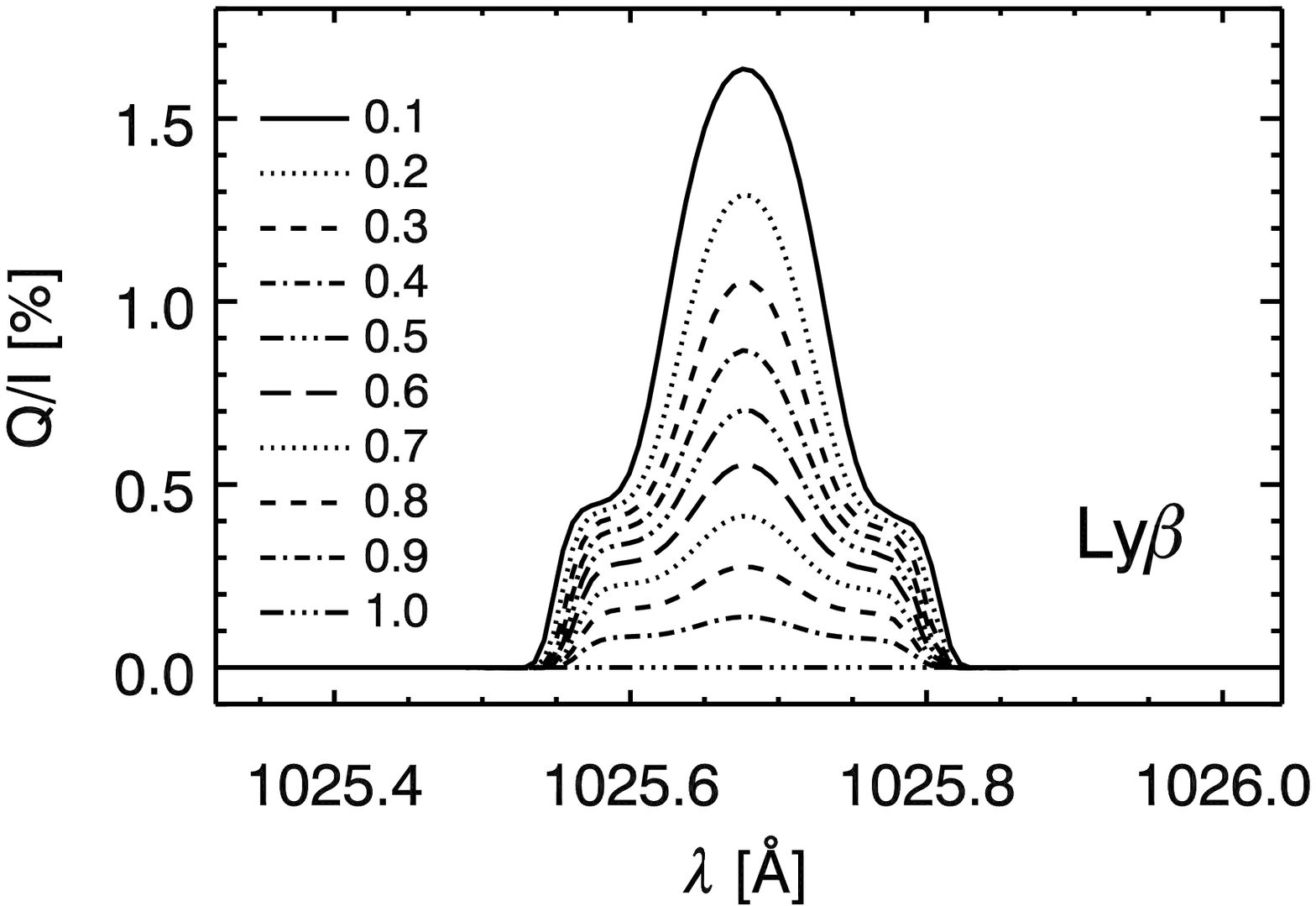}{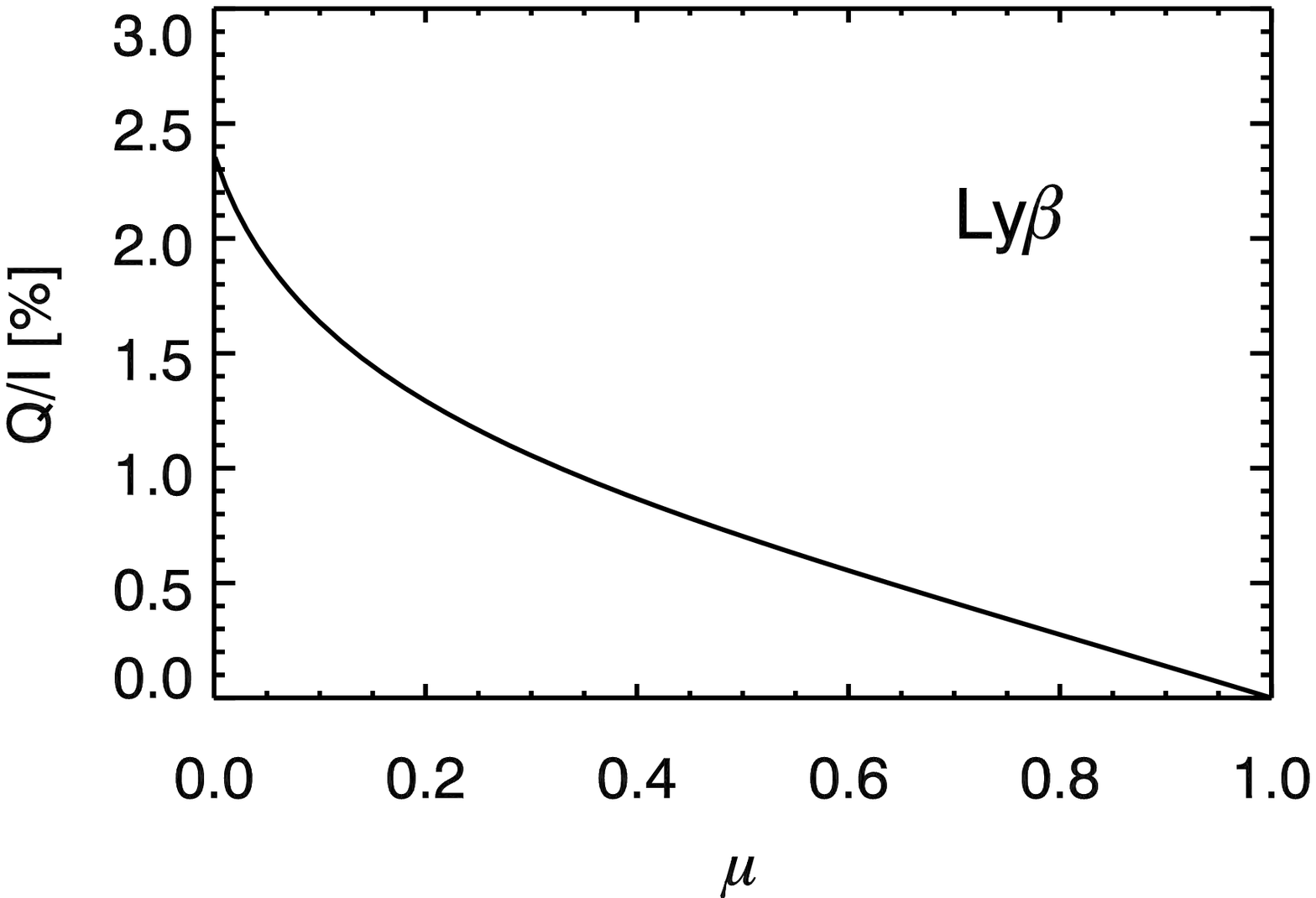}
\plottwo{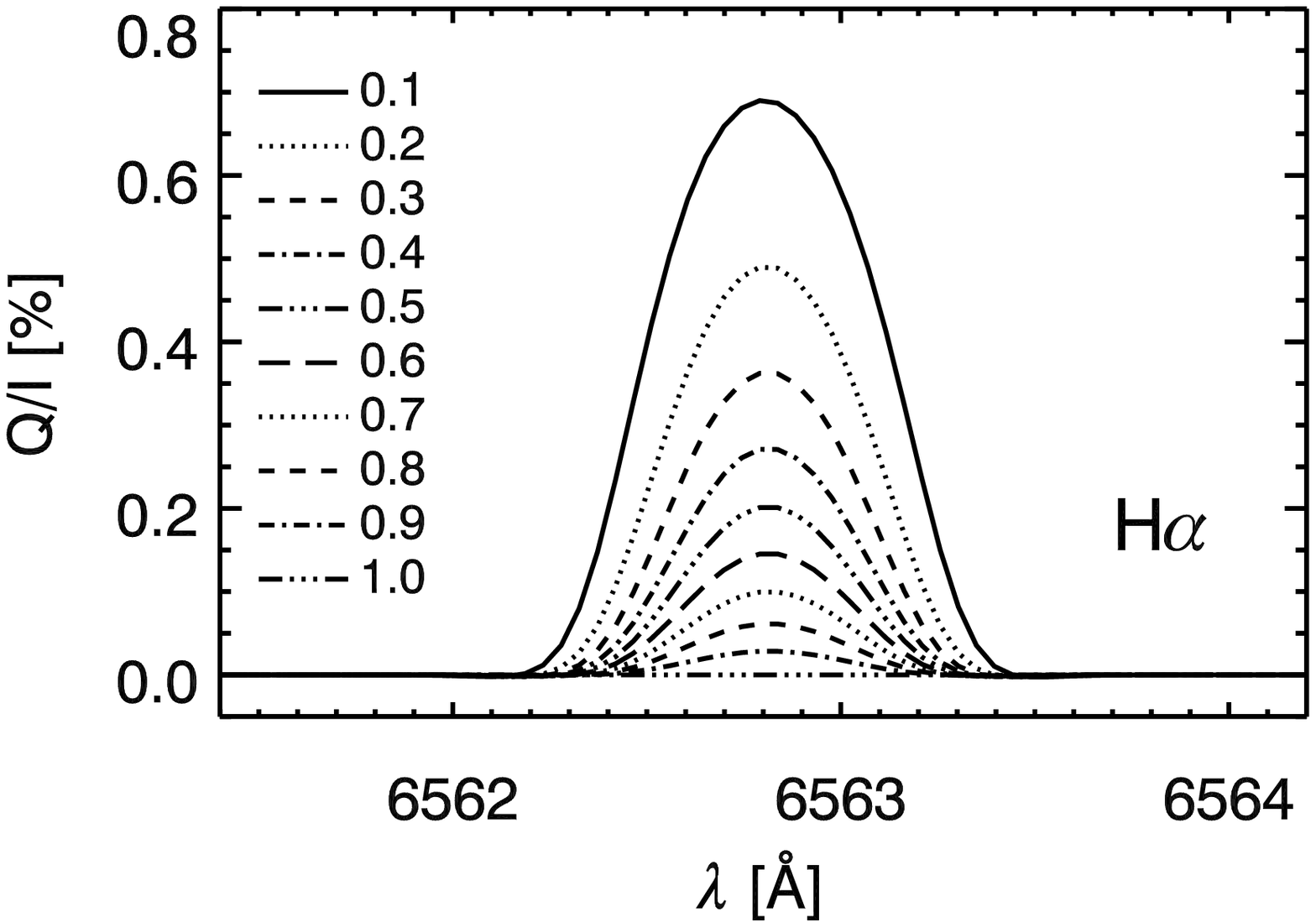}{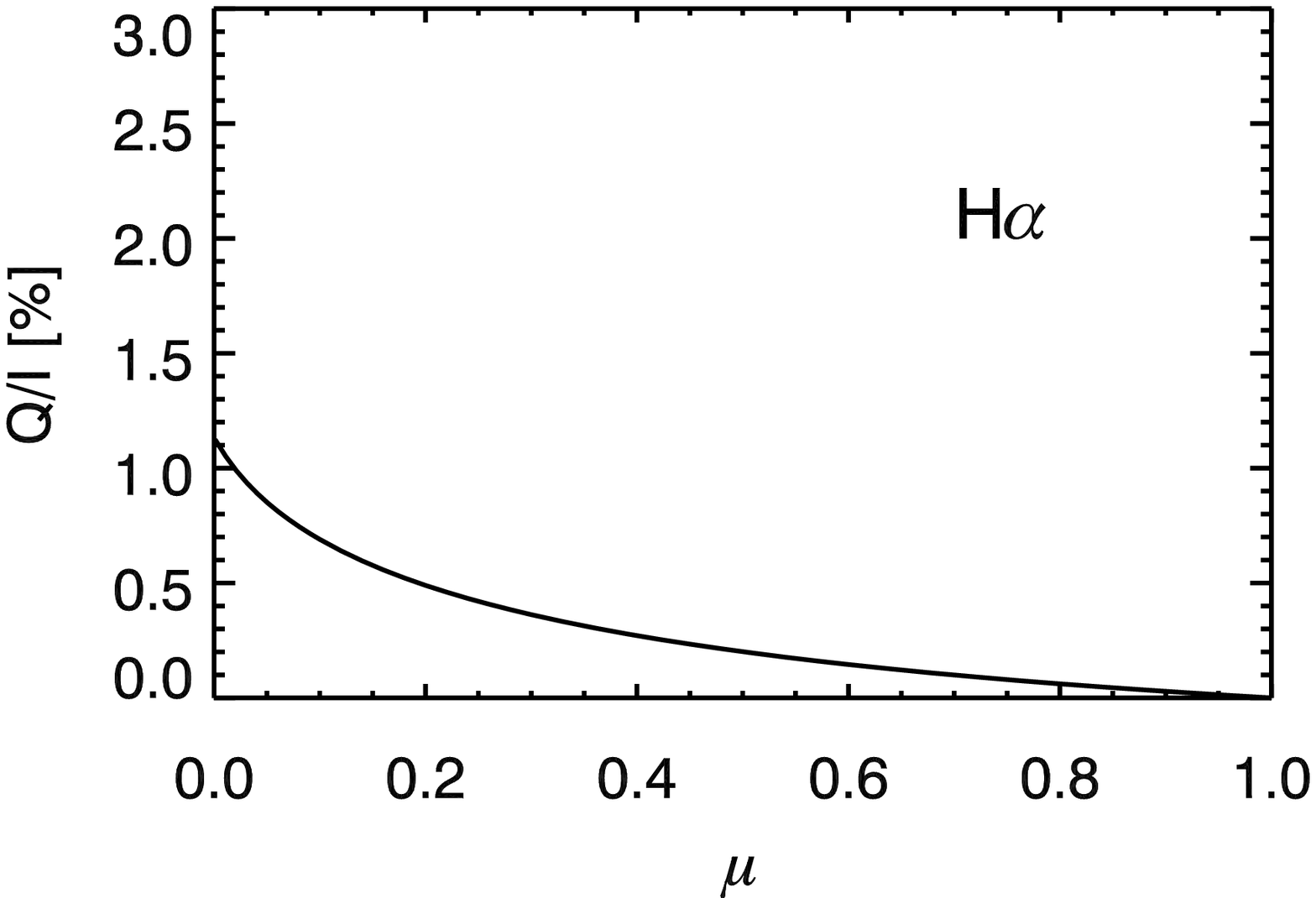}
\caption{
{\it Left panels}: CLV of the emergent $Q/I$ profiles of \La\/ ({\it top}), \Lb\/ ({\it middle}), and \Ha\/ ({\it bottom}). The LOS direction cosine, $\mu=\cos\theta$, varies from 0.1 ({\it solid lines}) to 1 ({\it dash-triple dotted lines}). {\it Right panels}: CLV of the ensuing line-center polarization amplitudes.}
\label{fig:clv-nm1}
\end{figure*}

The left panels of Fig.~\ref{fig:clv-nm1} show the fractional linear polarization profiles calculated for various line-of-sight inclinations using our nominal value for the perturber's density (i.e., $N_{\rm pert}=4\times 10^{10}\,{\rm cm^{-3}}$), while the right panels show the center to limb variation of the line-center amplitudes. The scattering polarization of the \La\/ line is produced in the uppermost layers of the model atmosphere by the atomic alignment of the $2p_{3/2}$ level that is induced by anisotropic radiation pumping in the \La\/ transition itself (i.e., $1s_{1/2}$--$2p_{3/2}$). The amplitude of the emergent $Q/I$ signal can be estimated using Eq.~\ref{eq:qieddb} from the fractional alignment $\rho^2_0(2p_{3/2})/\rho^0_0(2p_{3/2})$ at the height where the line-center optical depth along the LOS is unity (cf. the top left panel of Fig.~\ref{fig:dc1dp1}). Given that the fractional alignment increases with height, the maximum polarization amplitude is found close to the limb (see the top right panel of Fig.~\ref{fig:clv-nm1}).

The formation of the \Lb\/ line is affected by its coupling with \Ha\/ through the $n=3$ levels. This affects mainly the wings of the \Lb\/ line which show an increase of the $Q/I$ signal (see the middle left panel of Fig.~\ref{fig:clv-nm1}). The wings are formed in the region where the line center optical depth of \Lb\/ is much higher than unity, but the polarization of the level $3p_{3/2}$ is non-zero (see the discussion in \S\ref{ssec:anis1}). This effect is more obvious for a LOS with a large $\mu$ value, i.e., closer to the disk center, where the ratio of the $Q/I$ signal in the line center and in the wings becomes smaller due to a relatively higher contribution of the deeper layers between 1100\,km and 1600\,km. Since we only consider thermal Doppler broadening in our model, the resulting  profiles are quite narrow. If the natural and Stark broadening were taken into account, one would obtain broader $Q/I$ wings in \Lb. We note, however, that even though the partial redistribution effects in \Lb\/ are weaker than those of \La\/ \citep[e.g.,][]{vernazza81,hubeny95}, coherent scattering in the wings could significantly modify the resulting line profiles.


\section{The magnetized case}
\label{sec:iso-uni-mag}

\begin{figure*}
\plottwo{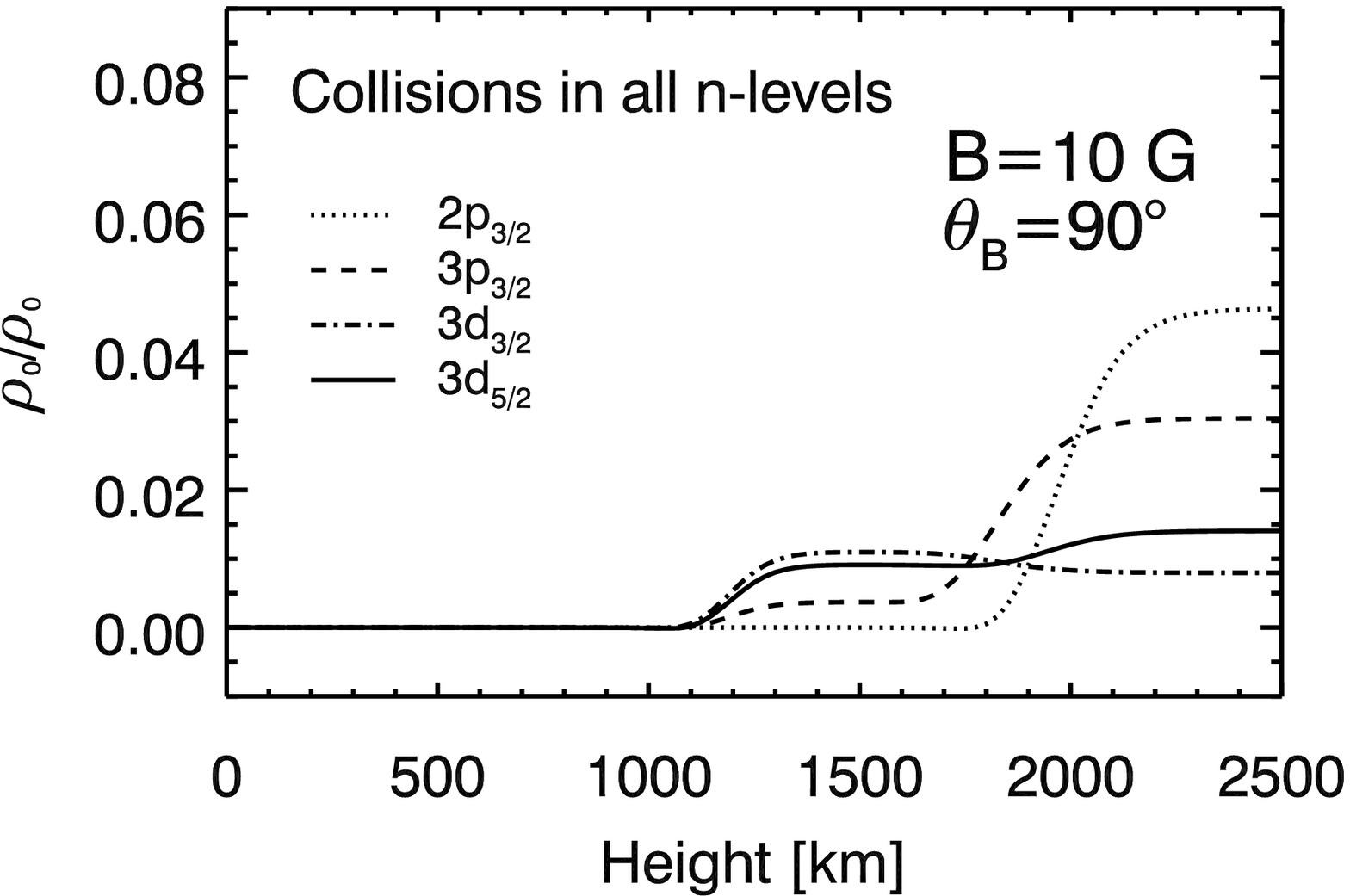}{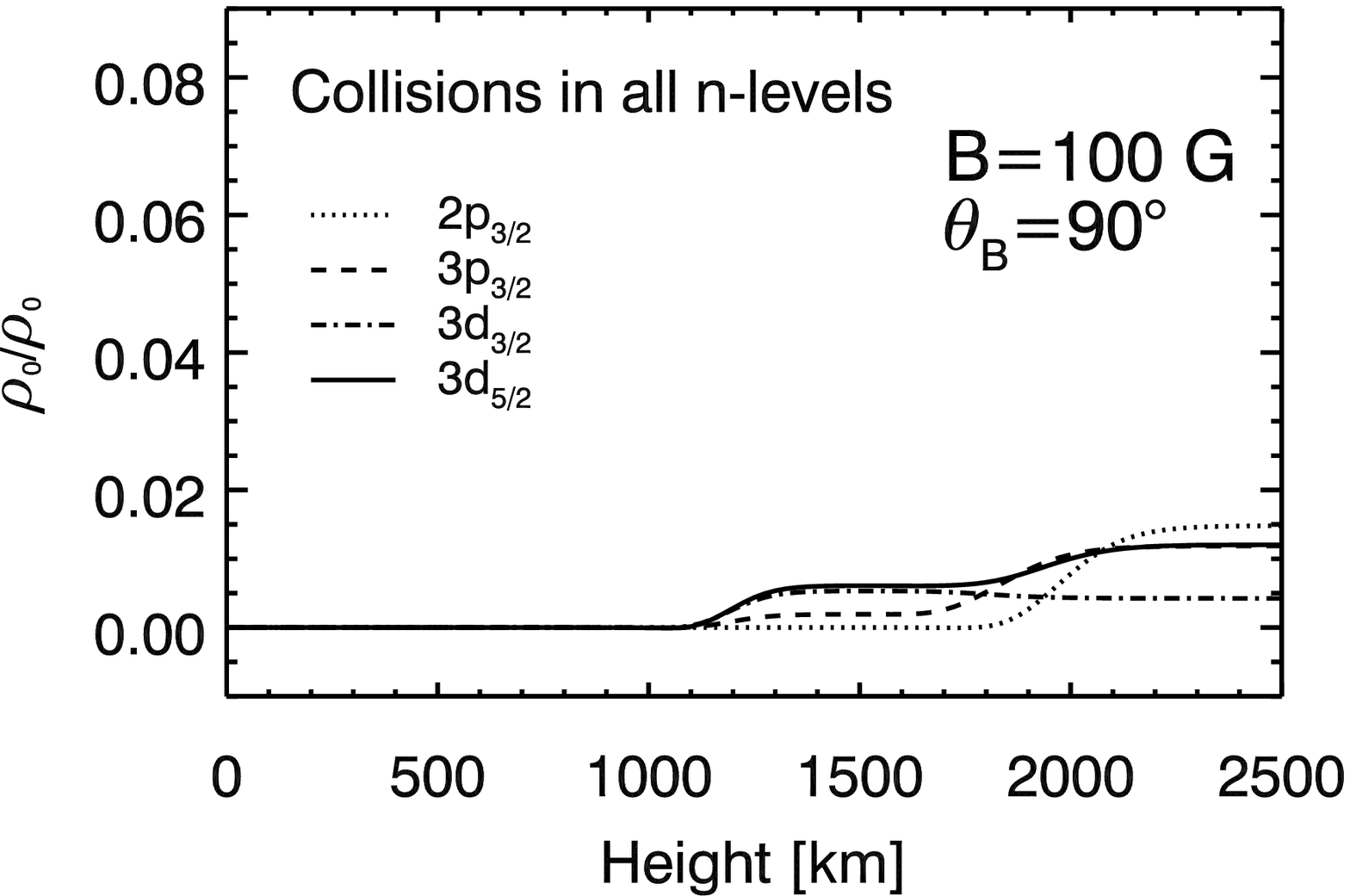}
\caption{
Fractional alignment of the fine-structure levels in a uniformly magnetized model atmosphere. The perturber's density is $N_{\rm pert}=4\times 10^{10}\,{\rm cm^{-3}}$. {\it Left}: Horizontal magnetic field with random azimuth and strength $B=10$\,G. {\it Right}: Horizontal magnetic field with random azimuth and strength $B=100$\,G.
}
\label{fig:ra2ra3}
\end{figure*}

The modification of the emergent linear polarization by the Hanle effect provides a tool for diagnostics of solar and stellar magnetic fields \citep{stenflobook,ll04,jtb01}. In this section, we discuss the  effects of magnetic fields, both deterministic and with random azimuth, on the scattering polarization of hydrogen lines. 


\subsection{Micro-structured magnetic field}
\label{ssec:mg-align}

Here we consider the micro-structured field case in which the intensity $B$ and the inclination $\theta_B$ are fixed at each height of the atmosphere but the azimuth $\chi_B$ is random with a uniform distribution in $[0,2\pi)$ at scales smaller than the mean free path of the line-center photons. The fact that under such circumstances the magnetic field is cylindrically symmetric with respect to the vertical axis implies that the rotational symmetry of the problem is preserved, so that the $\rho^K_{Q\neq 0}$ coherences vanish throughout the atmosphere. The problem is thus numerically similar to the case of a non-magnetic atmosphere (see Appendix~\ref{app:microturb} for details). The difference is that the micro-structured field decreases the alignment of the levels. The $Q/I$ profiles are thus depolarized and the Stokes-$U$ parameter remains identically zero.

\begin{figure}
\begin{center}$
\begin{array}{ccc}
\includegraphics[width=2.3in]{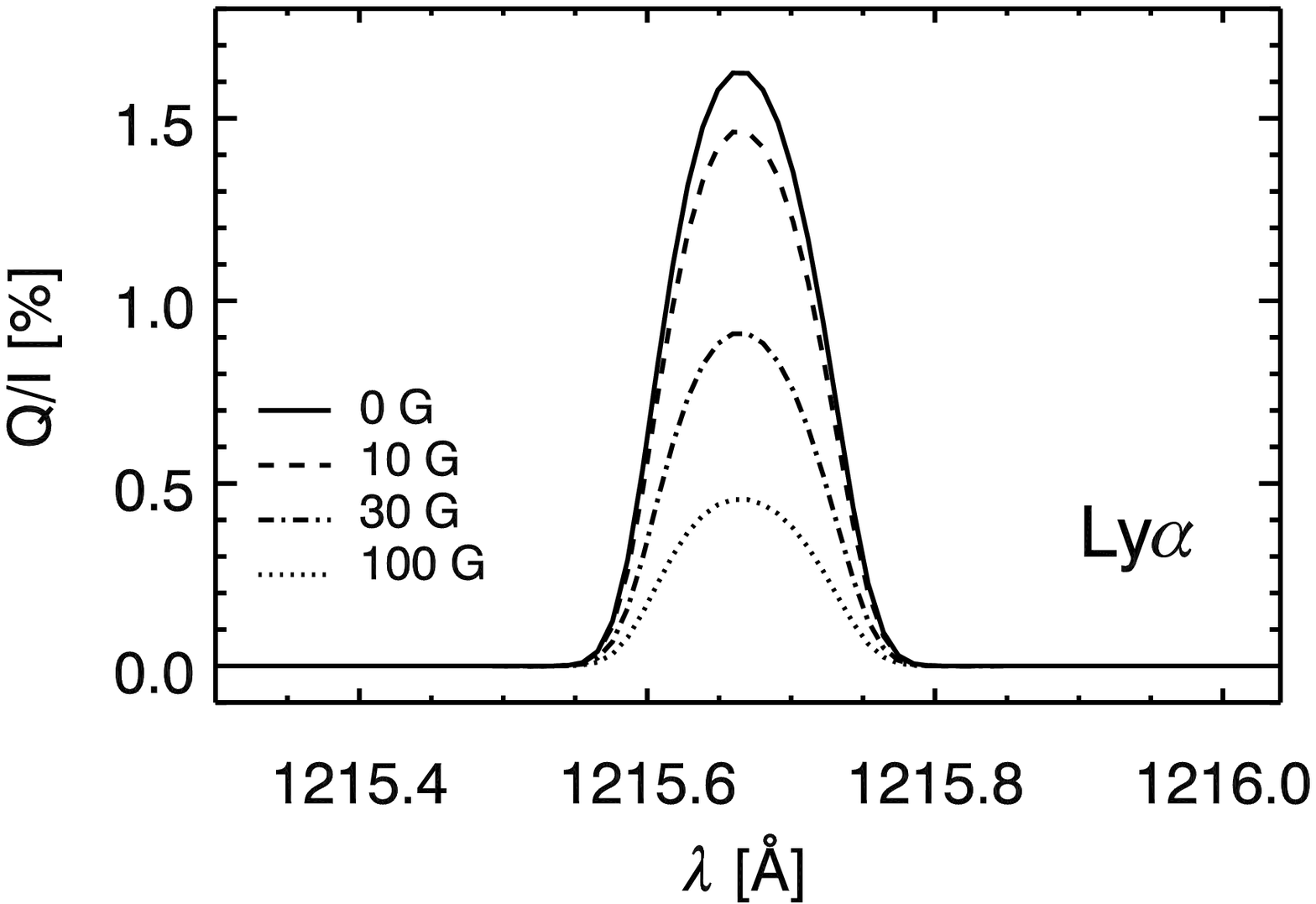} &
\includegraphics[width=2.3in]{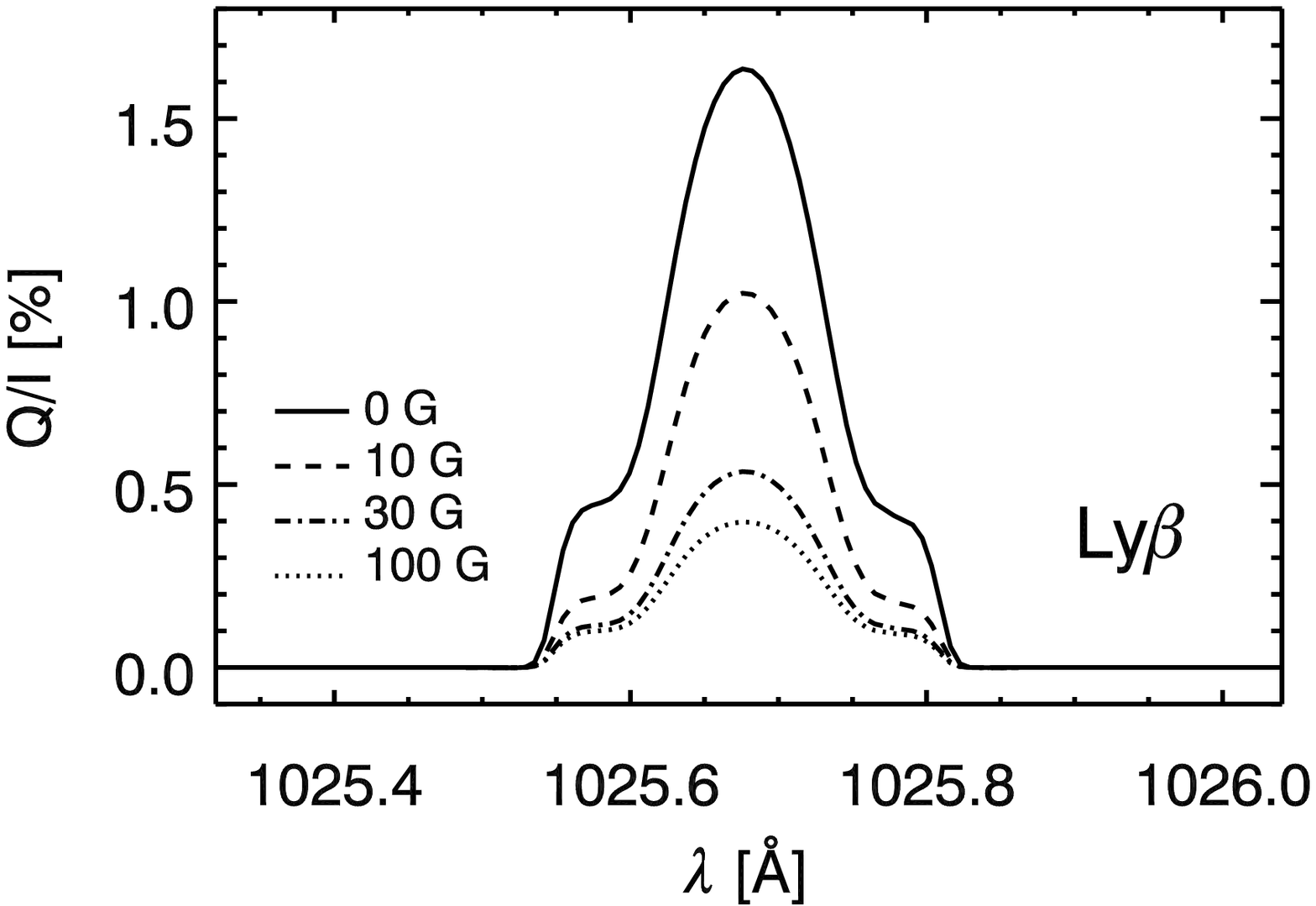} &
\includegraphics[width=2.3in]{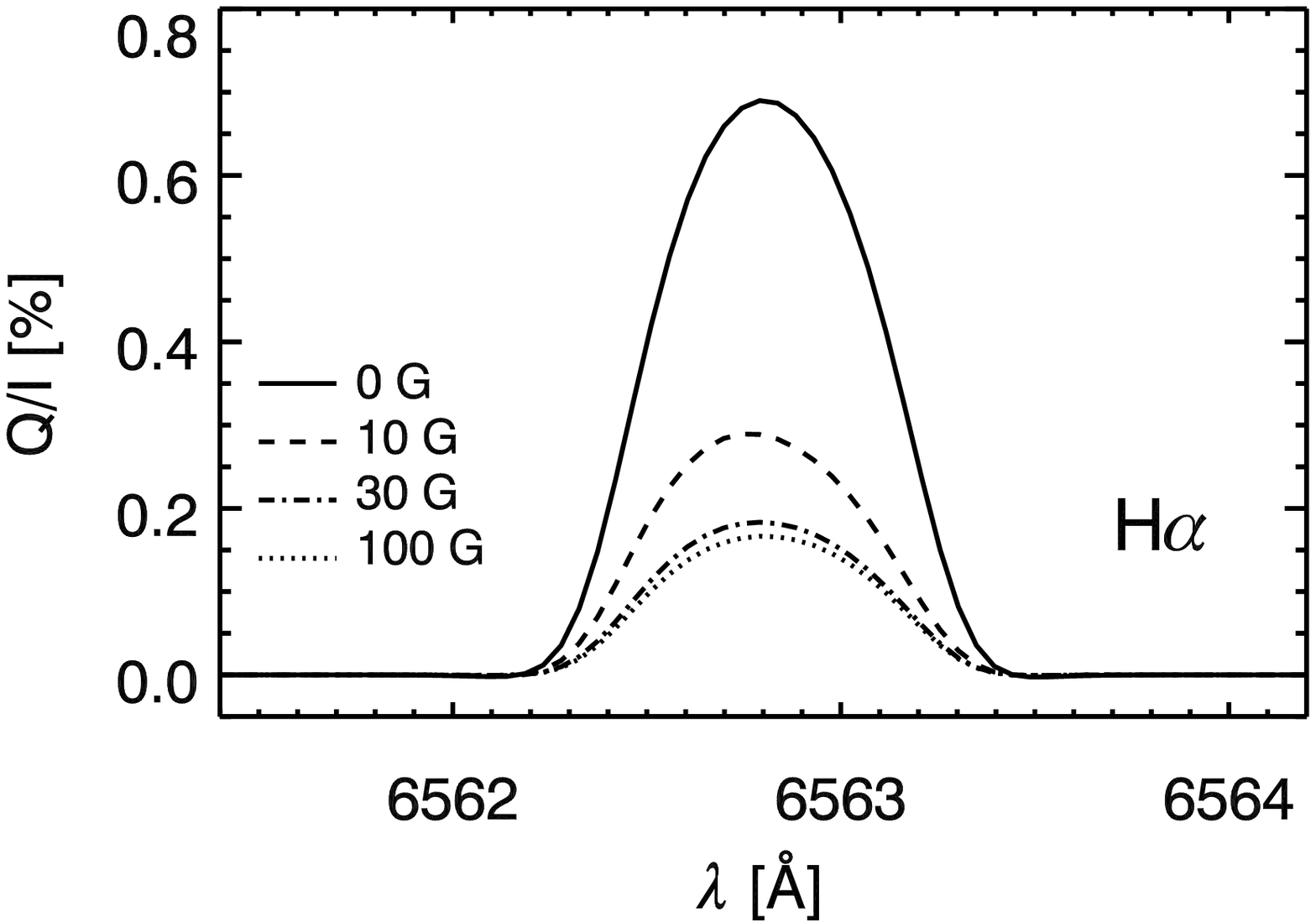}
\end{array}$
\end{center}
\caption{
The emergent $Q/I$ profiles of \La\/ ({\it left}), \Lb\/ ({\it middle}), and \Ha\/ ({\it right}) calculated for a LOS with $\mu=0.1$. The magnetic field is uniform, with a random azimuth $\chi_B$ and a constant inclination $\theta_B=90^\circ$. Three magnetic field strengths are considered: 0\,G ({\it solid lines}), 10\,G ({\it dashed lines}), 30\,G ({\it dash-dotted lines}), and 100\,G ({\it dotted lines}).
}
\label{fig:lmb-um1}
\end{figure}

It is sufficient to consider two magnetic field strengths, namely 10 and 100\,G. It follows from Table~\ref{tab:critical} that a 10\,G field is sufficient to modify the alignment of the $3p_{3/2}$ level and especially those of the $3d_{3/2}$ and $3d_{5/2}$ levels. The $2p_{3/2}$ level should be only slightly modified because its critical Hanle field, $B_H$, is about 53\,G (i.e., well above 10\,G). A 100\,G field will significantly alter the atomic polarization of the $2p_{3/2}$ level while the $3\ell j$ levels should already be in the Hanle effect saturation regime. Fig.~\ref{fig:ra2ra3} shows the variation with height in our stellar atmosphere model of the fractional alignment of the hydrogen levels, for the case of a random-azimuth horizontal magnetic field with such field strengths.

The left panel of Fig.~\ref{fig:ra2ra3} shows the results for the 10\,G case, using the same scale as in Fig.~\ref{fig:dc1dp1}. As expected, $\rho^2_0(2p_{3/2})/\rho^0_0(2p_{3/2})$ is decreased but since the critical field of this level is five-times larger than the applied magnetic field, the change is small. The strongest depolarization occurs for the $3d_j$ levels whose alignment is decreased by about a factor two with respect to the non-magnetic model. At heights lower than about 1800\,km we have $|\rho^2_0(3d_{5/2})/\rho^0_0(3d_{5/2})|<|\rho^2_0(3d_{3/2})/\rho^0_0(3d_{3/2})|$, i.e., the fractional polarization of the $3d_{3/2}$ level exceeds that of $3d_{5/2}$. The reason is that the $3d_{5/2}$ level is sensitive to weaker magnetic fields than the $3d_{3/2}$ level (see Table~\ref{tab:critical}). The dashed lines of Fig.~\ref{fig:lmb-um1} show the emergent $Q/I$ profiles corresponding to the 10\,G case. As expected, the $Q/I$ profile of \La\/ is similar to the non-magnetic one (see the solid lines). On the other hand, the \Lb\/ and \Ha\/ lines become noticeably depolarized.

The right panel of Fig.~\ref{fig:ra2ra3} considers the 100\,G case. Note that now the levels are significantly more depolarized, to the extent that the Hanle effect in the $n=3$ levels becomes virtually saturated. The alignment of the $2p_{3/2}$ level (i.e., the only \La\/ level that can be aligned) is also significantly reduced. The fractional alignment of the 3$d_j$ level and of the $3p_{3/2}$ level is approximately a factor four smaller than in the non-magnetic model \citep[as expected for the saturation field regime, see][]{mansosainz06} and the relation $|\rho^2_0(3d_{5/2})/\rho^0_0(3d_{5/2})|>|\rho^2_0(3d_{3/2})/\rho^0_0(3d_{3/2})|$ is satisfied again throughout the atmosphere because the field intensity is much larger than the critical fields of both of $3d_j$ levels. The emergent $Q/I$ profiles of the 100\,G model are given by the dotted lines in Fig.~\ref{fig:lmb-um1}. As expected, the amplitudes of the \Ha\/ and \Lb\/ signals are a factor four smaller than in the non-magnetic model.


\subsection{Deterministic magnetic field}

\label{ssec:mg-fwd}
\begin{figure*}
\plottwo{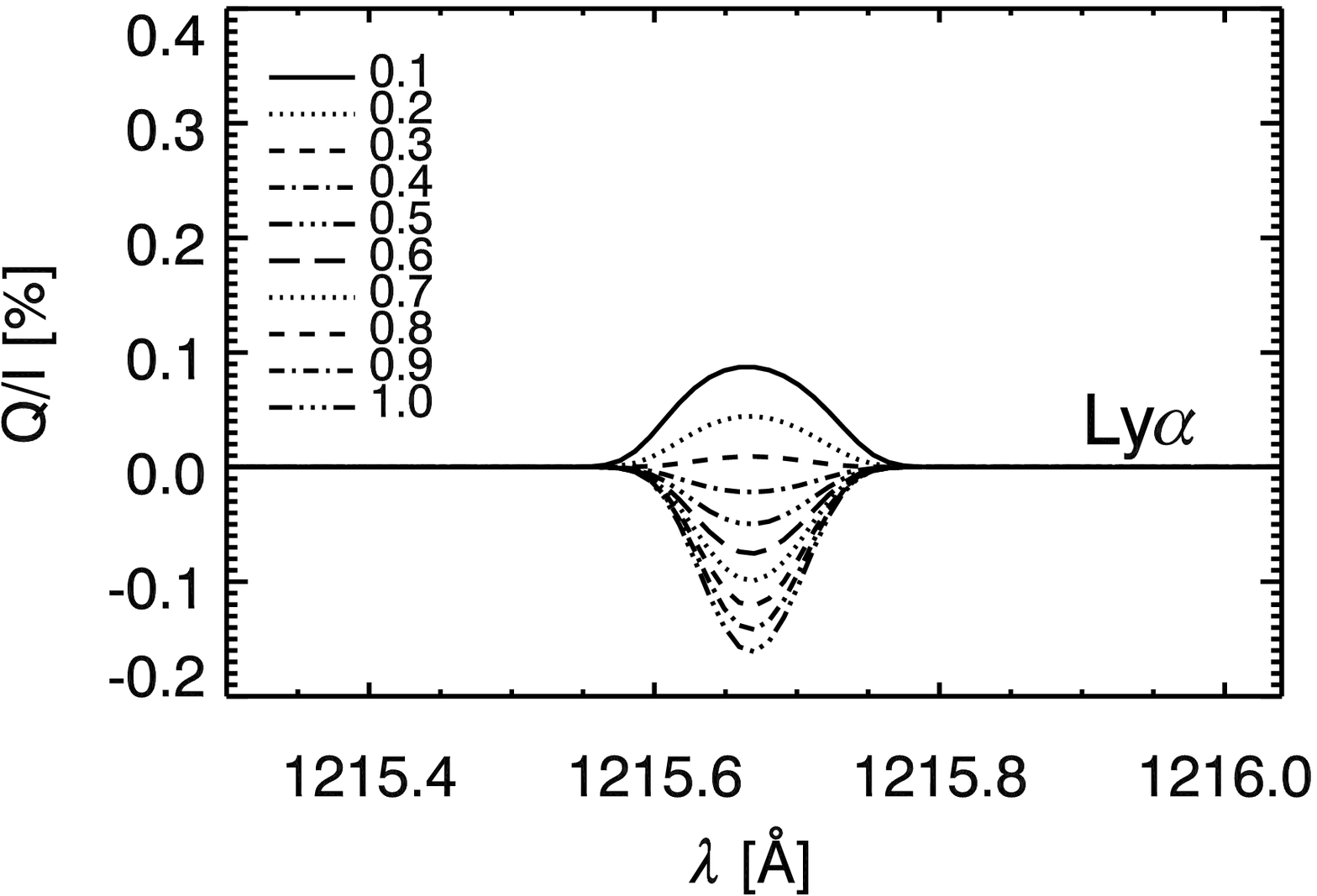}{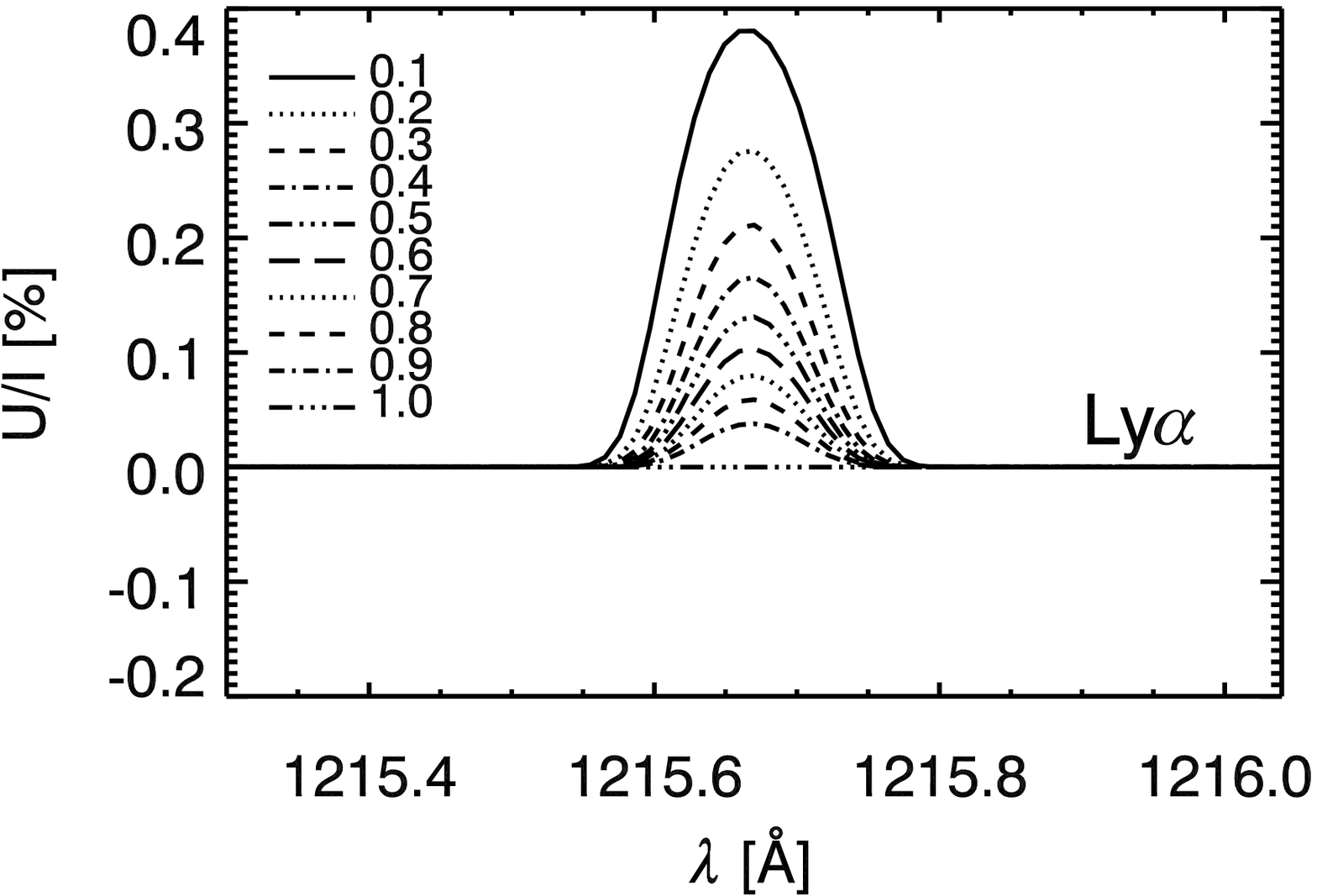}
\plottwo{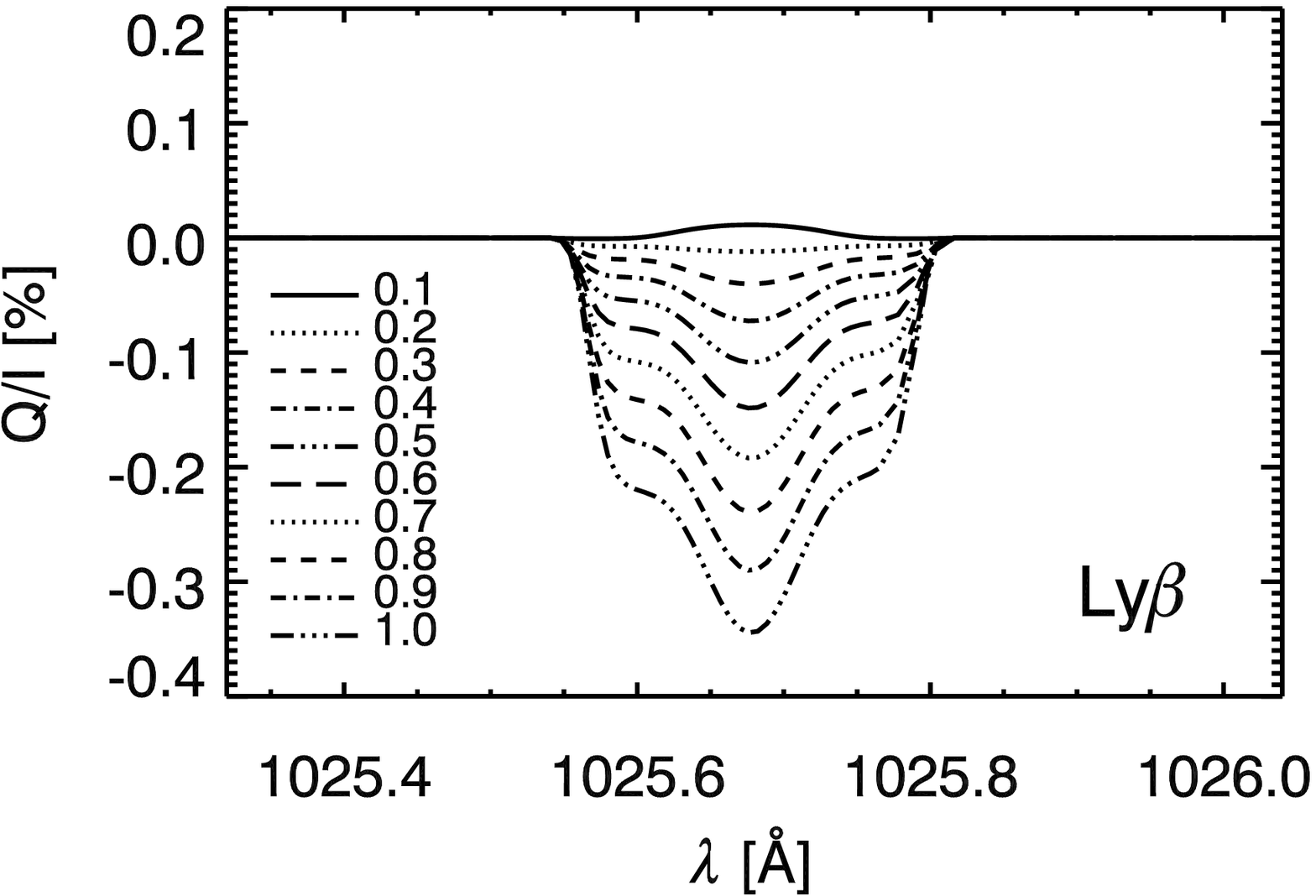}{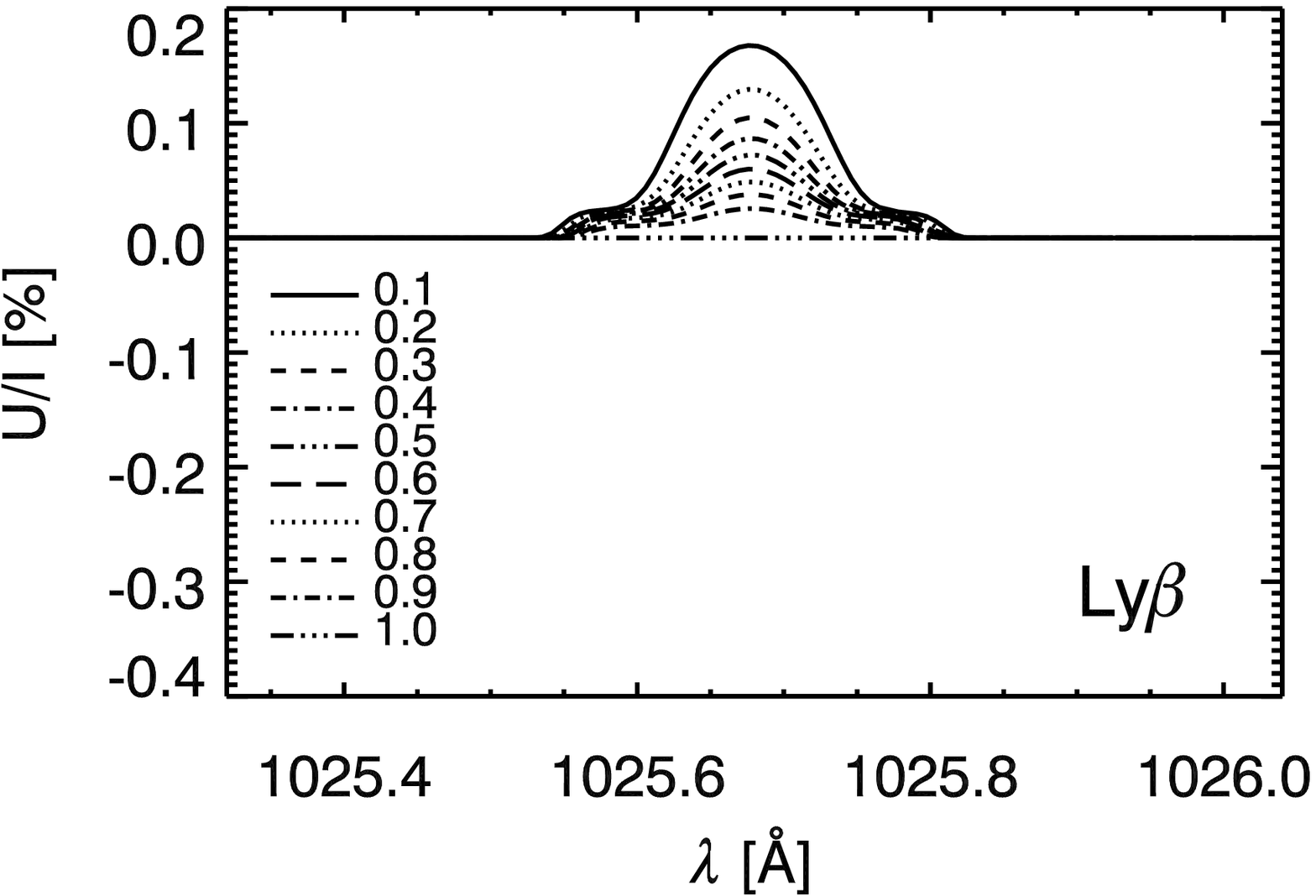}
\plottwo{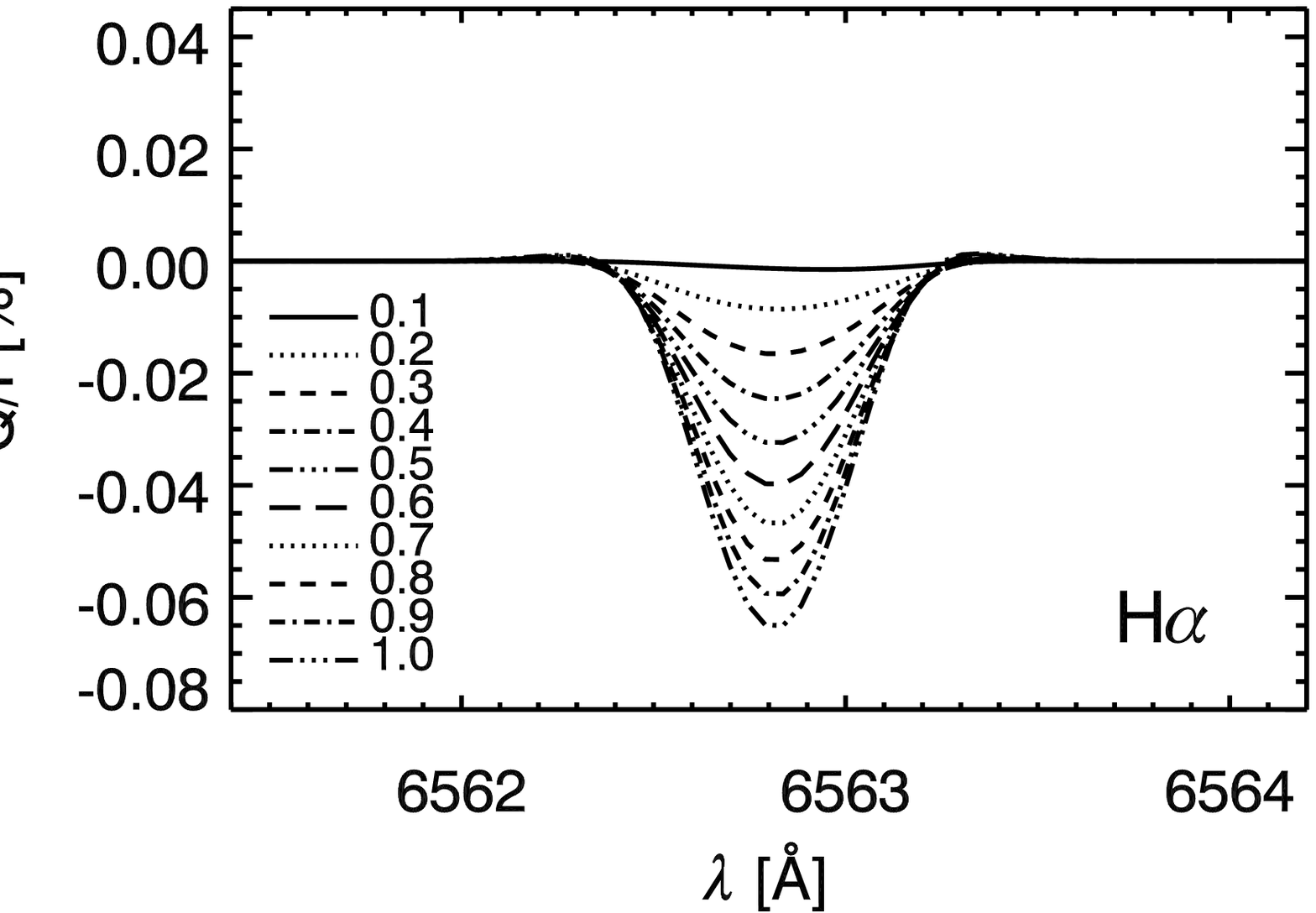}{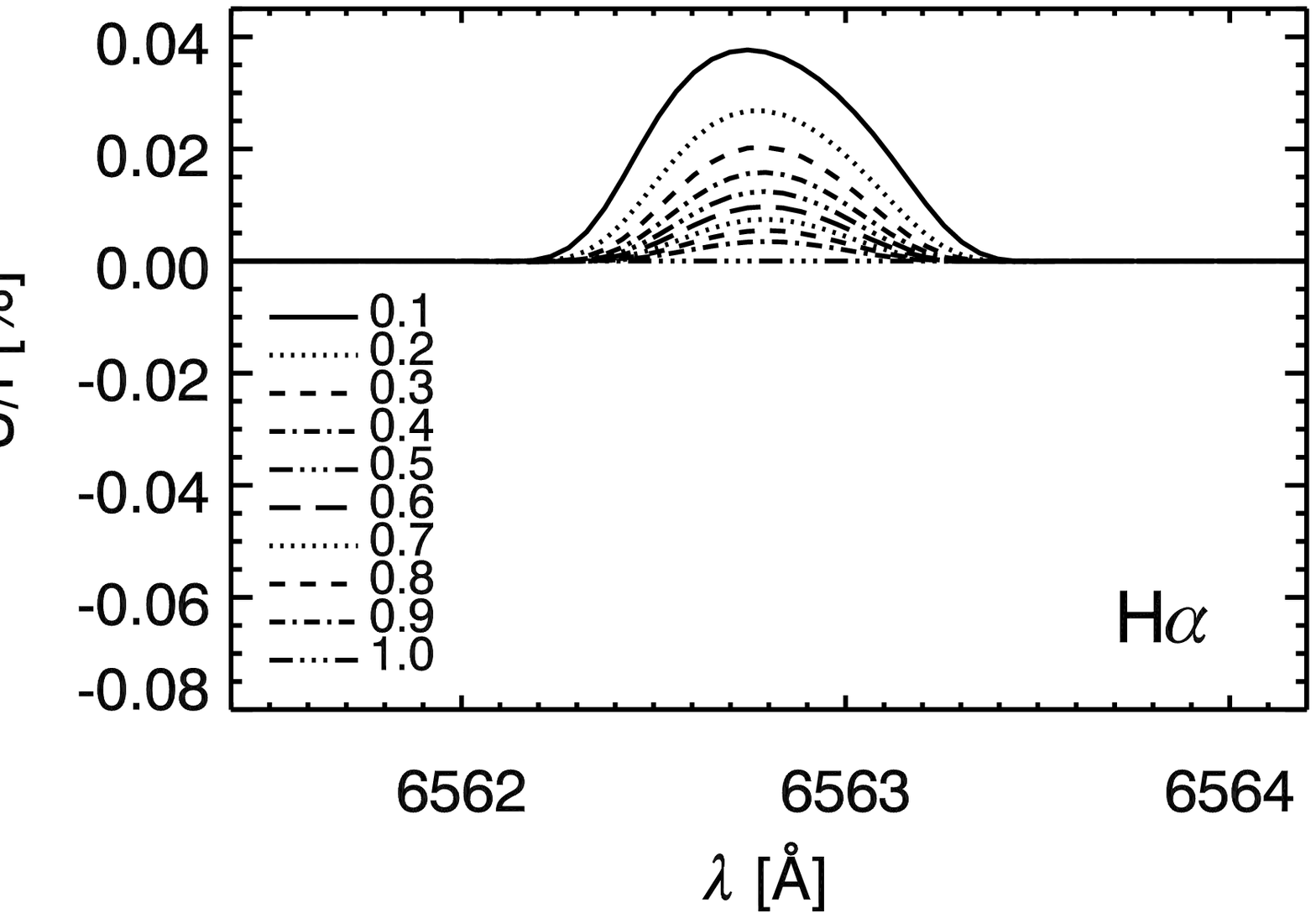}
\caption{
CLV of the $Q/I$ and $U/I$ line profiles for the case of a horizontal magnetic field with $B=100$\,G and azimuth $\chi_B=0^\circ$. From top to bottom: \La, \Lb, and \Ha\/ at the indicated $\mu$-value of the LOS. The positive $Q$-direction is the perpendicular to the stellar radius through the observed point. Note that for the forward scattering case the positive $Q$-direction is the perpendicular to the horizontal magnetic field.
}
\label{fig:clv-qu}
\end{figure*}

The positive reference direction we have chosen for Stokes $Q$ is the perpendicular to the stellar radius through the observed point. Therefore, it is clear that the Stokes $U$ parameter is zero if the magnetic field has a random-azimuth distribution, but non-zero in general if the atmosphere is permeated by a deterministic magnetic field vector (i.e., with a well-defined inclination and azimuth). This is illustrated in Fig.~\ref{fig:clv-qu}, which shows the center to limb variation of the $Q/I$ and $U/I$ profiles for the case of a horizontal magnetic field of 100\,G with azimuth $\chi_B=0^{\circ}$ (i.e., pointing towards the observer for a LOS with $\mu=0$ and perpendicular to the LOS for a $\mu=1$ disk center observation). 

\begin{figure}
\begin{center}$
\begin{array}{ccc}
\includegraphics[width=2.3in]{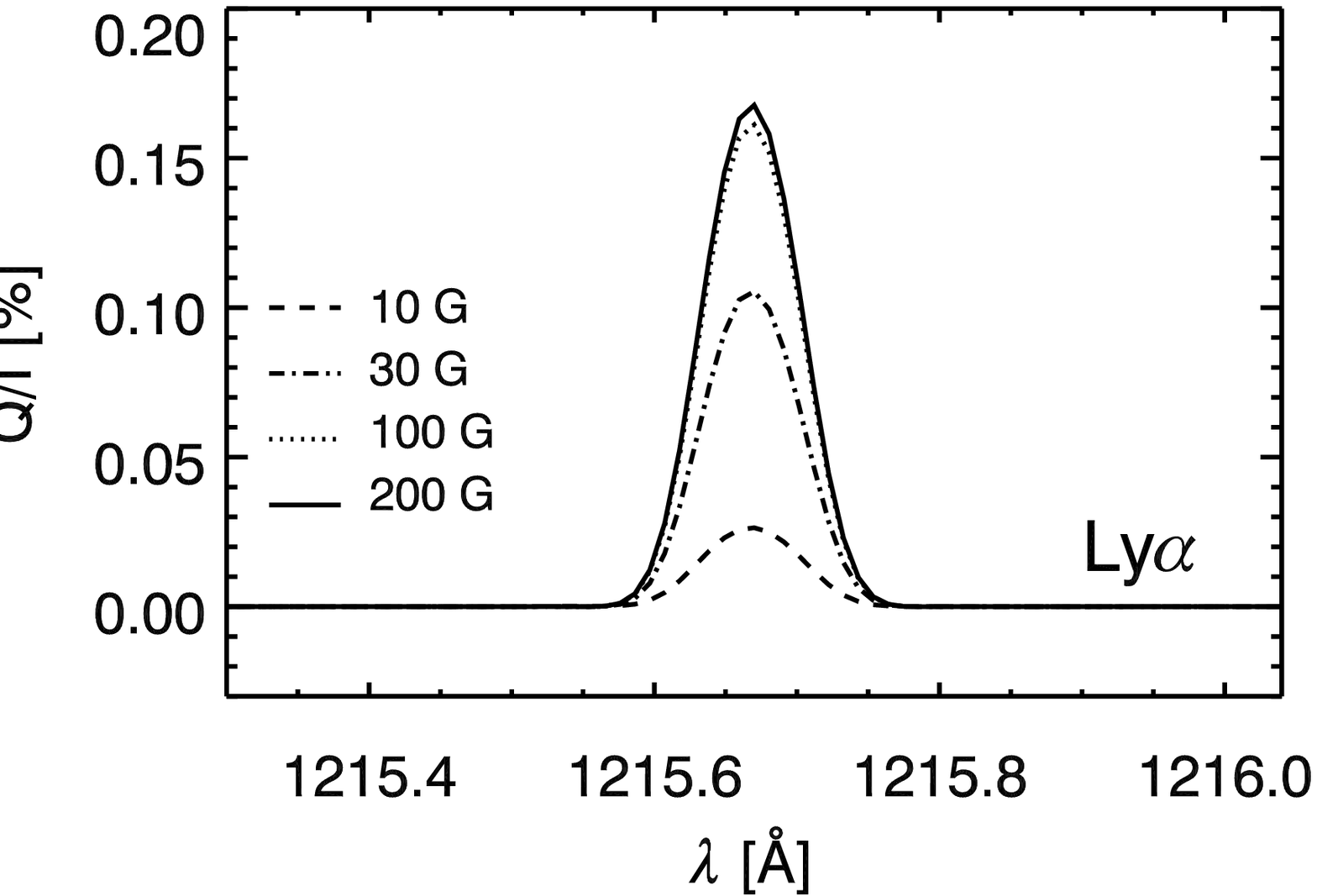} &
\includegraphics[width=2.3in]{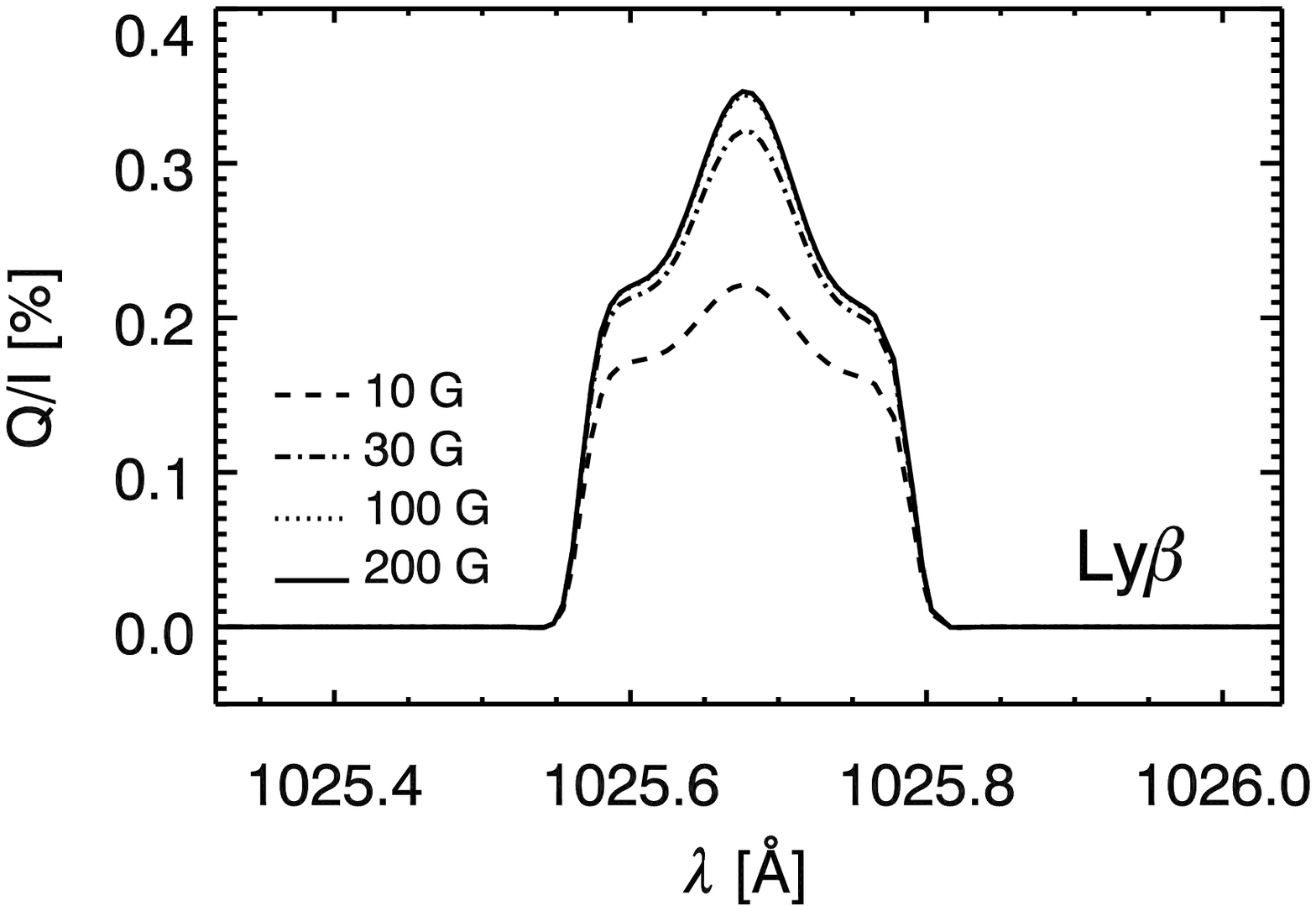} &
\includegraphics[width=2.3in]{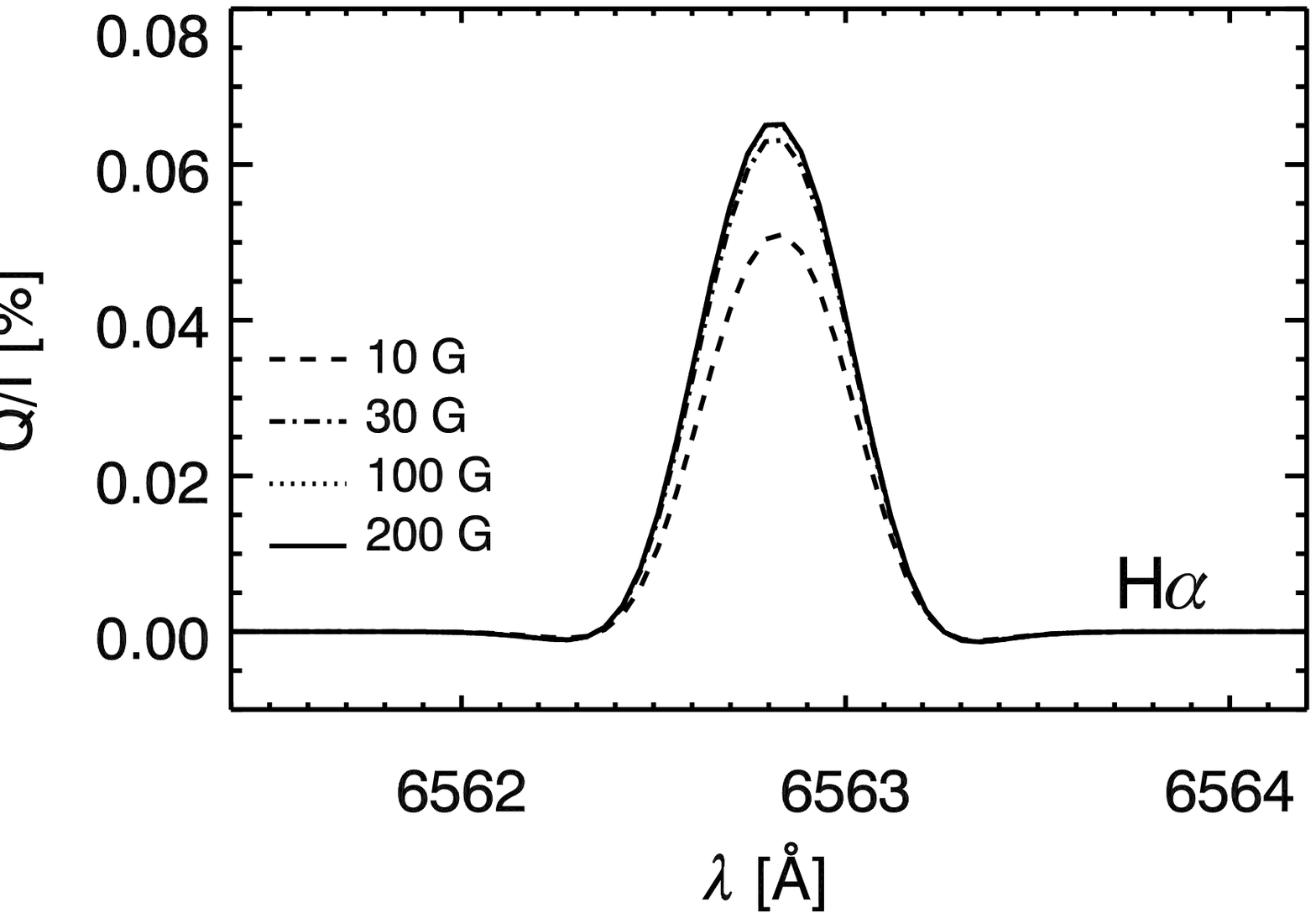}
\end{array}$
\end{center}
\caption{
The emergent $Q/I$ profiles of \La\/ ({\it left}), \Lb\/ ({\it middle}), and \Ha\/ ({\it right}) calculated at the disk center ($\mu=1$). The magnetic field in the atmosphere is uniform with the fixed azimuth $\chi_B=90^\circ$ and the constant inclination $\theta_B=90^\circ$. Three magnetic field strengths are considered: 10\,G ({\it dashed lines}), 30\,G ({\it dashed-dotted lines}), 100\,G ({\it dotted lines}), and 200\,G ({\it solid lines}).
}
\label{fig:fwd-um1}
\end{figure}

Note in Fig.~\ref{fig:clv-qu} that for the particular case of forward scattering geometry (i.e., LOS with $\mu=1$) we have $U/I=0$, while the $Q/I$ signal is the largest. The fact that for this case $U/I=0$ can be easily understood by symmetry reasons. Fig.~\ref{fig:fwd-um1} illustrates how the forward scattering polarization signals change with the field strength. As expected, the \Ha\/ forward scattering signal is rather small, due to the low radiation anisotropy around the atmospheric height where the line center optical depth is unity. On the other hand, the fractional linear polarization of \Lb\/ is very significant and the polarization profile is quite broad due to the significant alignment of the $3p_{3/2}$ level between 1200 and 1700\,km. In the limit of strong magnetic field, the polarization of \La\/ is about a factor two smaller than that of \Lb\/ and, in contrast to \Ha\/ and \Lb\, it appears at higher field strengths. 


\subsection{The impact of magnetic field gradients}
\label{ssec:gradients}

\begin{figure*}
\plottwo{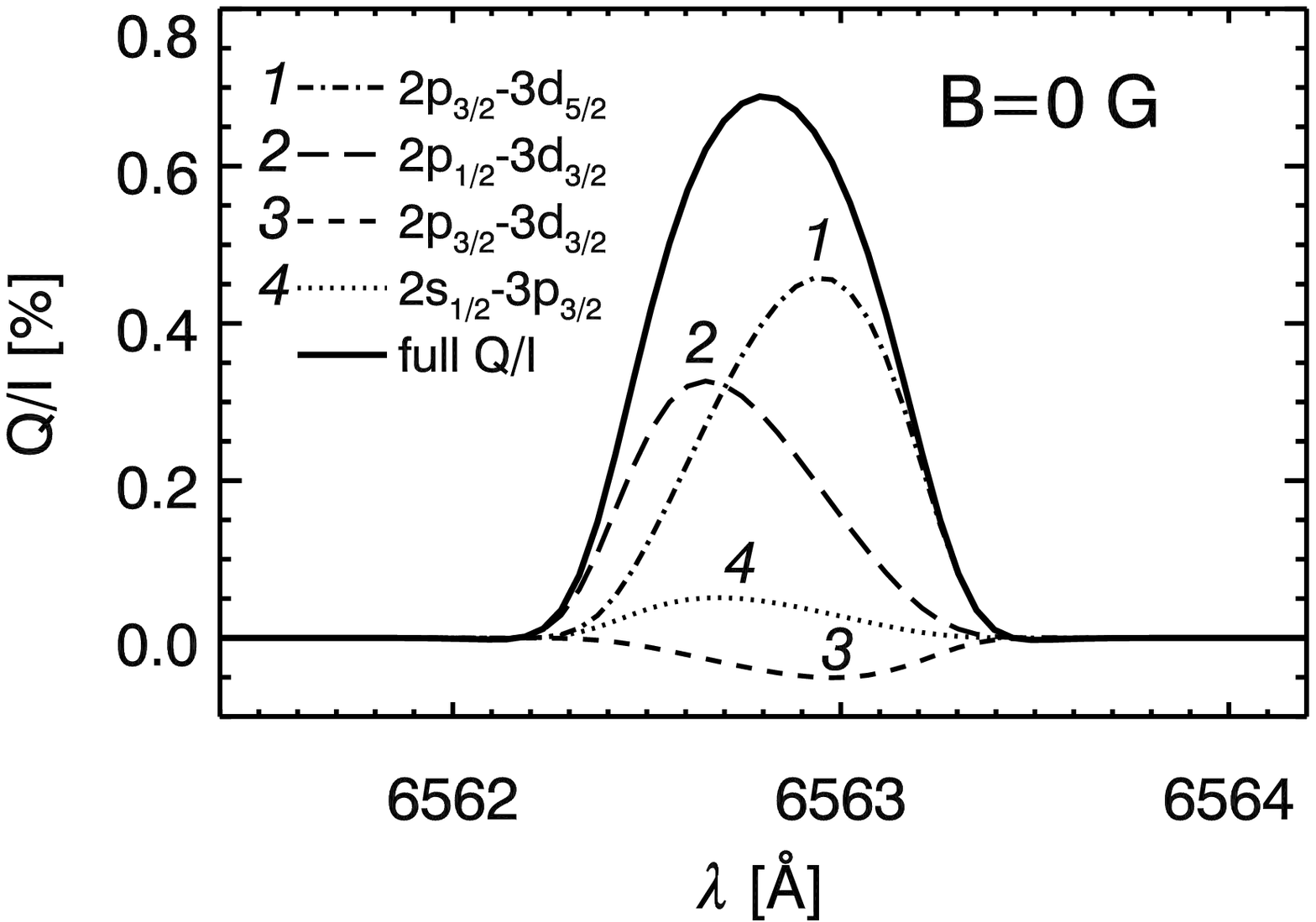}{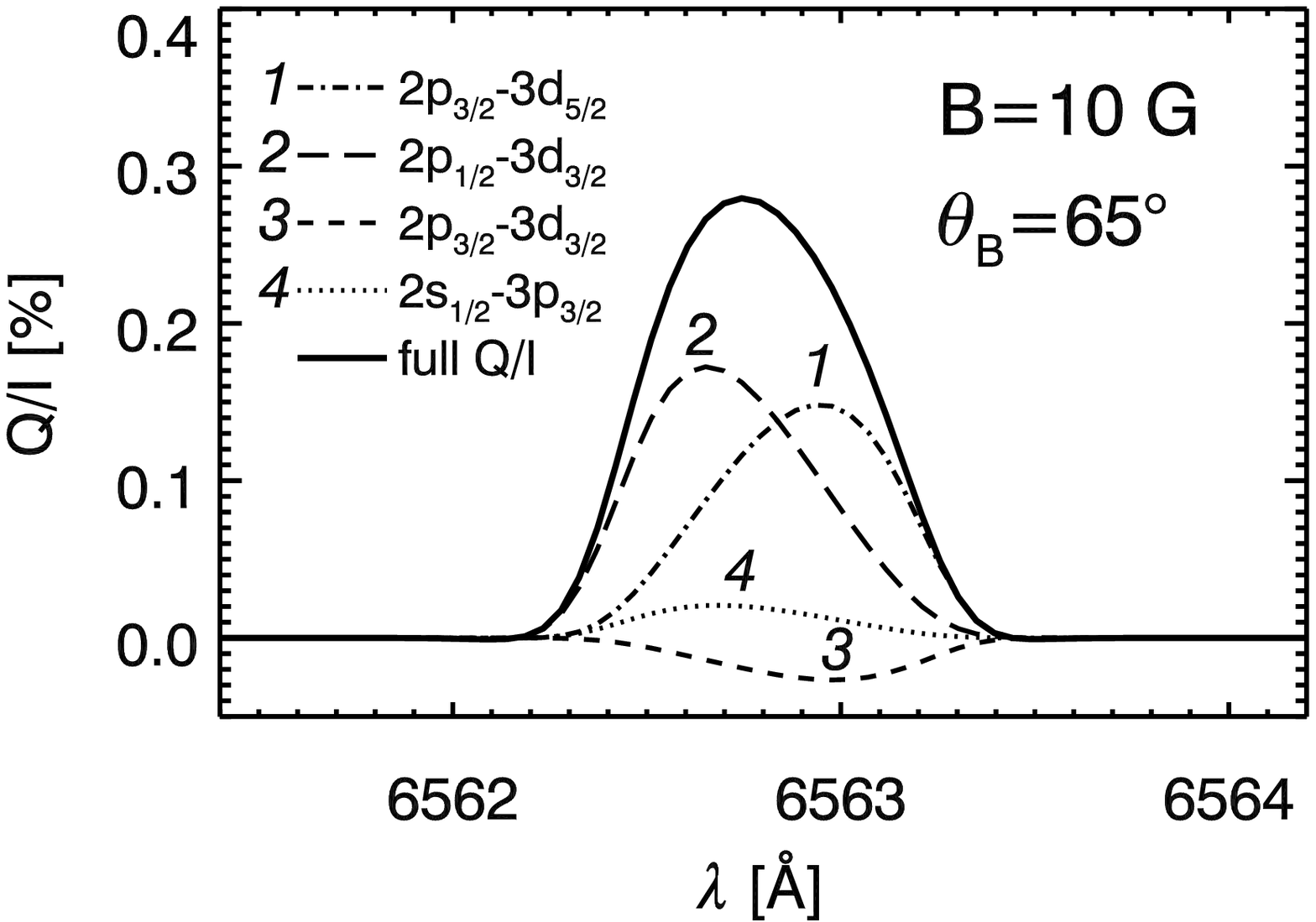}
\plottwo{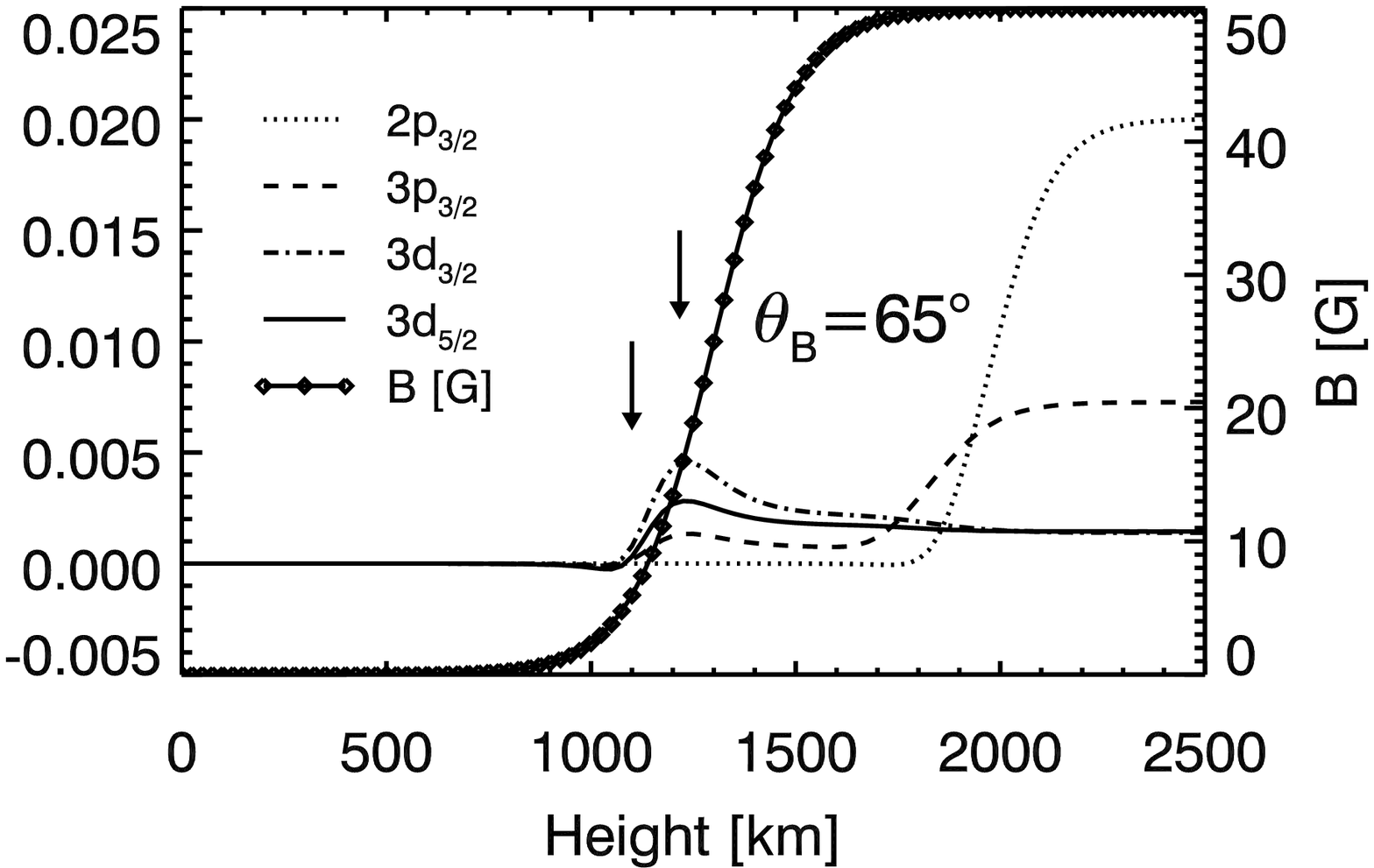}{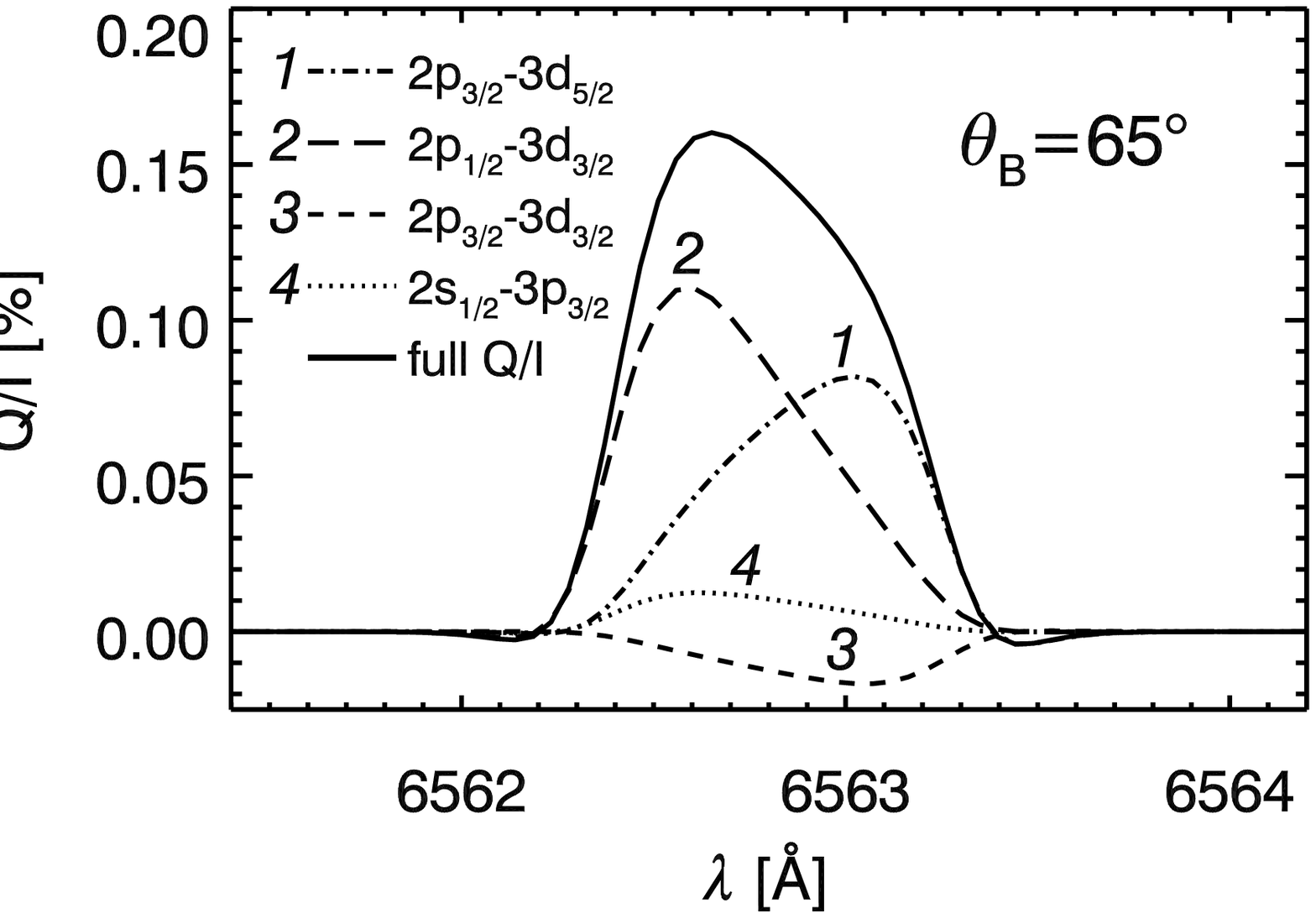}
\caption{
The $Q/I$ profiles of \Ha\/ and of its four main polarizing components calculated for a LOS with $\mu=0.1$ and the influence of a magnetic field. {\it Top left}: The case of a non-magnetic atmosphere. {\it Top right}: The case of a micro-structured magnetic field with $B=10$\,G and inclination $\theta_B=65^\circ$. The two bottom panels illustrate the effect of gradients in the magnetic field intensity. {\it Bottom left}: Variation of the magnetic field strength and fractional alignment of the levels against height in the model atmosphere. The chosen magnetic field strength increases with height ($Z$) according to the sigmoid function $B(Z)=B_0/[1+\exp(-(Z-Z_0)/w)]$, where $B_0=50$\,G, $Z_0=1300$\,km, and $w=100$\,km.  {\it Bottom right}: $Q/I$ profile of the emergent \Ha\/ line.
}
\label{fig:ragrad}
\end{figure*}

Whereas the \La\/ and \Lb\/ Stokes $Q$ signals result from just one polarized transition (1$s_{1/2}$--2$p_{3/2}$ and 1$s_{1/2}$--3$p_{3/2}$, respectively, cf. Fig.~\ref{fig:lst}), the scattering polarization of \Ha\/ is the result of the  superposition of 5 fine-structure transitions between the $n=2$ and $n=3$ $j$-levels that can, in principle, contribute to the line polarization. As we have seen, four of these transitions (i.e., the ones whose upper levels can be aligned) play the dominant role on the \Ha\/ scattering polarization. In solar-like atmospheres the contribution of the dichroic-only transition $2p_{3/2}$--$3s_{1/2}$ can be ignored because the atomic polarization of the $2p_{3/2}$ level is negligible in the \Ha\/ formation region (see \S\ref{ssec:dichr}). The wavelength separation of the \Ha\/ line components is much larger than those of the Lyman lines (see Fig.~\ref{fig:lst}). In the presence of magnetic field gradients the non-negligible wavelength separation of the \Ha\/ components gives rise to a wavelength-dependent Hanle effect, which is of diagnostic interest because it may produce asymmetries in the emergent linear polarization profile \citep{stepanjtb10asym}. In this subsection, we emphasize further the role of these overlapping line components on the radiation transfer problem of scattering polarization. 

The individual \Ha\/ components are displaced from the line center because of the fine-structue splitting and the Lamb shift of the levels. A typical separation of the \Ha\/ components is 0.1\,\AA\/ (see Fig.~\ref{fig:lst}). The thermal Doppler width of the line (which is typically about 0.3\,\AA\/ in the solar chromosphere) is significantly larger and, consequently, the \Ha\/ intensity profile resembles the Gaussian bell. Since the fine-stucture transitions $i$--$j$ of the \Ha\/ line are shifted from line center, the optical depth scale resulting from the opacity due to all possible absorptions at a given frequency is not symmetric with respect to the central wavelength of any given component, $\lambda_{ij}$. Rougly speaking, the photons at wavelengths $\lambda_{ij}-\Delta\lambda$ and $\lambda_{ij}+\Delta\lambda$ originate from different geometrical depths in the atmosphere. Neglecting selective absorption (because $\rho^2_0(2p_{3/2})\approx 0$), the Stokes $Q$ signal of the emergent radiation along a given LOS can be expressed as
\begin{equation}
Q(\lambda)=\int_0^\infty dt\, S_Q(\lambda,t)e^{-t}\,,
\end{equation}
where $t$ is the optical path along the LOS at the wavelength $\lambda$, and $S_Q=\epsilon_Q/\eta_I$ is the source function of Stokes $Q$. The emergent Stokes $Q$-profile can be decomposed as
\begin{equation}
Q(\lambda)=\int_0^\infty dt\,\sum_{ij} S^{ij}_Q(\lambda,t){\rm e}^{-t}=\sum_{ij}Q_{ij}(\lambda)\,,
\end{equation}
where $Q_{ij}$ and $S^{ij}_Q=\epsilon^{ij}_Q/\eta_I$ are the profile components of the individual transitions and their source functions, respectively. We emphasize that $\eta_I$ is the total absorption coefficient resulting from all the \Ha\/ transitions, whereas $\epsilon^{ij}_Q$ is the emissivity of the $i$--$j$ transition only. 

Let us now consider a particular point on the LOS parameterized by the geometrical path $s$. The emission coefficient $\epsilon^{ij}_Q$ is symmetric with respect to the component central wavelength $\lambda_{ij}$, i.e., $\epsilon^{ij}_Q(\lambda_{ij}-\Delta\lambda,s)=\epsilon^{ij}_Q(\lambda_{ij}+\Delta\lambda,s)$. From now on, we will denote this quantity by $\epsilon^{ij}_Q(\lambda_{ij}\pm\Delta\lambda,s)$. The total absorption coefficient $\eta_I$ is only approximately symmetric with respect to the line-center wavelength. On the other hand, it is not symmetric with respect to the $\lambda_{ij}$ wavelengths, i.e., $\eta_I(\lambda_{ij}-\Delta\lambda,s)\neq\eta_I(\lambda_{ij}+\Delta\lambda,s)$. Changing the integration variable from $t$ to $s$, it follows that the difference
\begin{eqnarray}
Q_{ij}(\lambda_{ij}-\Delta\lambda)-Q_{ij}(\lambda_{ij}+\Delta\lambda)=
\int_0^\infty ds\, \epsilon^{ij}_Q(\lambda_{ij}\pm\Delta\lambda,s)\nonumber\\
\times\left(e^{-\int_0^s ds'\, \eta_I(\lambda_{ij}-\Delta\lambda,s')}-e^{-\int_0^s ds'\, \eta_I(\lambda_{ij}+\Delta\lambda,s')}\right)\,,
\end{eqnarray}
is generally non-zero and that the component $Q_{ij}$ of the line is asymmetric with respect to its central wavelength $\lambda_{ij}$. We may thus conclude that (1) the $Q_{ij}/I$ profiles are asymmetric even in the absence of magnetic field and (2) the formation depth of the $Q_{ij}/I$ components is a non-trivial function of wavelength.

The top left panel of Fig.\,\ref{fig:ragrad} shows the $Q_{ij}/I$ components and the full $Q/I$ profile calculated for a LOS with $\mu=0.1$ in our isothermal model atmosphere, assuming $B=0$ gauss. A comparison with the transitions central wavelengths (see right panel of Fig.~\ref{fig:lst}) shows that the positions of the $Q_{ij}/I$ maxima are shifted from the line center towards the wings. This fact can be understood by considering the component's source function $S^{ij}_Q(\lambda)=\epsilon^{ij}_Q(\lambda)/\eta_I(\lambda)$ at any point in the atmosphere. One realizes that the maximum of $S^{ij}_Q(\lambda)$ is always shifted towards the wing with respect to the maximum of $\epsilon^{ij}_Q(\lambda)$ because the denominator $\eta_I(\lambda)$ decreases towards the wing. Note that in the absence of magnetic fields the $Q/I$ profile of \Ha\/ is clearly dominated by transitions ``1'' and ``2''.

The top right panel of Fig.\,\ref{fig:ragrad} shows the results of a similar calculation but taking into account the Hanle effect of a micro-structured magnetic field of 10 G with an inclination $\theta_B=65^\circ$. The $Q/I$ profile is now skewed towards the blue part of the spectrum. This effect can be understood by considering the $Q_{ij}/I$ components. In contrast to the non-magnetic case (see left panel of Fig.~\ref{fig:ragrad}), transition ``2'' now dominates the $Q/I$ profile because the alignment of the $3d_{5/2}$ level is strongly depolarized by the Hanle effect. Component ``2'' contributes mainly in the blue part of the line. The contribution from transition ``1'' in the red part of the profile is smaller and transition ``3'' decreases the $Q/I$ signal (note that transitions ``2'' and ``3'' have the same upper level $3d_{3/2}$, which is less depolarized than the upper level of transition ``1''). Consequently, the whole profile is slightly skewed towards the blue wavelengths. For increasingly stronger fields, the importance of the difference between the critical field values of $3d_{3/2}$ and $3d_{5/2}$ becomes increasingly negligible and the relative $Q_{ij}/I$ contributions corresponding to the non-magnetic model are eventually restored for field strengths larger than the \Ha\/ Hanle saturation field.

From the previous discussion related to the top panels of Fig.\,\ref{fig:ragrad} we point out that neither the emergent $Q$ profile nor $Q/I$ show any significant asymmetry, even though the individual components are asymmetric. The reason is that the shapes of the four $Q_{ij}/I$ components of \Ha\/ are smooth ``skewed gaussians'' located around the line center, whose superposition gives rise to a rather smooth $Q/I$ profile without any noteworthy asymmetry. 

However, something interesting happens if we have magnetic field gradients in the line-core formation region of the  \Ha\/ line (see the bottom panels of Fig.\,\ref{fig:ragrad}). The above-mentioned second conclusion provides the clue for understanding why the emergent $Q/I$ profile shows now a line core asymmetry (LCA). In agreement with our numerical experiments, the LCA can be created by a modification of the $Q_{ij}/I$ profiles of some components (mainly the ``1'' and ``2'' transitions) induced by a spatially varying magnetic field. The superposition of the blended linear polarization profiles can then give rise to a sizable LCA whose shape depends sensitively on the spatial distribution of the chromospheric magnetic field. The interpretation of the $Q/I$ profile observed by \citet{gandorfer00} in the \Ha\/ line led us to suggest that there is a significant and abrupt magnetization in the upper chromosphere of the quiet Sun \citep{stepanjtb10asym}.


\section{Concluding comments}
\label{sec:concl}

In this first paper of a series on the scattering polarization of hydrogen lines in weakly magnetized stellar atmospheres we have considered the case of an exponentially stratified isothermal model atmosphere, which has allowed us to gain physical insight  on the atomic level polarization produced by optical pumping processes and the operation of the Hanle effect in the \La\/, \Lb\/ and \Ha\/ lines. In order to be able to investigate this problem taking into account multilevel radiative transfer effects, the overlapping of the fine-structure transitions contributing to each spectral line and the impact of collisions and the Hanle effect, we have applied the efficient numerical methods outlined in Appendix~\ref{app:methods} and the method explained in Appendix~\ref{app:microturb} when dealing with micro-structured magnetic fields.

It may be useful to emphasize the following points:
\begin{itemize}
\item The fractional scattering polarization of \La, which is solely due to the selective emission of polarization components resulting from the atomic alignment of the $2p_{3/2}$ level, can be safely calculated by solving the radiative transfer problem assuming a three-level atomic model, with the $1s_{1/2}$ ground level and the $2p_{1/2}$ and $2p_{3/2}$ upper levels, taking into account the overlapping between the two possible radiative transitions and neglecting collisions between the $n=2$ fine-structure levels (if the perturber's density $N_{\rm pert}{\lesssim}10^{11}\,\,{\rm cm}^{-3}$). Via the Hanle effect the scattering polarization of the \La\/ line is sensitive to inclined magnetic fields with strengths $10\,{\rm G}\lesssim B\lesssim 250$\,G, approximately.
\item Reliable calculations of the fractional scattering polarization of \Lb\/, which is solely due to the selective emission of polarization components resulting from the atomic alignment of the $3p_{3/2}$ level, require taking into account the fine-structure levels of the first three hydrogen $n$-levels, including the collisional transfer of aligment between the $n=3$ levels. The Hanle effect sensitivity of \Lb\/ lies between 3\,G and 80\,G, approximately.
\item Modeling the scattering polarization of \Ha\/, which is in general the result of the contributions from five polarizing blended transitions, requires taking into account also the fine-structure levels of the first three hydrogen $n$-levels, including the collisional transfer of aligment between the $n=3$ levels. In solar-like atmospheres the emergent linear polarization has nothing to do with lower-level polarization, because in the atmospheric region where the \Ha\/ polarization is produced the atomic alignment of the $2p_{3/2}$ is negligible. Interestingly, the wavelength separation between the overlapping \Ha\/ transitions makes possible a wavelength-dependent Hanle effect which may give rise to asymmetric $Q/I$ profiles when in the presence of magnetic field gradients. The scattering polarization of \Ha\/ reacts to fields between 1\,G and 50\,G, approximately. 
\end{itemize}

In a forthcoming Letter and in the next paper of this series we will describe in great detail the results of our calculations in semi-empirical models of the solar atmosphere, with predictions on the scattering polarization signals produced by atomic level polarization and the Hanle effect.


\acknowledgments
We are grateful to Luca Belluzzi (IAC) and Egidio Landi Degl'Innocenti (University of Firenze) for carefully reviewing the paper and for suggesting various useful improvements. Financial support by the Spanish Ministry of Science and Innovation through project \mbox{AYA2010--18029} (Solar Magnetism and Astrophysical Spectropolarimetry) and CONSOLIDER INGENIO \mbox{CSD2009-00038} (Molecular Astrophysics: The Herschel and Alma Era) is gratefully acknowledged.

\appendix


\section{The numerical method of solution}
\label{app:methods}

In order to obtain the self-consistent solution of the hydrogen scattering polarization problem taking into account the Hanle effect in multilevel atomic models we have applied a computer program \citep{thesis08} based on an accurate formal solver of the radiative transfer equation \citep{jtb03} and on a generalization of a very efficient iterative scheme \citep{jtb99spw} to the case of overlapping transitions. 

The Stokes-vector transfer equation (see  Eq.~\ref{eq:rte}) can be also written as follows:
\begin{equation}
{{d}\over{d{\tau}}}{\vec I}\,=\,{\vec I}\,-\,{\vec S}_{\rm eff},  
\end{equation}
where $d\tau=-{\eta_I}\,ds$ and the effective source-function vector $\vec S_{\rm eff}=\vec S-\vec K'\vec I$ being $\vec K'={\vec K}/{\eta_I}-\vec 1$ (with $\vec 1$ the unit matrix and $\vec S={\vec\epsilon}/{\eta_I}$). Therefore, the formal solution of the Stokes-vector transfer equation can be expressed as
\begin{equation}
\vec I=\vec\Psi[\vec S-\vec K'\vec I]+\vec T\,,
\label{eq:fs1}
\end{equation}
where $\vec\Psi$ denotes an operator which reduces to that introduced by \citet{rybicki92} when atomic level polarization is neglected. In contrast to the standard $\vec\Lambda$ operator \citep{rybicki91}, $\vec\Psi$ acts on the emission coefficients of a line rather than on its source function. This enables us to treat more easily the present complex case of hydrogen spectral lines composed of multiple overlapping transitions. In Eq.~(\ref{eq:fs1}), we have used the modified propagation matrix, $\vec K'=\vec K/\eta_I-\vec 1$, and the Stokes source function vector $\vec S=\vec\epsilon/\eta_I$. The coefficient $\eta_I$ is the standard absorption coefficient of the radiative transfer theory, which resides on the diagonal of $\vec K$. The vector $\vec T$ denotes the contribution resulting from the Stokes vector illuminating the boundary of the integration domain.

Since $\vec\Psi$ is a linear integral operator, we can formally rewrite Eq.~(\ref{eq:fs1}) in the form
\begin{equation}
\vec I=\vec\Psi[\vec S]-\vec\Psi[\vec K'\vec I]+\vec T\,,
\label{eq:fs2}
\end{equation}
and following the ideas of operator splitting, we write $\vec\Psi=\vec\Psi^*+(\vec\Psi-\vec\Psi^*)$ and we use the diagonal of the exact $\vec\Psi$ operator for the approximate operator $\vec\Psi^*$ \citep[c.f.,][]{rybicki92}. We label the ``old'' quantities (i.e., those calculated using the density matrix elements calculated in the previous iteration) by the superscript ``+'' and we replace the first term in the right-hand side of Eq.~(\ref{eq:fs2}) by
\begin{equation}
\vec\Psi[\vec S]\to\vec\Psi[\vec S^+]+\vec\Psi^*[\vec S-\vec S^+]\,.
\label{eq:split}
\end{equation}
The quantities in the remaining terms are taken from the previous iteration, hence we can rewrite the iterative scheme as
\begin{equation}
\vec I=\vec I^+ +\vec\Psi^*[\vec S-\vec S^+]\,.
\label{eq:fsit}
\end{equation}
The source function vectors in the current and in the previous iteration are defined as
\begin{equation}
\vec S=\frac{\vec\epsilon}
{\eta_I^+}\,,\label{eq:souceS}\qquad
\vec S^+=\frac{\vec\epsilon^+}
{\eta_I^+}\,.\label{eq:sourceSplus}
\end{equation}
Note that the $\vec\Psi$-operator only acts linearly on the emission coefficient and that the highly non-linear dependence on the absorption coefficient is suppressed \citep[see][for an analogous strategy]{jtb99spw,mansosainz03}. In this way the formal solution of the Stokes-vector transfer equation at a given frequency can be expressed as a linear function of the density matrix elements of the upper levels of the active transitions.

\begin{figure}
\epsscale{0.7}
\plotone{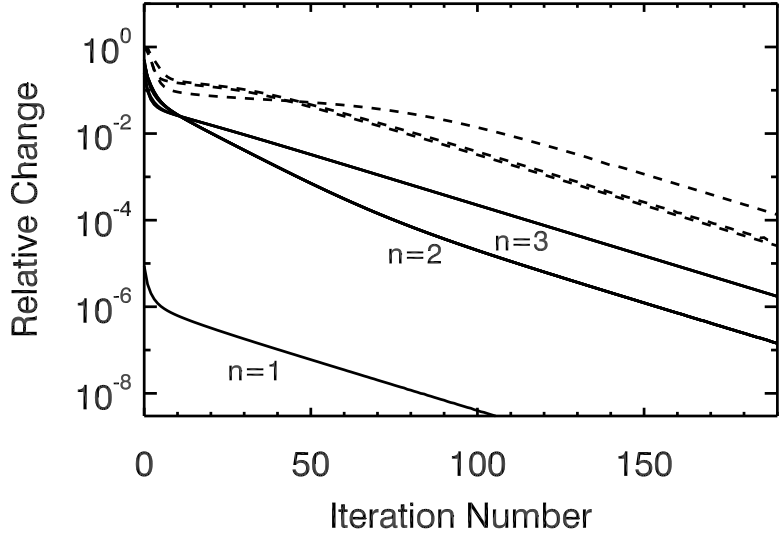}
\epsscale{1.0}
\caption{
Relative change versus iteration number $it$ defined as $|\vec\rho^K_Q(it)-\vec\rho^K_Q(it-1)|/|\vec\rho^K_Q(it)|$,  where $|\vec\rho^K_Q(it)|=\sqrt{\sum_i |\rho^K_Q(i,it)|^2}$ is the euclidian 2-norm of the vector formed by the density matrix multipoles at all spatial grid points $i$ in the unmagnetized model atmosphere. The solid lines correspond to the $\rho^0_0$ quantities, while the dashed lines to $\rho^2_0$. Note that the relative changes corresponding to the fine structure levels pertaining to the same level $n$ are very similar.
}
\label{fig:converg}
\end{figure}

The next step is to use Eq.~(\ref{eq:fsit}) in the calculation of the radiation field tensors of Eq.~(\ref{eq:jkqij}) and to use the resulting expressions in the statistical equilibrium equations (SEE). The source function vectors $\vec S$ and $\vec S^+$ have to be explicitly expressed in terms of the density matrix elements of the upper levels of the radiative transitions. For the sake of simplicity, we denote in the following the multipole components of the density matrix by $\rho_k$. The population multipoles $\rho^0_0$ are the convergence-driving quantities. We will identify them by a bar over the index, for example $\rho_{\pruh m}$. Using only the terms related to the $J^0_0$ tensor,\footnote{These are responsible for the convergence rate.} the preconditioning scheme of SEE is of the form
\begin{equation}
J^0_0(ij)\rho_k\to J^0_0(ij)^+\rho_k+\sum_{\pruh m}L_{ij\pruh m}(\rho_{\pruh m}\rho_k-\rho_{\pruh m}^+\rho_k)\,.
\label{eq:it1}
\end{equation}
The summation runs over all the population multipoles $\rho_{\pruh m}$ of the levels emitting at the frequencies of the $i$--$j$ transition under consideration. The $\rho_k$ elements correspond to the lower-level multipole absorbing at the frequencies of that transition. The coefficients $L_{ij\pruh m}$ can be straightforwardly calculated from the approximate $\vec\Psi^*$ operator while doing the formal solution via the DELOPAR method proposed by \citet{jtb03}. The advantage of using the diagonal approximate operator is that the SEE remain local and it is easy to evaluate $L_{ij\pruh m}$ from the interpolation coefficients of the DELOPAR formal solution method of the transfer equation (see \citet{mansosainz02}, and \citet{thesis08} for the explicit form of the $L_{ij\pruh m}$ coefficients of the multi-level atom).

Whereas the second term in the bracket in the right-hand side of Eq.~(\ref{eq:it1}) is linear in the ``new'' density matrix element, the first product is not. In order to preserve linearity of SEE, the term $\rho_{m}\rho_k$ has to be replaced by $\rho_{m}\rho_k^+$ \citep[see the analogous linearization by][]{jtb99spw,mansosainz03}. Finally, we end up with the iterative scheme
\begin{eqnarray}
J^0_0(ij)\rho_k&\to& J^0_0(ij)^+\rho_k+\sum_{\pruh m}L_{ij\pruh m}(\rho_{\pruh m}\rho_k^+-\rho_{\pruh m}^+\rho_k)\,,\\
J^K_Q(ij)\rho_k&\to& J^K_Q(ij)^+\rho_k\,,\qquad {\rm for}\qquad K>0\,.
\end{eqnarray}

As an example, consider the \La\/ line which consists of two overlapping transitions: $2p_{1/2}$--$1s_{1/2}$ and $2p_{3/2}$--$1s_{1/2}$. The product $J^0_0(2p_{3/2}\,-\,1s_{1/2})\rho^0_0(1s_{1/2})$ will thus become
\begin{eqnarray}
J^0_0(2p_{3/2}-1s_{1/2})^+\rho^0_0(1s_{1/2})+L_{2p_{3/2}-1s_{1/2},\,2p_{1/2}}\nonumber\\
\times[\rho^0_0(2p_{1/2})\rho^k_q(1s_{1/2})^+
-
\rho^0_0(2p_{1/2})^+\rho^k_q(1s_{1/2})]\nonumber\\
+L_{2p_{3/2}-1s_{1/2},\,2p_{3/2}}\nonumber\\
\times[\rho^0_0(2p_{3/2})\rho^k_q(1s_{1/2})^+
-
\rho^0_0(2p_{3/2})^+\rho^k_q(1s_{1/2})]\,.
\end{eqnarray}
The resulting SEE, linear in all $\rho^k_q$, can be solved using the standard methods of numerical linear algebra. Figure~\ref{fig:converg} illustrates the convergence behavior of the iterative method for the density matrix elements corresponding to the hydrogen model atom of Fig. 1, considering radiative transfer effects in an isothermal model atmosphere.

In the case of negligible polarization, the previously described technique is equivalent to the full-preconditioning strategy of \citet{rybicki92}. For multilevel atomic models without overlapping transitions, the method is equivalent to the one applied by \citet{mansosainz03}. We note that an analogous iterative scheme can be used for modeling atomic lines with even more complicated structure, such as those involving levels in the incomplete Paschen-Back regime or levels with hyperfine structure  \citep{stepanspw5,thesis08}.


\section{Collisions}
\label{app:colls}

We treat collisions within the framework of the irreducible components of the density matrix \citep[e.g.,][]{ll04}. The collisional rates of transfer and relaxation of the components of the atomic density matrix due to interactions with the ambient thermal particles have the form $C^{(k)}(n'\ell'j',n\ell j)$. In this paper, we do not explicitly distinguish between inelastic (i.e., exciting) and superelastic (i.e., de-exciting collisions). The $C^{(0)}$ rates are responsible for transfer and relaxation of the level populations and they are equal to the familiar collisional rates of the unpolarized theory, which will be denoted by $C_{n\ell j\to n'\ell'j'}$ in the following. The $C^{(2)}$ rates are responsible for transfer of alignment between the atomic levels. The effect of elastic depolarizing collisions, quantified by the $D^{(k)}$ rate \citep[see Eqs. 7.101 in][]{ll04}, is negligible for the case of hydrogen levels at the densities and temperatures typical of the upper solar chromosphere. The depolarization of the atomic levels is almost exclusively due to the ``weakly inelastic'' transitions $n\ell j\to n(\ell\pm 1)j'$ between the close-lying fine structure levels of a given level $n$, which are induced mainly by collisions with the ambient protons.


\subsection{Inelastic collisions}
\label{sapp:inel}

\begin{deluxetable}{cccc}
\tablewidth{150pt}
\tablecaption{Excitation and de-excitation collisional rates $[\mathrm{ s^{-1}\,cm^{-3} } ]$ for $T=10^4$\,K, $n_{\rm e}=4\times 10^{10}\,{\rm cm^{-3}}$. Data obtained from \citet{przybilla04}.\label{tab:crates}}
\tablehead{\colhead{$n_l$} & \colhead{$n_u$} & \colhead{$C_{n_l\to n_u}$} & \colhead{$C_{n_u\to n_l}$}}
\startdata
1   &    2   &    1.01(-2)    &    3.49(+2)\\
1   &    3   &    3.62(-4)    &    5.01(+1)\\
2   &    3   &    1.63(+3)    &    6.48(+3)\\
\enddata
\tablecomments{$a(b)$: $a\times 10^{b}$.}
\end{deluxetable}


A careful consideration of the collisional processes is needed in the unpolarized non-LTE problem and even more in the present case of scattering polarization.

The role of inelastic collisions with thermal electrons is indeed similar in both kinds of models. Unfortunately, the inelastic excitation cross-sections $\sigma(n\to n',E)$ by electrons are typically poorly known. Hydrogen cross-sections are notoriously difficult to calculate because of the existence of high number of levels and resonances. From author to author, the uncertainty of cross-sections and, consequently, the collisional rates, can vary by tens of percent \citep[see][and references therein]{przybilla04}.

If one considers the fine structure of the levels, the situation with the cross-section data is even more uncertain. To our knowledge, only the excitation cross-sections for terms from the ground level, $\sigma(1s\to nl,E)$, are currently available from laboratory measurements \citep[e.g.,][]{janev93}. At low energies, which play the most important role in the solar atmosphere, these transitions do not obey the strong selection rules of the dipolar optical transitions, $|\Delta\ell|=1$, and the non-dipolar excitations such as $1s\to 2s$, $1s\to 3s$, $1s\to 3d$, have rates comparable to the dipolar ones. No experimental data for the cross-sections between the fine structure levels of the  subordinate lines are available and we have to rely on theoretical calculations.

The approach we have chosen in this work is the following. We have adopted the most reliable data of \citet{przybilla04} available nowadays for the excitation rates, $C_{n_l\to n_u}$, in a wide range of temperatures and we have used the approximation that the collisional rates between the Zeeman sublevels do not depend on the orbital angular momentum, nor on the magnetic quantum number of the $n_l\to n_u$ transition under consideration. The normalization condition implies the following equation for the population, $N_{n_u}$, of the upper Bohr level,\footnote{For the sake of simplicity we only explicitly write the rates between two $n$-levels.}
\begin{equation}
\frac{dN_{n_u}}{dt}=\sum_{j_l}\sum_{j_u} C_{j_l\to j_u}N_{j_l}
-\sum_{j_u}\left(\sum_{j_l}C_{j_u\to j_l}\right)N_{j_u}\,,
\end{equation}
It is easy to show that the excitation collisional rate between the two fine-structure levels is 
\begin{equation}
C_{j_l\to j_u}=\frac{g_{j_u}}{g_{n_u}}C_{n_l\to n_u}\,,
\label{eq:excrate}
\end{equation}
and the relaxation rate 
\begin{equation}
C_{j_u\to j_l}=\frac{g_{j_l}}{g_{n_u}}C_{n_l\to n_u}{\rm exp}\left(\frac{E_{n_u}-E_{n_l}}{kT}\right)\,,
\label{eq:dexrate}
\end{equation}
where $g_{j_l}$ ($g_{j_u}$) and $g_{n_u}$ is the statistical weight of the lower (upper) fine structure level and of the full $n$-level, respectively.\footnote{It is $g_j=2j+1$ and $g_n=2n^2$. We have neglected the energy differences among the different fine structure levels pertaining to the same $n$-level in comparison to the separation between  different $n$-levels.} Note that the rates (\ref{eq:excrate}) and (\ref{eq:dexrate}) connect the fine-structure levels whose populations $N_j$ are related to the zero-rank of the density matrix element $\rho^0_0(nlj)=N_j/\sqrt{2j+1}$ (see \S\ref{ssec:densm}). Reformulation of the previous expressions in the framework of irreducible tensorial operators is straightforward \citep[see Sect.~7.13 of][for details]{ll04}. The collisional rates between different $n$-levels that have been used in this paper are tabulated in Table~\ref{tab:crates}.

We use the approximation described above in order to guarantee that the collisional rates of different lines are consistent. We have numerically verified that replacing the $C_{1s_{1/2}\to n_u \ell_u j_u}$ rates by those calculated from the available excitation cross-sections $\sigma(n_l \ell_l\to n_u \ell_u)$ of \citet{janev93} does not modify the Lyman line intensities nor the polarization by more than a few percent of their amplitude. Such error is quite acceptable taking into account the uncertainty of the available collisional data. It is also worth to mention that the Balmer lines are even less affected by a modification of the collisional rates \citep[cf.][]{przybilla04}, especially in semi-empirical chromospheric models where photoionization dominates over the collisional transitions.


\subsection{Collisions connecting nearby $j$-levels}
\label{sapp:depol}

\begin{figure*}
\plottwo{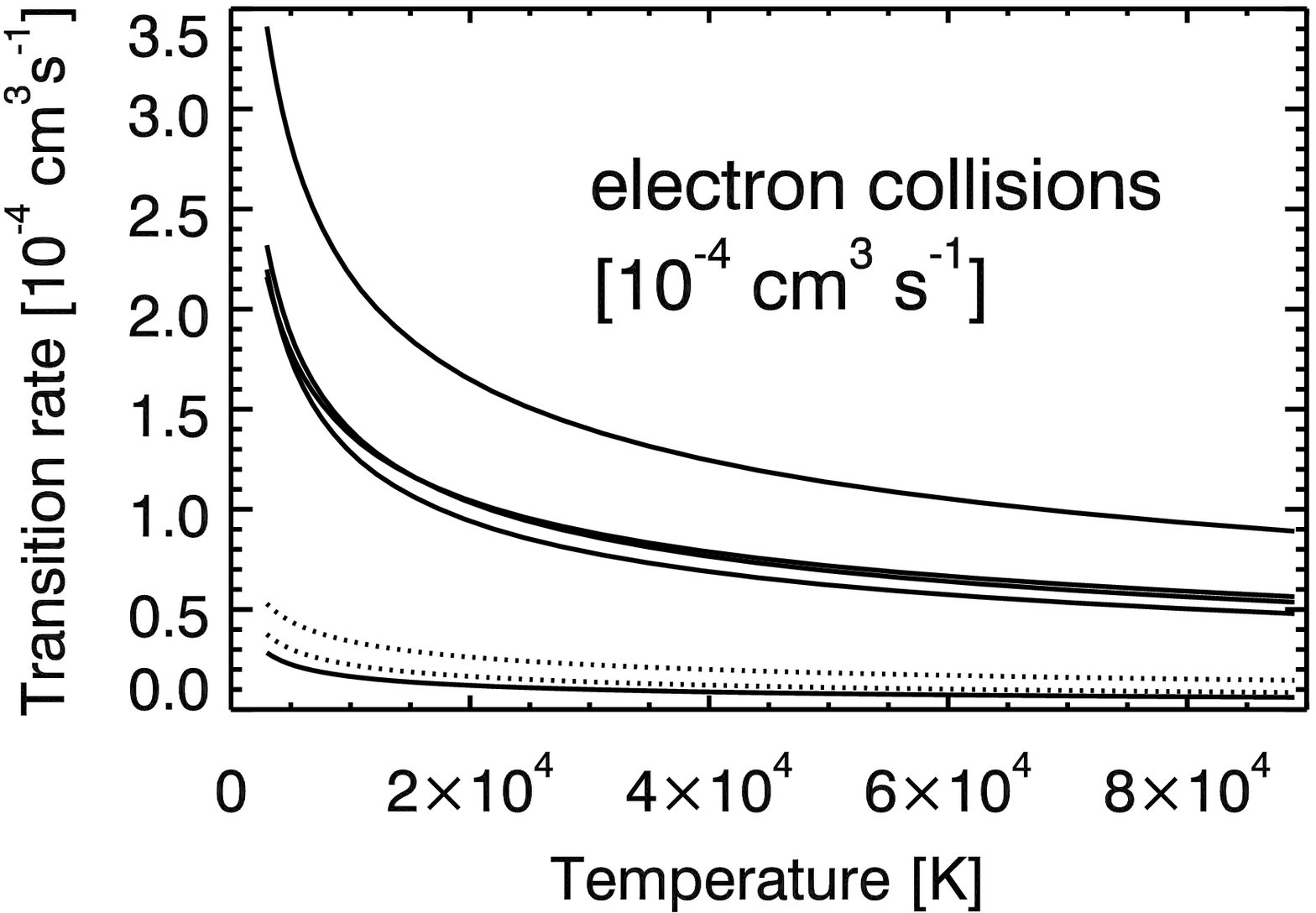}{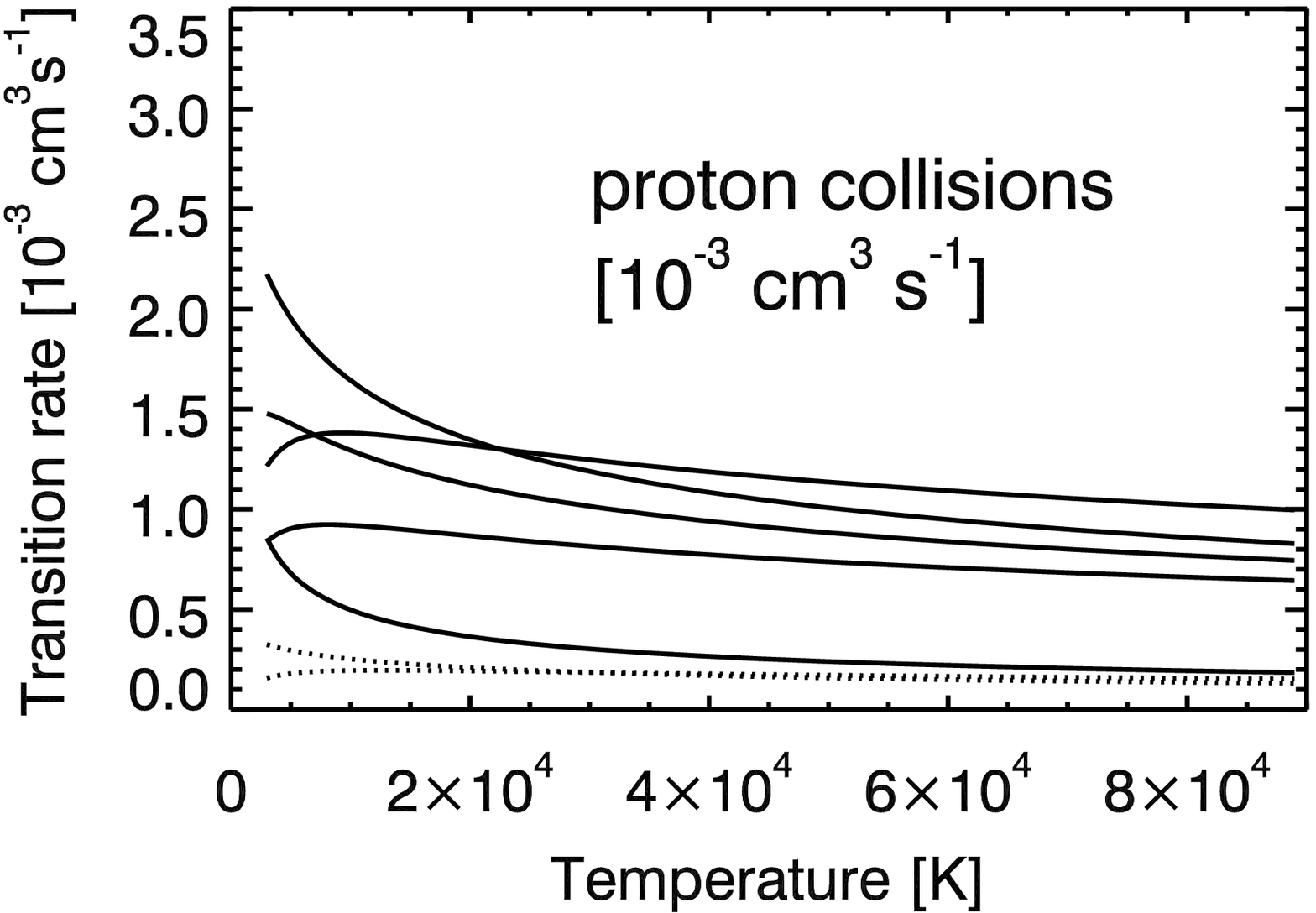}
\caption{
Collisional transition rates of the population transfer per perturber as a function of the plasma temperature for all dipolar transitions $2\ell j\to 2(\ell\pm 1)j'$ ({\it dashed lines}) and $3\ell j\to 3(\ell\pm 1)j'$ ({\it solid lines}). The rates have been calculated for a proton and electron density of $10^{10}\,{\rm cm^{-3}}$. The curves show the rates calculated using the semi-classical perturbation method of \citet{sahal96} between 3\,000\,K and 90\,000\,K, which is based on the impact approximation (see \cite{bommier91} for a discussion  on the validity conditions.  {\it Left}: collisions with electrons. {\it Right}: collisions with protons. Note the different scales in the vertical axes.
}
\label{fig:depolep}
\end{figure*}

While the intensity spectrum of the hydrogen lines can be modeled neglecting the fine structure of the levels, the fine structure and the effects of depolarizing collisions with different atmospheric species (electrons, ions, etc.) have to be taken into account when it comes to modeling the linear polarization produced by atomic level polarization.

The levels of neutral hydrogen show a high degree of accidental (quasi)degeneracy. The fine-structure levels $n\ell j$ and $n\ell'j'$ pertaining to the same Bohr level $n$ are very close to each other, with a typical separation of $10^{-7}$ to $10^{-5}$\,eV for $n=3$. It can be shown that, at the typical densities of the upper solar chromosphere, collisional transitions induced by electrons and mainly by protons, both having a long-range interaction potential, play a significant role on the depolarization process, the collisional rates due to protons being about one order of magnitude higher than those due to electrons \citep{bommier86b,sahal96}. Collisions with neutral hydrogen \citep{griem65,barklem00}, which play a significant role in depolarizing other lines of the second solar spectrum, can be safely neglected in the case of hydrogen lines at the considered densities. Collisions with protons and to a lesser extent with electrons also play a significant role in broadening of the hydrogen lines via the Stark effect \citep[e.g.,][]{stehle96}.\footnote{Note that collisional depolarization and the Stark broadening are intimately related processes \citep[see][and references therein]{sahal09}.} Here we only consider the depolarizing role of collisions, with protons and electrons, connecting nearby $j$-levels pertaining to each $n$-level. In general, the depolarizing effect of collisions with He\,{\sc ii} ions should also be taken into account, although its significance is expected to be much smaller than that due to collisions with protons \citep[cf.,][]{bommier86b}. All these additional ingredients will be considered in subsequent papers.

Collisions of hydrogen atoms with perturbers having a Maxwellian velocity distribution tend to produce partial or full equilibration of the populations of the Zeeman sublevels, i.e., to a reduction of the atomic level polarization. Since the thermal energy of the perturbers ($\sim 1$\,eV) is by several orders of magnitude larger than the transition thresholds, the most efficient transitions are the dipolar ones, $n\ell j\to n(\ell\pm 1)j'$. Strictly speaking, these transitions are inelastic but they actually have a depolarizing effect and they are also responsible for transfer of alignment between the fine structure levels. These collisions reduce the lifetime of the excited levels. Consequently, the width of the fine structure levels increases \citep[see Tables~4 and 5 of][]{sahal96}. If the density of protons becomes too high (roughly about $10^{13}\,{\rm cm^{-3}}$ for the case of $n=3$), the levels start to partially overlap and the quantum coherences between them should be taken into account in calculations of the polarization produced by radiation scattering. Moreover, the impact approximation breaks at high densities \citep{stehle83}. But it is worth mentioning that at such a high density of perturbers the atomic polarization of $n=3$ level would be practically destroyed (cf. Fig.~\ref{fig:qinenp}) regardless of the theory used. In fact, such high densities are only expected in the deep layers of the atmosphere (where also H\,{\sc i}\,--\,H\,{\sc i} collisions have to be taken into account) and affect  mainly the line wings. In the upper chromosphere the fine structure levels can be considered as being well separated.

The dipolar cross-sections can be calculated using the semi-classical perturbation theory (see \citet{seaton62}, \citet{bommier86b}, \citet{bommier91}, and mainly \citet{sahal96} for a detailed derivation of the collisional rates in the impact approximation\footnote{Note that the authors use a notation slightly different from that of \citet{ll04}.}). Using this approach, we have calculated the multipolar components of the collisional transition rates $C^{(0)}$ and $C^{(2)}$ at each atmospheric height and for every allowed transition. The collisional rates $C^{(0)}$ by electrons and protons (normalized to the unit density of the perturbers) between the $j$-levels of $n=2$ and 3 calculated using the theory of \citet{sahal96} can be found in Fig.~\ref{fig:depolep}. Multiplication of these quantities by the actual perturbers density gives the standard collisional rate in units of ${\rm s}^{-1}$. When compared to the inverse lifetime of the level, it quantifies the degree of collisional depolarization of the level. As seen in Fig.~\ref{fig:depolep}, collisions with protons are about one order of magnitude more efficient than those with electrons. It is also clear that the $j$-levels of $n=3$ are depolarized significantly more than those of level $n=2$ \citep[cf. Table~2 of][]{sahal96}. This is why the \La\/ scattering polarization is less affected by the depolarizing collisions than \Ha\/ and \Lb\/ (see right panels of Fig.~\ref{fig:clv-nm2}). Using the data of the right panel of Fig.~\ref{fig:depolep} reveals the reason why collisional depolarization of \Ha\/ becomes critical around a proton density $n_{\rm p}\approx 10^{11}\,{\rm cm^{-3}}$. Using $10^{-3}\,{\rm cm^3\,s^{-1}}$ as a typical value of the collisional rate per perturber we obtain $10^{11}\,{\rm cm^{-3}}\times 10^{-3}\,{\rm cm^3\,s^{-1}}=10^{8}\,{\rm s^{-1}}$ which is comparable to the inverse radiative lifetime of the upper levels of \Ha.

Another important phenomenon related to the depolarizing effect of such collisions connecting nearby $j$-levels is a modification of the critical Hanle field of any given level. Since the value of the critical Hanle field, $B_H$, increases with the inverse lifetime of the level (cf. Eq.~\ref{eq:bhcritical}) the higher collisional rates make the level less sensitive to the magnetic field.  As mentioned above, the collisional rates between the $j$-levels of level $n=3$ at a typical chromospheric density can be of the order of the inverse radiative lifetime. Consequently, the critical Hanle field can easily be doubled by collisional quenching. As a result, the amplitude of the emergent polarization is smaller due to collisional depolarization and the onset of the Hanle depolarization is shifted towards higher intensities of the magnetic field.


\section{Multilevel atom in the presence of a micro-structured magnetic field with random azimuth}
\label{app:microturb}

A well-known strategy for modeling the Hanle depolarization of various solar spectral lines is the microturbulent magnetic field approximation in which the spatial scale of the field variation is smaller than the mean free path of the line-center photons \citep[e.g.,][]{stenflobook,ll04}. Such a model has been developed further by several authors for the case of a two-level atom model \citep{jtb99}, for multilevel atoms in the Hanle effect saturation regime \citep{mansosainz06} and in the optically thin regime \citep{belluzzi07}. In this appendix, we consider instead the case of a single-valued micro-structured magnetic field having a fixed (but in general height-dependent) inclination $\theta_B$ and a uniformly distributed azimuth $\chi_B$ at microscopic scales. We show how to handle easily such a micro-structured magnetic field when solving the non-LTE problem of the second kind for a multilevel atom in a cylindrically symmetric atmosphere with any degree of radiation field
  anisotropy.

Let us consider a statistical ensemble of multi-level atoms in a magnetized plasma. We suppose that the magnetic field azimuth $\chi_B$ is random with a uniform distribution while both the inclination $\theta_B$ and the intensity $B$  are fixed at any given point. Physically, this approximation corresponds to a situation in which the magnetic field varies rapidly on a geometrical scale much smaller than the mean free path of the photons. The density matrix of an ensemble of $N$ non-interacting atoms is simply an average of the density matrices $\rho_i$ of the individual atoms \citep{fano57},
\begin{equation}
\rho=\frac 1N\sum_i\rho_i\,.
\end{equation}
For the sake of simplicity, we do not explicitly consider the time dependence of the quantities. Every single atom is under the influence of a deterministic magnetic field. The density matrix of a subset of atoms being embedded in the magnetic field with azimuth $\chi_B$ will be denoted by $\rho(\chi_B)$. The density matrix of the whole ensemble is a superposition of the density matrices of these sub-ensembles weighted over the field distribution. We make a transition from discrete to continuous quantities and express the density matrices in the basis of irreducible tensorial operators in the atmospheric reference frame with the quantization axis $Z$ being parallel to the vertical axis of the atmosphere. Assuming that the distribution of the field's azimuth $\chi_B$ is uniform in the $[0,2\pi)$ interval, we can calculate the local density matrix
\begin{equation}
\left[\rho^K_Q\right]_{\rm A}=
\frac 1{2\pi}\int_0^{2\pi}d\chi_B\,\left[\rho^K_Q(\chi_B)\right]_{\rm A}
\,,\label{eqA3}
\end{equation}
where the subscript A refers to quantities expressed in the atmospheric reference frame.

Rotation of an ``old'' reference frame to a ``new'' one by an angle $\chi$ around the $Z$ axis leads to the following transformation of the density matrix elements \citep{ll04}
\begin{equation}
[\rho^K_Q]_{\rm new}=\sum_P[\rho^K_P]_{\rm old}\mathcal{D}^K_{PQ}(R)\,,
\end{equation}
where $\mathcal{D}^K_{PQ}(R)$ is the rotation matrix corresponding to the rotation of the reference frame parametrized by the Euler angles, $R=(0,0,\chi)$. In our case, the rotation matrix elements are simply $\mathcal{D}^K_{PQ}(R)=e^{-i\chi Q}\delta_{PQ}$; thus
\begin{equation}
[\rho^K_Q]_{\rm new}=e^{-i\chi Q}[\rho^K_Q]_{\rm old}\,.\label{eqA2}
\end{equation}
Using this transformation in Eq.~(\ref{eqA3}) leads to
\begin{equation}
\left[ \rho^K_Q \right]_{\rm A}=
\frac 1{2\pi}\int_0^{2\pi}d\chi_B\,\left[\rho^K_Q(\chi_B)\right]_{\rm A}=
\frac 1{2\pi}\int_0^{2\pi}d\chi_B\,e^{-i\chi_BQ}\left[\rho^K_Q(\chi_B)\right]_{\chi_B}\,,\label{eqA5}
\end{equation}
where $[\rho^K_Q(\chi_B)]_{\chi_B}$ is expressed in the reference frame in which the magnetic field has zero azimuth.

Taking into account that any one-dimensional model atmosphere permeated by such a micro-structured field implies that the radiation field is cylindrically symmetric with respect to the vertical, it follows that the system is cylindrically symmetric as a whole and that $[\rho^K_Q(\chi_B)]_{\chi_B}\equiv\rho^K_Q(0)$, being a solution of the statistical equilibrium equations in the case of zero-azimuth magnetic field, is independent of $\chi_B$. The integral of the exponential in Eq.~(\ref{eqA5}) produces the Kronecker $\delta_{Q0}$. Finally we have
\begin{equation}
\left[ \rho^K_Q \right]_{\rm A}=\rho^K_Q(0)\delta_{Q0}\,.\label{eq:Af}
\end{equation}

Eq.~(\ref{eq:Af}) gives a simple strategy for the numerical solution of the problem. At every point in the atmosphere, the statistical equilibrium equations have to be constructed as if the magnetic field was deterministic with azimuth $\chi_B=0$.\footnote{In fact, the azimuth can be arbitrary because the $\rho^K_0$ components do not depend on it.} Solution of the equations gives the $\rho^K_Q(0)$ density matrix elements. The desired density matrix is then given by Eq.~(\ref{eq:Af}), i.e., by the $\rho^K_0(0)$ elements. These elements can then be used for calculating the local radiative transfer coefficients and the RTE can be solved to obtain the emergent fractional polarization profile.


\bibliography{bibs}{}

\end{document}